\documentclass[fleqn,usenatbib]{mnras}

\usepackage[T1]{fontenc}

\DeclareRobustCommand{\VAN}[3]{#2}
\let\VANthebibliography\thebibliography
\def\thebibliography{\DeclareRobustCommand{\VAN}[3]{##3}\VANthebibliography}

\usepackage{ae,aecompl}

\usepackage{graphicx}	
\usepackage{amsmath}	
\usepackage{amssymb}	
\usepackage{times}
\usepackage{newtxtext,newtxmath}

\newcommand{\bs}{\boldsymbol}
\newcommand{\percent}{\,\mathrm{per\,cent}}
\defcitealias{Matsunaga2009}{M09}
\usepackage[table]{xcolor}
\definecolor{lightgray}{gray}{0.9}
\definecolor{darkred}{rgb}{0.76, 0.23, 0.13}

\makeatletter
\newcommand\footnoteref[1]{\protected@xdef\@thefnmark{\ref{#1}}\@footnotemark}
\makeatother

\title[Mira variables in the nuclear stellar disc]{Mira variables in the Milky Way's nuclear stellar disc: discovery and classification}
\author[J. L. Sanders et al.]{Jason L. Sanders,$^{1,2}$\thanks{E-mail: jason.sanders@ucl.ac.uk (JLS)}
Noriyuki Matsunaga,$^{3}$
Daisuke Kawata,$^{4}$
Leigh C. Smith,$^{1,5}$
Dante Minniti,$^{6,7}$
\newauthor
and
Philip W. Lucas$^{5}$
\\
$^{1}$Department of Physics and Astronomy, University College London, London WC1E 6BT, UK\\
$^{2}$Institute of Astronomy, University of Cambridge, Madingley Road, Cambridge CB3 0HA, UK\\ 
$^{3}$Department of Astronomy, School of Science, The University of Tokyo, 7-3-1, Hongo, Bunkyo-ku, Tokyo 113-0033, Japan\\
$^{4}$Mullard Space Science Laboratory, University College London, Holmbury St. Mary, Dorking, Surrey, RH5 6NT, UK\\
$^{5}$Centre for Astrophysics, University of Hertfordshire, College Lane, Hatfield AL10 9AB, UK\\
$^{6}$Instituto de Astrof\'isica, Facultad de Ciencias Exactas, Universidad Andr\'es Bello, Fern\'andez Concha 700, Las Condes, Santiago, Chile\\
$^{7}$Vatican Observatory, Vatican City State, V-00120, Italy
}

\date{Accepted XXX. Received YYY; in original form ZZZ}

\pubyear{2022}

\begin{document}
\label{firstpage}
\pagerange{\pageref{firstpage}--\pageref{lastpage}}
\maketitle

\begin{abstract}
The properties of the Milky Way's nuclear stellar disc give crucial information on the epoch of bar formation. 
Mira variables are promising bright candidates to study the nuclear stellar disc, and through their period--age relation dissect its star formation history. 
We report on a sample of $1782$ Mira variable candidates across the central $3\times3\,\mathrm{deg}^2$ of the Galaxy using the multi-epoch infrared VISTA Variables in Via Lactea (VVV) survey. 
We describe the algorithms employed to select candidate variable stars and then model their light curves using periodogram and Gaussian process methods. 
By combining with WISE, 2MASS and other archival photometry, we model the multi-band light curves to refine the periods and inspect the amplitude variation between different photometric bands. 
The infrared brightness of the Mira variables means many are too bright and missed by VVV. 
However, our sample follows a well-defined selection function as expected from artificial star tests. 
The multi-band photometry is modelled using stellar models with circumstellar dust that characterise the mass loss rates. 
We demonstrate how $\gtrsim90$ per cent of our sample is consistent with O-rich chemistry. 
Comparison to period--luminosity relations demonstrates that the bulk of the short period stars are situated at the Galactic Centre distance. 
Many of the longer period variables are very dusty, falling significantly under the O-rich Magellanic Cloud and solar neighbourhood period--luminosity relations and exhibit high mass-loss rates of $\sim2.5\times10^{-5}M_\odot\,\mathrm{yr}^{-1}$.
The period distribution appears consistent with the nuclear stellar disc forming $\gtrsim8\,\mathrm{Gyr}$ ago although it is not possible to disentangle the relative contributions of the nuclear stellar disc and the contaminating bulge.
\end{abstract}




\section{Introduction}
In studies of the Milky Way, we are often interested in piecing together the series of events that resulted in what we observe today. In this way, we can study the Milky Way as a detailed exemplar galaxy in the cosmological context of star-forming galaxies across the Universe \citep{BHG,Barbuy2018}. One key component of the Milky Way is the bar(-bulge) \citep{BlitzSpergel1991,WeggGerhard2013}. The bar is an important dynamical driver in the Milky Way responsible for significant restructuring of both the stars and the gas within the disc. Knowledge of the time over which it has had a dynamical impact on the Milky Way is crucial to understanding the dynamical history of the entire Galaxy. 

One structure intimately linked to the formation of the Milky Way's bar-bulge is the nuclear stellar disc (NSD). The NSD is a flattened distribution of stars with radius $\sim250\,\mathrm{pc}$ \citep{Launhardt2002}, aspect ratio $\sim5:1$ \citep{Nishiyama2013,GallegoCano2020} and mass $\sim10^9M_\odot$ \citep{Sormani2022} that rotates at approximately $100\,\mathrm{km}\,\mathrm{s}^{-1}$ as confirmed through both radial velocity \citep{Lindqvist1992,Schoenrich2015,Matsunaga2015,Schultheis2021} and proper motion studies \citep{Shahzamanian2021}. The NSD sits between the larger scale Galactic bar/bulge and the nuclear stellar cluster \citep[see the review of][]{Schoedel2014}, and coincides with the central molecular zone (CMZ), a region of significant interstellar dust and gas \citep{Morris1996}.

Beyond the Milky Way, nuclear stellar discs are often observed in barred spiral galaxies \citep{Erwin2002,Pizzella2002,Gadotti2018,Gadotti2020}. A consistent picture for their formation has been built up from hydrodynamical simulations \citep[e.g.][]{Athanassoula1992}: once a bar forms in a disc galaxy, gas is readily funnelled along the bar  towards the centre of the galaxy where it can settle on central `x2' orbits forming nuclear gas rings. 
The gas then begins forming stars which approximately inherit the `x2' orbital geometry and so the resulting stellar population resembles a disc. This paradigm is supported by observations of external galaxies that show nuclear stellar discs are younger, more metal-rich and of a lower velocity dispersion than the surrounding bar stars \citep{Gadotti2020,Bittner2020}. In the Milky Way, this picture is supported by observations that the NSD and CMZ overlap spatially and kinematically \citep{Schoenrich2015,Schultheis2021}, gas is visibly being funnelled along the bar \citep{Hatchfield2021} and the CMZ is star-forming today \citep{Morris1996}.
\cite{Baba2019} have suggested this connection between the history of the NSD and the bar is a way to pin down the epoch of bar formation in the Milky Way. Their simulations demonstrated that the formation of a bar is rapidly followed by a $\sim1\,\mathrm{Gyr}$ long intense period of star formation that forms the NSD. The oldest NSD stars then give the bar's formation time. Note that there are studies of the age of stars within the Galactic bar-bulge \citep[e.g.][]{Bovy2019,Hasselquist2020} but crucially the dynamical age of the bar can be quite different from the age of the bar stars.

There have been relatively few studies of the detailed star formation history of the NSD. \cite{Figer2004} argued from Hubble Space Telescope photometry that the star formation history of the NSD was quite continuous over time particularly when compared to the Galactic bulge fields that were more consistent with ancient bursts of star formation. \citet{Schultheis2020,Schultheis2021} have similarly demonstrated that the metallicity distribution of NSD stars differs from that of the NSC and the Galactic bulge, giving further evidence of its separate formation channel (and possibly epoch). The more extended GALACTICNUCLEUS photometric survey \citep{GALACTICNUCLEUS} allowed a fuller analysis of the colour--magnitude diagrams across the NSD from which \cite{Nogueras-Lara2020} demonstrated that the low number of stars in the earlier analysis of \cite{Figer2004} did not enable clear discrimination between a bursty and continuous star formation history, and instead the giant-branch luminosity functions across a more extended range of fields were consistent with a star formation history with an early ($>8\,\mathrm{Gyr}$) burst and then lower levels until a recent ($<1\,\mathrm{Gyr}$) burst. A very recent burst is corroborated by observations of classical Cepheids in this region with periods indicating ages of $\sim25\,\mathrm{Myr}$ \citep{Matsunaga2011} and confirmation that at least some fraction of the population in these regions must be very old ($\gtrsim10\,\mathrm{Gyr}$) comes from the detection of RR Lyrae there \citep{Minniti2016,Molnar2022}.
However, for probing the detailed star formation at intermediate ages, the differences in the giant branch luminosity function with age are quite subtle \citep[see figure 8 of][]{Nogueras-Lara2020}. For example, when one goes beyond ages of $\sim2\,\mathrm{Gyr}$ the red clump has a weak $\sim0.015\,\mathrm{mag}\,\mathrm{Gyr}^{-1}$ gradient with age \citep{Girardi2016,Chen2017,Huang2020} and one must instead rely on the relative fraction and location of red clump giants to red-giant-branch-bump stars.

Alternative age tracers for the NSD are Mira variables. Mira variables are thermally pulsating asymptotic giant branch stars, and are typically recognised as the final stages of the giant branch life of a low to intermediate mass star 
\citep{Catelan2015}. Nearly all asymptotic giant branch stars pulsate to some degree through a mechanism driven by convection \citep{Freytag2017,Xiong2018}. The range of pulsation modes form an entire family of different long period variables \citep{Wood2015} of which Mira variables have been identified as those pulsating in the fundamental mode with the highest amplitudes, $\Delta V>2.5\,\mathrm{mag}$, and periods in the range $80$ to $1000$ days. Their light curve shapes are distinguished from the similar, but lower amplitude, semi-regular variables (SRV) and OGLE small amplitude red giants (OSARG) by a more regular, near sinusoidal nature although long-term trends and variations in the period are observed \citep{ZijlstraBedding2002,He2016,Ou2022} possibly related to thermal pulses \citep{VassiliadisWood1993}, the interactions of pulsation with convective flow \citep{Freytag2017} or the presence of circumstellar dust \citep{Whitelock2003,Ou2022}. Additionally, as evidenced clearly in observations of the Large Magellanic Cloud (LMC), the classes of long period variables lie on distinct period--luminosity sequences \citep{Wood1999,Wood2000,Soszynski2009}, with the Mira variables lying along a single sequence \citep{GlassLloydEvans1981,Feast1989,Ita2004,Groenewegen2004,Fraser2008,Riebel2010,Ita2011,Yuan2017,Yuan2018,Bhardwaj2019,Iwanek2021b}. The tight period--luminosity relation and high luminosities of Mira variables have made them ideal standard candles both for cosmological studies \citep{Huang2018,Huang2020_Mira} and Local Group and Galactic structure studies \citep{Menzies2011,Whitelock2013,Menzies2015,Catchpole2016,Deason2017,Menzies2019,Grady2019,Grady2020}. Having well-understood Mira variables across a range of local environments will enable their precise calibration as a cosmological tracer, particularly in the era of the James Webb Space Telescope and the Vera Rubin Observatory.

It has long been observationally known that solar neighbourhood Mira variables with shorter periods have hotter kinematics \citep{Merrill1923} and more extended profiles perpendicular to the Galactic plane \citep{Feast1963}. This behaviour is indicative of shorter period variables belonging to older populations that have undergone more dynamical heating. Furthermore, older LMC and Milky Way clusters are hosts to shorter period Mira variables \citep{Grady2019}. The period of a Mira variable is largely governed by the mass and radius of the star and Mira-like pulsations only begin once a star has reached a narrow radial range at a given mass \citep{Trabucchi2019}. It is therefore expected that the period is a direct indicator of mass, and hence age of the star. However, there are relatively limited theoretical studies of the Mira variable period--age relation \citep{WyattCahn1983,FeastWhitelock1987,Eggen1998,Trabucchi2022}, and stellar population work has largely been done using period--age relations empirically calibrated from the solar-neighbourhood correlations with kinematics \citep{FeastWhitelock1987,FeastWhitelock2000,Feast2006,FeastWhitelock2014,Catchpole2016,LopezCorredoira2017,Grady2020,Nikzat2022}.
A typical period--age ($P-\tau$) relation is $\tau\approx13\,\mathrm{Gyr}\tfrac{1}{2}(1+\tanh((330\,\mathrm{day}-P)/250\,\mathrm{day}))$ where we see Mira variables of $(200,300,400)$ day periods have ages of $\sim(9.5,7.5,4.5)\,\mathrm{Gyr}$.
However, it is anticipated that there is a significant spread in age at each period with \cite{Trabucchi2022} reporting a $3\,\mathrm{Gyr}$ range for the age of a $350\,\mathrm{day}$ period Mira variable using a set of theoretical pulsation models.
Nonetheless, due to their potentially excellent resolution at intermediate ages and their high intrinsic brightness, Mira variables offer ideal age tracers for the inner Galaxy.

The first step in using Mira variables to constrain the star formation history of the NSD is to reliably identify them. Early work in this area targeted OH/IR maser stars in the very central regions of the Galaxy \citep[e.g.][]{Blommaert1998,Wood1998}. These searches are biased towards longer period stars. Both \cite{Glass2001} and \citet[][M09]{Matsunaga2009} undertook broader searches for variable stars and have presented samples of Mira variables around the Galactic Centre. The VISTA Variables in Via Lactea (VVV) survey is a multi-epoch infrared survey that has taken observations of the Galactic bulge over a $\sim$10-year baseline. This makes it an ideal survey for extending the sample of  long-period variables in the NSD. This is the goal of our work. Our paper is structured as follows: we begin by describing the selection and light-curve modelling of Mira variable candidates in Section~\ref{Section::Data}. Additional details on the specifics of the light curve modelling are given in Appendix~\ref{appendix::period}. We go on to inspect the properties of our sample in Section~\ref{section::properties} focusing on the spatial distribution, the selection effects that impact our sample, the photometric classification and the period--age distribution. We close with our conclusions in Section~\ref{section::conclusions}. This is the first of two papers on this sample. Our second paper will focus on the kinematic properties of the sample probed through the proper motion data provided by VIRAC2 \citep[][Smith et al., in prep.]{VIRAC,Sanders2019a}.

\section{Discovery of new NSD Mira variables}\label{Section::Data}
We describe the sequence of steps taken to extract a sample of NSD Mira variable candidates. First, we begin by describing the primary dataset employed, the VIRAC2 $K_s$ light curve set. We go on to describe the initial sample of likely variable stars and the methods used for modelling their light curves. From this sample, we define a series of cuts to isolate the Mira variable candidates. We further check the quality of our light curve modelling by comparison to overlap Mira variables in the literature. We finally model the multi-band light curves of a set of Mira variable candidates to further refine the periods and perform a visual inspection to weed out any remaining contaminants.

\subsection{Primary light curve sample}\label{sec:primary_lcdata}
\begin{figure}
    \centering
    \includegraphics[width=\columnwidth]{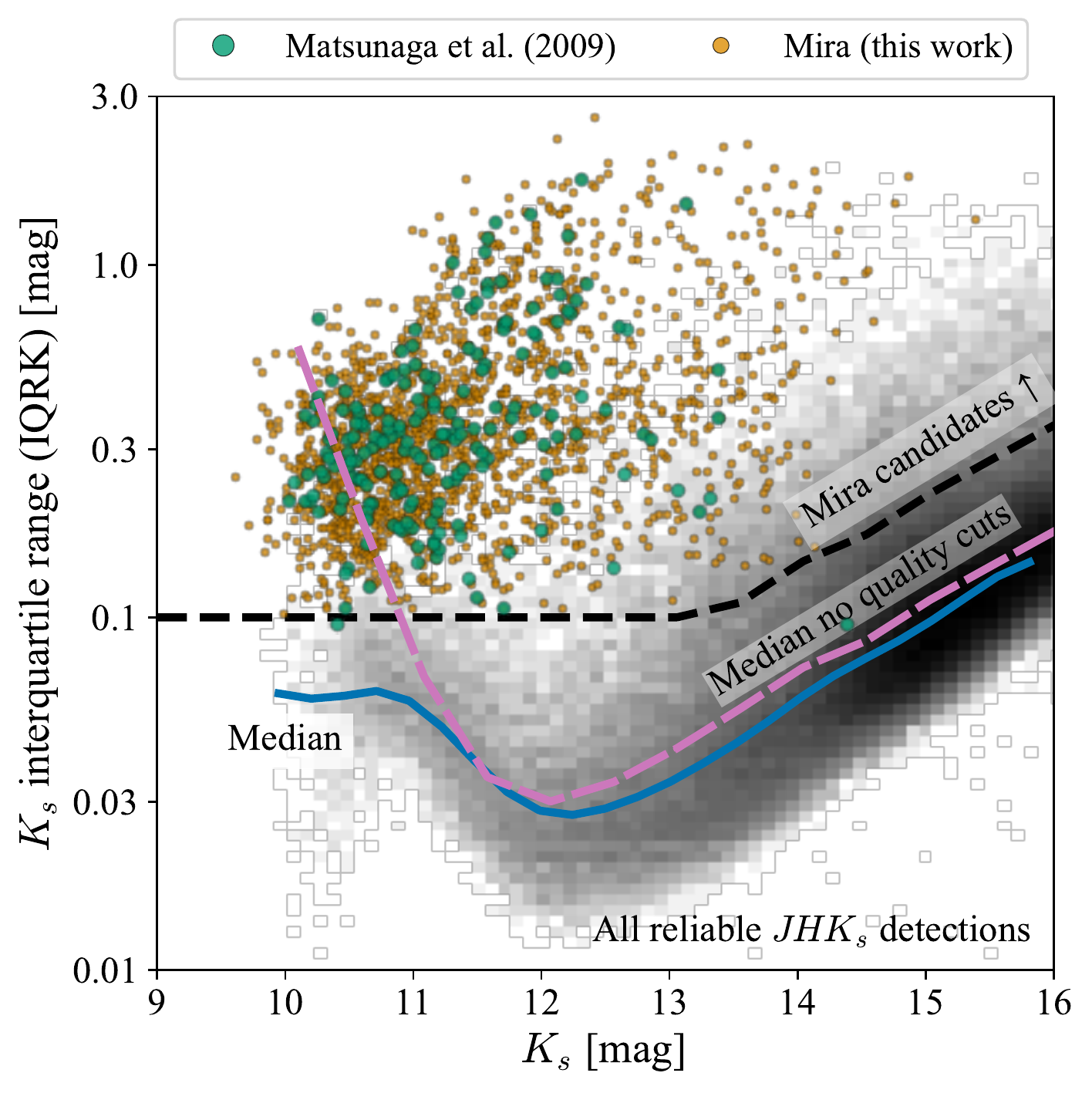}
    \caption{Selection of Mira candidates in $K_s$ magnitude against the interquartile range of $K_s$ (IQRK): background greyscale shows the distribution of all VIRAC2 sources with $J$, $H$ and $K_s$ detections in a field $10\times20\,\mathrm{arcmin}^2$ centred on $(\ell,b)=(0.08,0)\,\mathrm{deg}$ using only those detections in each light curve that satisfy our astrometric and photometric quality cuts. The blue line gives the median trend and the pink dashed line shows the median trend without applying the astrometric and photometric quality cuts. The black dashed marks the line above which stars are considered candidate Mira variables. The larger green points show the Mira variables from \protect\cite{Matsunaga2009} and smaller orange the final Mira variable sample in this paper.}
    \label{fig:mira_selection}
\end{figure}

Our primary source of data is the VISTA Variables in Via Lactea (VVV) survey \citep{Minniti2010,Saito2012}. The VVV survey is a multi-epoch near-infrared ($ZYJHK_s$) survey of the Galactic bulge and disc conducted using the $16$ detector VIRCAM camera \citep{vircam} mounted on the $4.1$m VISTA telescope \citep{Sutherland2015} at the Cerro Paranal Observatory. The initial $560\,\mathrm{deg}^2$ survey ran from 2010 to 2015 and covered the Galactic bulge ($|\ell|<10\,\mathrm{deg}$, $-10\,\mathrm{deg}<b<5\,\mathrm{deg}$) and the southern Galactic disc separated into $1.5 \times 1.1\,\mathrm{deg}^2$ tiles. The primary observations were taken in the $K_s$ band ($\sim160$ observations except for $8$ high cadence tiles with $\sim600$ observations) with additional $ZYJH$ observations typically taken at the beginning and end of the survey. In 2016, the VVVX survey commenced, extending the sky coverage of both the bulge and disc components and providing more epochs for the region covered by the initial survey. VVVX extended the $JHK_s$ coverage of the original VVV data resulting in at least $\sim200$ $K_s$ observations for each source and an average of $\sim30$ $J$ and $\sim20$ $H$ observations per source. In order to focus on the NSD region, we only use VVV and VVVX data within Galactic coordinates $|\ell|<1.5\,\mathrm{deg}$ and $|b|<1.5\,\mathrm{deg}$.

The VVV Infrared Astrometric Catalogue \citep[VIRAC,][]{VIRAC} was generated using VVV epoch aperture photometry \citep{GonzalezFernandez2018} and calculated relative proper motions (and parallaxes) for $\sim120$ million sources. Using the second Gaia data release \citep{Gaia1,Gaia2}, \cite{Sanders2019a} used bright overlapping sources between Gaia and VIRAC to anchor the relative proper motions to Gaia's absolute reference frame. The second version of VIRAC (VIRAC2, Smith et al., in prep.) utilises point spread function (PSF) photometry to deliver more accurate centroids and a deeper source catalogue, and calibrates astrometry to Gaia's reference frame for individual observations (rather than using a post hoc correction). Here we use a preliminary version of the final VIRAC2 dataset. Each VVV image has been processed with the PSF photometry fitting programme \textsc{DoPhot} \citep{dophot,dophot2} and the resulting photometry zero-point calibrated on a chip and time-dependent basis using a pool of 2MASS reference sources. Initial astrometric solutions were computed by grouping nearby detections and then improved by iteratively re-grouping detections based upon fitted astrometry (allowing for detections to be included in multiple groupings). The final set of detections grouped using the derived astrometry is our set of \emph{light curves}. 

The deeper photometry provided by PSF fitting comes at the expense of spurious sources detected in the wings of bright objects. Many of these spurious sources have similar magnitudes and variability to our target Mira variables. An initial list of reliable sources was obtained by requiring the sources are non-duplicate (defined as the source having less than $20\percent$ of detections shared with other sources), have $10$ or more epochs (these sources are fitted with a full five-parameter astrometric solution in VIRAC2) and are detected in more than $20\percent$ of observations. Spurious sources are approximately randomly distributed around bright sources and given a large number of observations occasionally there is random alignment and grouping of the spurious sources. Requiring the source is detected in more than $20\percent$ of observations is a trade-off between retaining genuine faint sources and removing these spurious sources. The cut is implemented in terms of a fraction of observations such that it is homogeneously applied across the entire VVV survey which can have quite large variations in epoch counts. For the $|\ell|<1.5\,\mathrm{deg}$, $|b|<1.5\,\mathrm{deg}$ region, this results in a primary catalogue of $27,714,023$ sources (light curves). We further clean up the light curves of individual sources by removing detections based on the quality of their individual astrometric and photometric fits. Some bright ($11\lesssim K_s\lesssim12$) spurious detections can be identified by the high $\chi^2$ of their PSF fits \citep[e.g. see][]{Braga2019}. We reject all individual detections brighter than $K_s=13.2$ with the \textsc{DoPhot} photometric fit chi-squared 
$>10$. Furthermore, we remove all ambiguous detections (detections shared with another reliable source) and all detections with astrometric chi-squared deviation of the detection relative to the astrometric solution
$>11.83$ (approximately $3\sigma$ outliers for a two degree-of-freedom fit). We also always employ a single $5\sigma$ clip for the light curves (estimating $\sigma$ from the $16$th and $84$th percentiles) to remove further outliers. Despite these quality cuts, we have found that variable seeing for blended sources can lead to spurious variability and periodicity. In Appendix~\ref{appendix::blended} we discuss how we calibrate the light curves of suspected blended sources. We refer to this set of resultant light curves as `cleaned'. 

\subsubsection{Complementary photometric data}
In addition to VVV photometry, we will also utilise some other near infrared and longer wavelength photometry. We cross-match all of our candidates to the DECAPS catalogue \citep[][with a $0.3\,\mathrm{arcsec}$ radius]{DECAPS}, the GALACTICNUCLEUS catalogue \citep[][with a $0.4\,\mathrm{arcsec}$ radius]{GALACTICNUCLEUS}, the 2MASS catalogue \citep[][with a $1\,\mathrm{arcsec}$ radius]{2MASS}, the AKARI catalogue \citep[][with a $5\,\mathrm{arcsec}$ radius]{AKARI}, the GLIMPSE catalogue \citep{GLIMPSE}\footnote{As the GLIMPSE-II catalogue puts a requirement on the sources having similar magnitudes at the two GLIMPSE-II epochs, some variable sources are not present in the combined catalogue and instead we use the results from the Epoch 1 catalogue. This affects $\sim200$ stars in our final sample.} and the Spitzer-IRAC GALCEN point source catalogue of \cite{Ramirez2008} (with a $0.4\,\mathrm{arcsec}$ radius), preferentially keeping the GALCEN data over GLIMPSE, the WISE catalogue \citep[][with a $1\,\mathrm{arcsec}$ radius]{Wright2010}\footnote{\label{footnote1}We correct the WISE photometry of bright stars using the tables from \url{https://wise2.ipac.caltech.edu/docs/release/neowise/expsup/sec2_1civa.html}.}, the $24\,\mu\mathrm{m}$ MIPSGAL catalogue \citep[][with a $1.5\,\mathrm{arcsec}$ radius]{Gutermuth2015}, the $7$ and $15\,\mu\mathrm{m}$ ISOGAL catalogue \citep[][with a $1.5\,\mathrm{arcsec}$ radius]{Omont2003} and the Herschel Infrared Galactic plane survey \citep[Hi-GAL,][with a $3\,\mathrm{arcsec}$ radius]{Molinari2016}. Although the point-spread function for some of the surveys is large and so contamination might be expected in the considered crowded regions, we assume the Mira variables are significantly brighter in the mid-infrared than any nearby stars such that contamination is minimal.

\subsection{Light curve modelling}\label{sec::light_curve_modelling}

\begin{figure}
    \centering
    \includegraphics[width=\columnwidth]{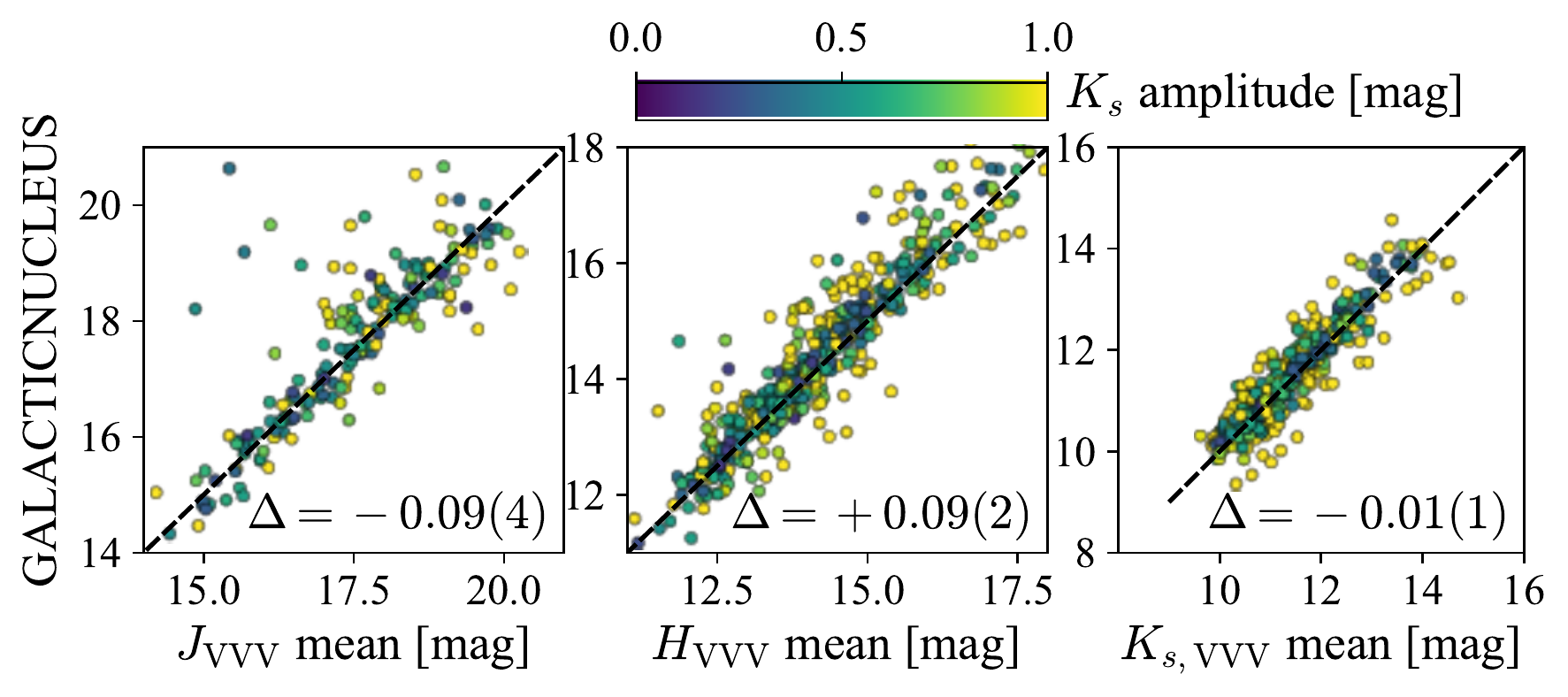}
    \caption{GALACTICNUCLEUS against VVV magnitudes (we plot the mean of our model fits) coloured by amplitude. $\Delta$ reports the magnitude offset (GALACTICNUCLEUS -- VVV) with the bracketed digit the uncertainty in the last decimal place. In the saturated regime ($K_s\lesssim 11.5$) there is no significant bias. The scatter correlates with Mira variable amplitude.}
    \label{fig:saturated}
\end{figure}
\begin{figure}
    \centering
    \includegraphics[width=.99\columnwidth]{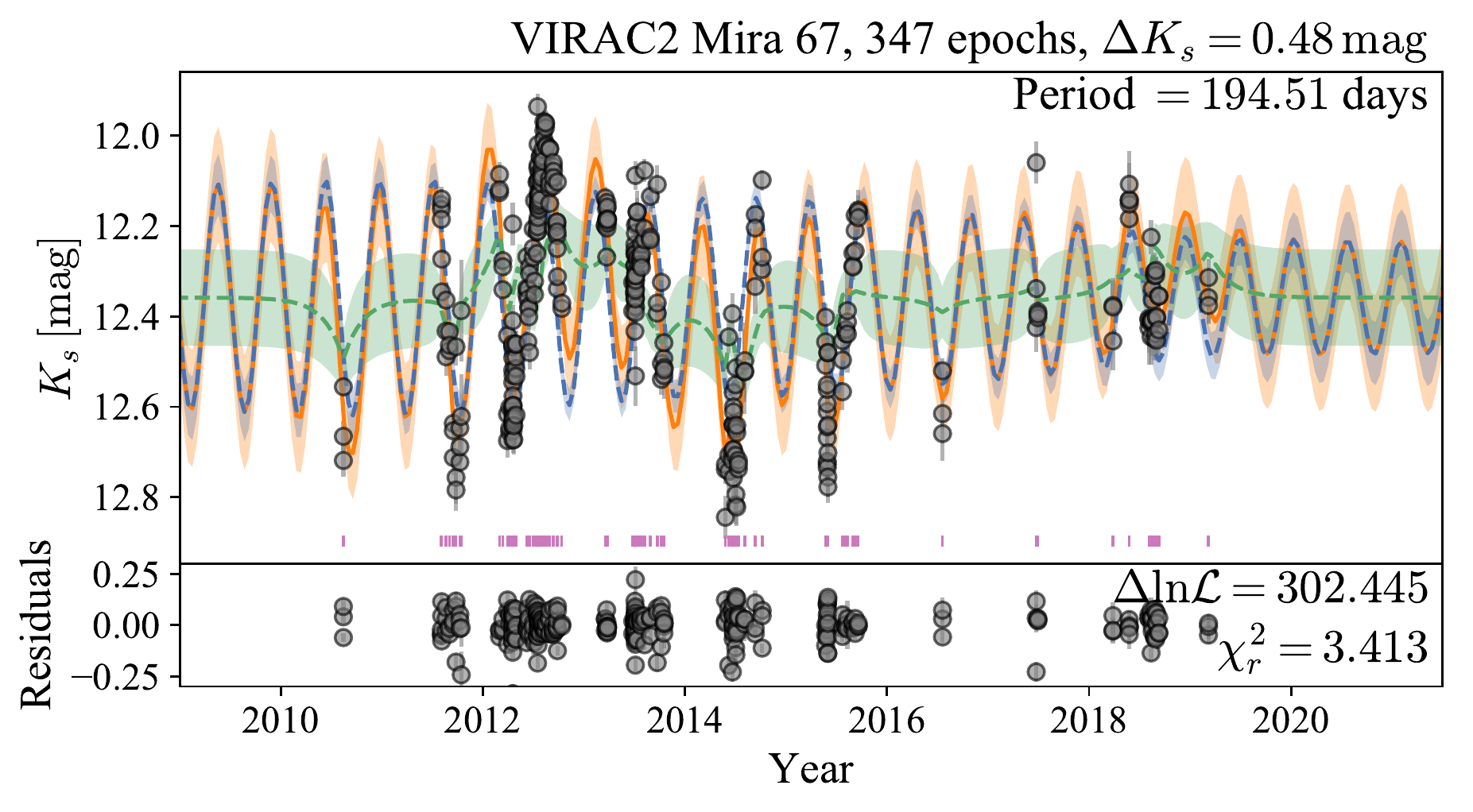}
    \includegraphics[width=.99\columnwidth]{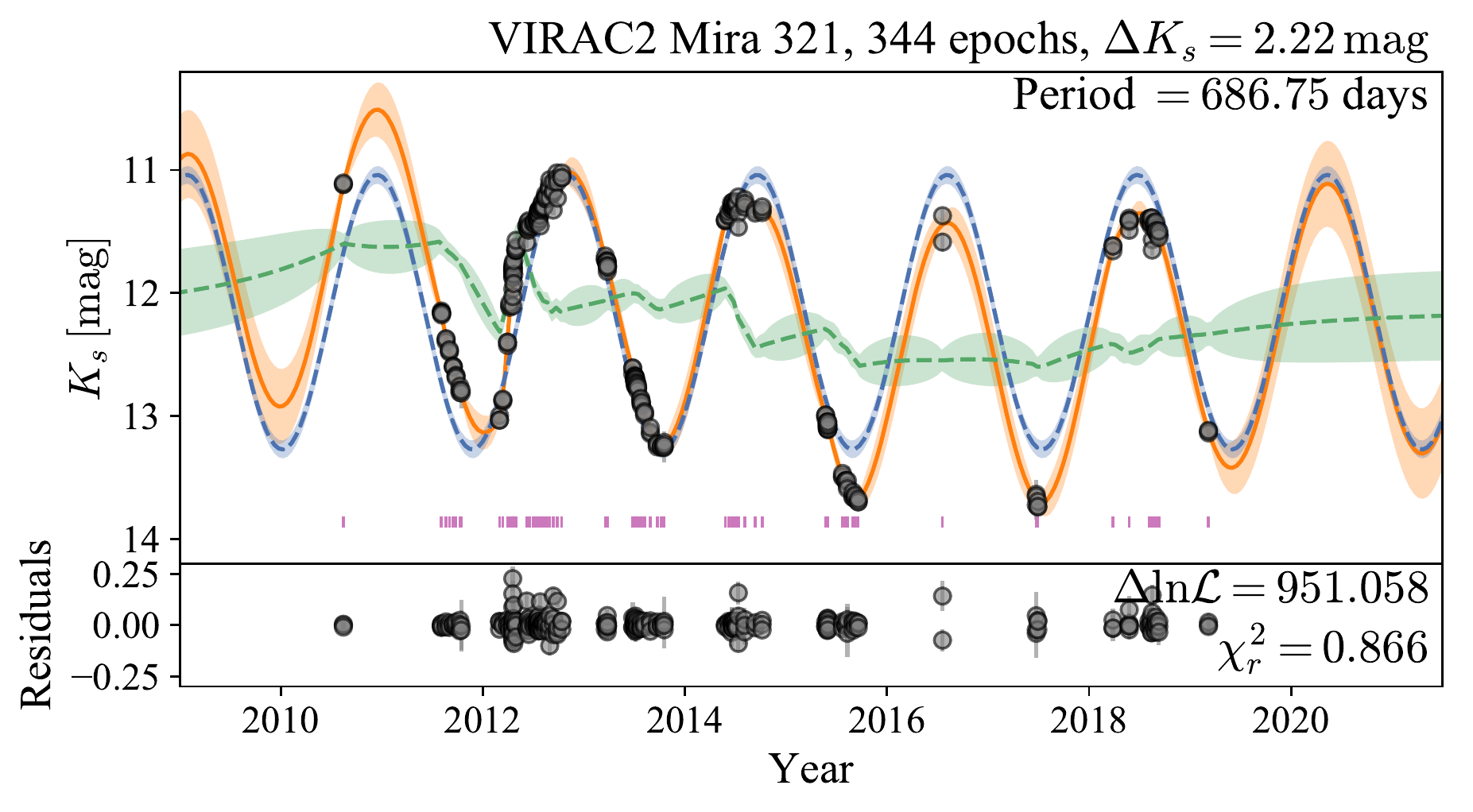}
    \includegraphics[width=.99\columnwidth]{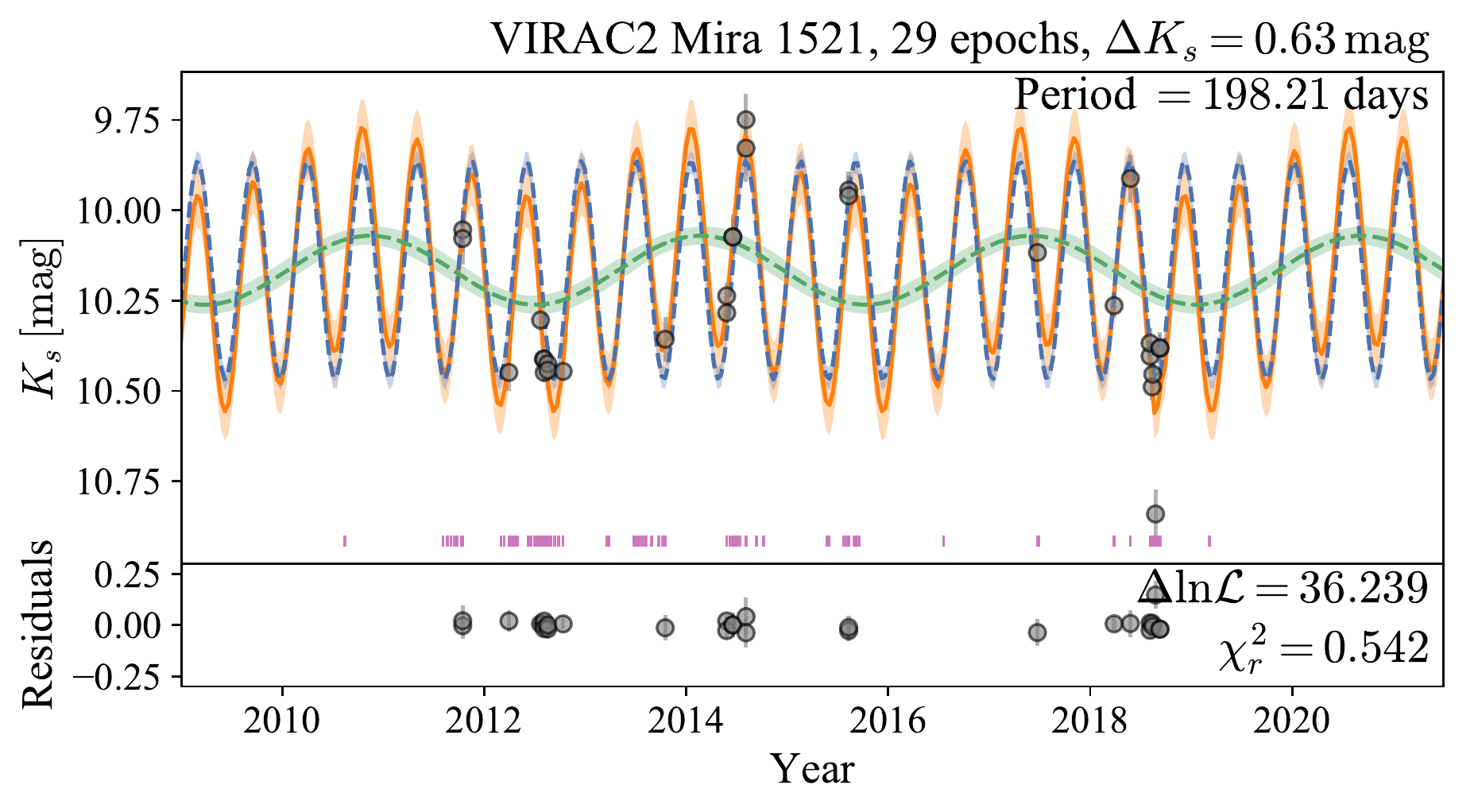}
    \includegraphics[width=.99\columnwidth]{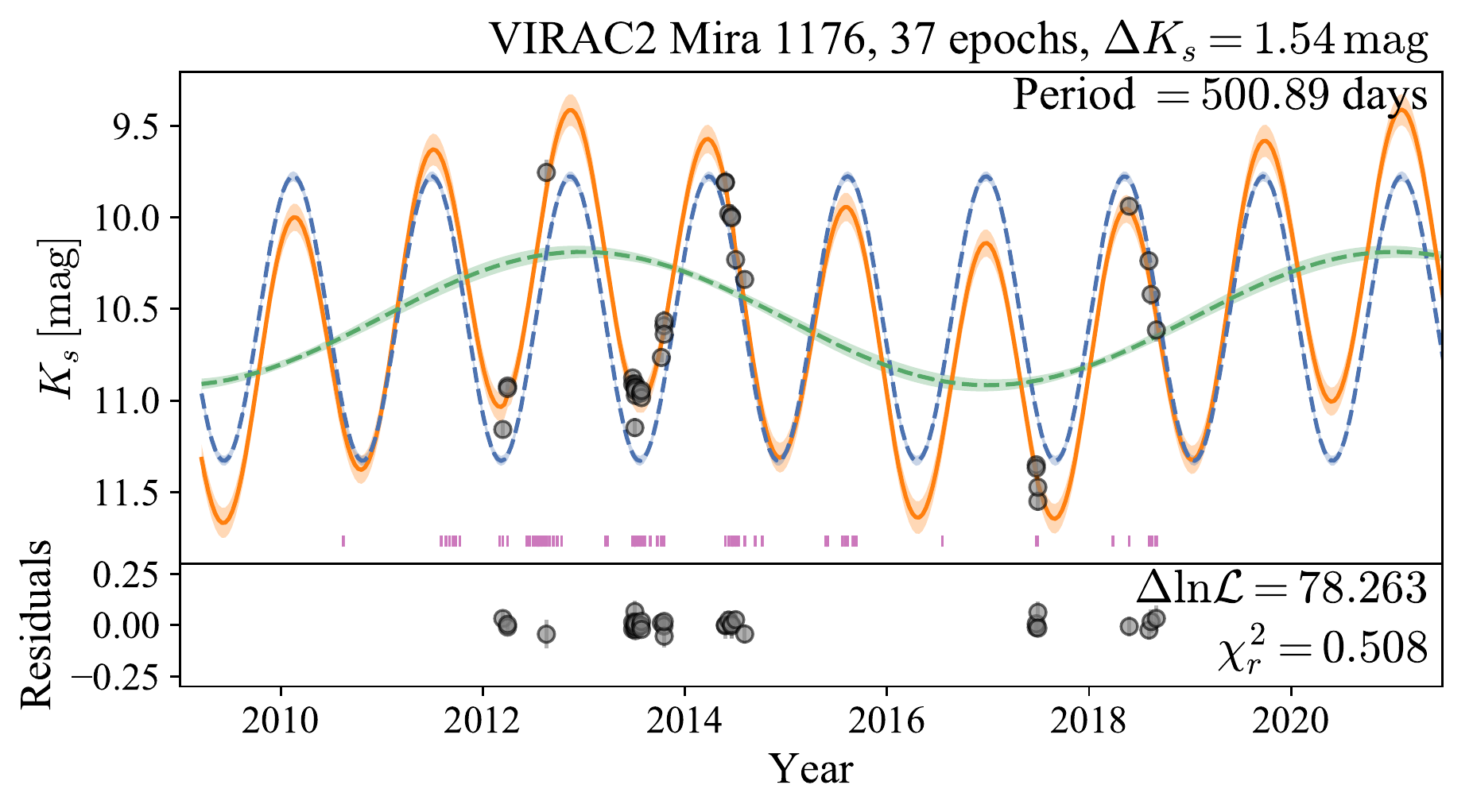}
    \caption{Example light curves for our Mira variable sample. Each set of panels show a different Mira light curve with the model residuals below. The grey points are the cleaned VIRAC2 light curves. The means and standard deviations of the Gaussian process models are shown in orange and the sub-components of the kernels in green and blue. The small pink ticks show the times at which the star could have been detected. $\Delta\ln\mathcal{L}$ gives the difference in log-likelihood with respect to a non-periodic model (with an additional variance) and $\chi_r^2$ the reduced chi-squared of the fit. Note that deviations from pure periodicity are sometimes fitted with the random walk part of the kernel (top two panels) and sometimes a longer periodic signal (other panels). The four examples are chosen as two well-sampled light curves (top -- the very top light curve is also in the \citetalias{Matsunaga2009} sample) and two poorly-sampled (bottom). The first and third light curves are low amplitude and short period, whilst the second and fourth are high amplitude and long period.}
    \label{fig:example_light curves}
\end{figure}

\begin{figure}
    \centering
    \includegraphics[width=\columnwidth]{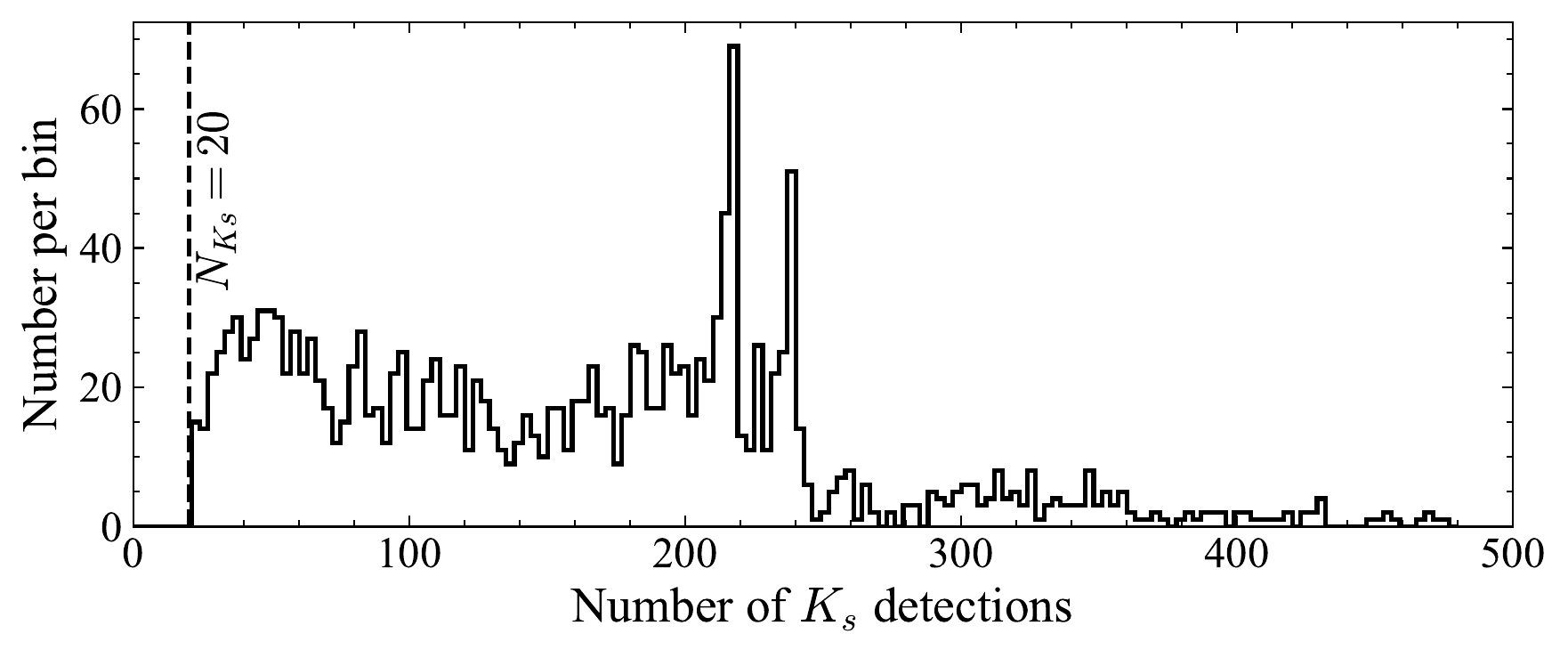}
    \caption{Distribution of the number of $K_s$ detections for our Mira variable sample. The sources have to have at least $20$ detections (vertical dashed line).}
    \label{fig:nepochs}
\end{figure}

\begin{figure*}
    \centering
    \includegraphics[width=\textwidth]{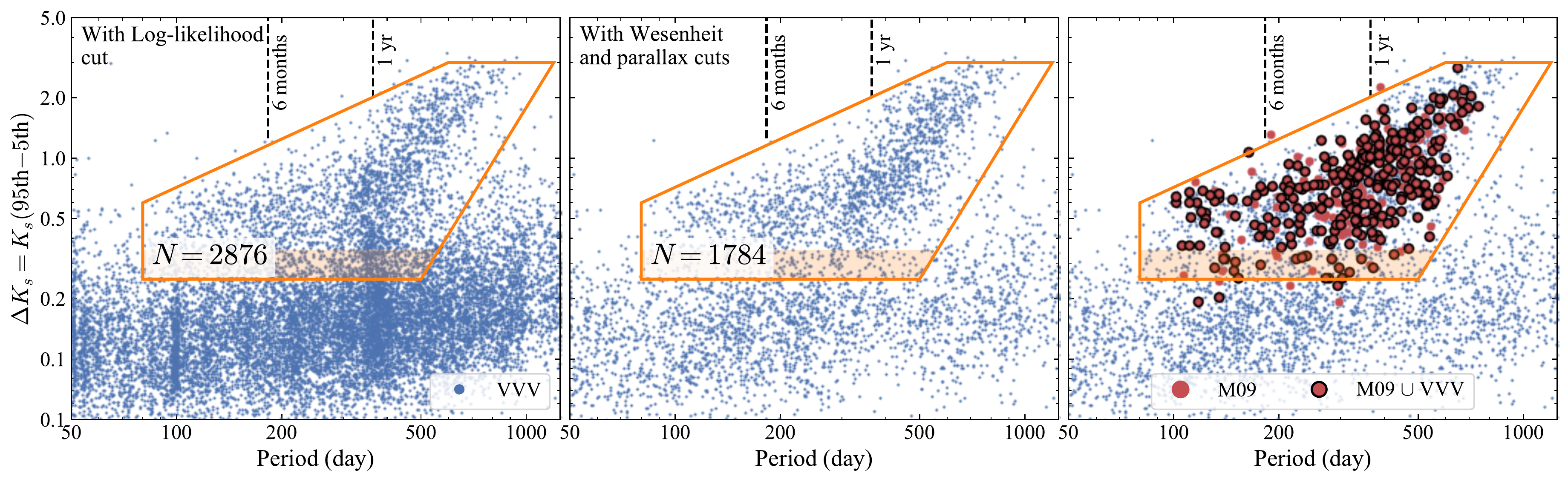}
    \caption{Period--amplitude plot for all processed variables. The left panel shows only sources satisfying our initial cut on log-likelihood whilst the central panel shows the addition of the cuts on Wesenheit magnitudes and parallax. In the right panel we overlay the \protect\citetalias{Matsunaga2009} Mira variables either with black outlines if they have been included in our processing, or no outline if not. The Mira variable period--amplitude sequence is visible and demarcated by the orange selection box, which contains $1784$ stars. The shaded region gives the amplitude range over which the variables are likely semi-regular variables. We mark on possible aliases of a year and $6$ months which do not appear as overdensities in our parallax-cut sample.}
    \label{fig:selection}
\end{figure*}

From the set of cleaned light curves, we form an initial Mira candidate list by finding highly variable sources in the $K_s$ band. Our search is guided by the previous \citetalias{Matsunaga2009} search for Mira variables in the very inner $20\times30\,\mathrm{arcmin}^2$ area around the Galactic centre. These authors first selected stars with photometry three times more variable than the median  variability at a given star's magnitude and found $1364$ long period variable candidates, of which $549$ were assigned periods. No detailed classification was performed such that some level of contamination from young stellar objects \citep[e.g.][]{Guo2022} is likely (although at the magnitude range this survey probed they would be foreground objects so likely dwarfed in number density by the background long-period variables) and also that the long-period variables themselves are not guaranteed Mira variables but could contain a mixture of other semi-regular variables. The precise definition of a Mira variable is slightly awkward. Physically they are often linked with high-amplitude fundamental mode pulsation and hence membership of a particular period--luminosity sequence. From an observational perspective, this definition is often approximated as a pure amplitude cut \citep[e.g][]{Soszynski2013} although \cite{Trabucchi2021} acknowledge that this simple consideration removes lower amplitude stars on the same fundamental period--luminosity relation as the higher amplitude systems. In the \citetalias{Matsunaga2009} analysis, stars are considered as non-Mira variables if the amplitude in any of $J$, $H$ or $K_s$ is less than $0.4$ although this removes only $\sim10\percent$ of the stars in their sample. Here we attempt to emulate the fuller selection of \citetalias{Matsunaga2009} keeping in mind that the lower amplitude variables could be semi-regular variable contaminants.

Guided by \citetalias{Matsunaga2009}, in each field we construct the median curve of the interquartile range of $K_s$ (IQRK) as a function of $K_s$ for sources detected in $J$, $H$ and $K_s$ (typically sources detected in all three bands are highly reliable although sources are lost due to no detections in the bluer bands in high extinction regions). No astrometric or photometric quality cuts were applied to this sample although it makes little difference to our selection. The expected IQRK varies significantly with location in the bulge due to both observation quality and crowding. For all cleaned light curves, we compute the median $K_s$ and IQRK. We retain all light curves with more than $20$ epochs and with IQRK $>0.1\,\mathrm{mag}$ and IQRK greater than $2$ times the median line for $K_s>12$ (see Fig.~\ref{fig:mira_selection} for the IQRK cut employed). As shown in Fig.~\ref{fig:mira_selection}, this cut encompasses nearly all of the long-period variables with periods presented by \citetalias{Matsunaga2009}. Although non-linearities from saturation begin at $K_s=12$ we still consider sources brighter than this as otherwise we would reject many Mira variables and we have found that period estimation from VVV is still reliable for saturated sources. From Fig.~\ref{fig:mira_selection}, we see that when astrometric and photometric quality cuts are not applied to the parent sample, the IQRK rises significantly at the bright end. Removal of the astrometric and photometric outliers causes the IQRK to plateau at bright $K_s$ at values significantly below the expected IQRK for Mira variables. This gives confidence that spurious variability from saturated sources is not a significant concern for the cleaned light curves. In Fig.~\ref{fig:saturated} we show a comparison of the VVV modelled mean magnitudes compared to the GALACTICNUCLEUS measurements \citep{GALACTICNUCLEUS} for our final Mira variable sample. Although both VVV and GALACTICNUCLEUS suffer from saturation effects for $K_s\lesssim11$, the effects are weaker in GALACTICNUCLEUS as it has shorter exposure times than VVV and the HAWK-I camera smaller pixels than VIRCAM \citep{GALACTICNUCLEUS}. There is no visible bias between the modelled VVV mean magnitudes and GALACTICNUCLEUS at the bright end giving some confidence in our use of the data in this regime.

From our set of candidate cleaned light curves, we must model the light curve properties to produce a set of Mira variables. The two key properties for identifying Mira variables are their long periods ($>80$ days) and high amplitudes $0.25<\Delta K_s<3$ \citepalias[][as discussed above, here we adopt a generous lower limit for $K_s$ amplitude to match the long-period variable selection of \citetalias{Matsunaga2009} although as acknowledged by \citetalias{Matsunaga2009} it is expected stars with $\Delta K_s\lesssim0.4$ are in fact semi-regular variables]{Matsunaga2009}. We construct Fourier models for each light curve finding the best period. However, Mira light curves tend to not be completely periodic and exhibit a range of other behaviour with both short and long-term amplitude and period variability \citep{ZijlstraBedding2002,He2016,Molina2018}. Therefore, as a second step we also use Gaussian process models that can capture quasi-periodic signals. We fully describe the details of these two methods, as well as their application to multi-band photometry, in Appendix~\ref{appendix::period}.

We first run a simple Lomb-Scargle periodogram searching periods from $0.5$ days to the time-span of the light curve. We continue to consider the star if one of the top three periods (excluding aliases identified through the Lomb-Scargle periodogram on the magnitudes replace by noise) is $>10$ days (with a false alarm probability less than $0.001$) or if the period found by the `string-length' method \citep{LaflerKinman1965} is $>10$ days. Aliases are defined as the top five peaks in a periodogram of the magnitudes replaced by a constant value \citep{Vanderplas2018}, as well as $1$ and $1/2$ day periods. On the remaining stars, we run a second Fourier fit with $N_\mathrm{F}=2$ Fourier terms and $N_\mathrm{P}=3$ polynomial terms using $10$ days as the minimum period. If the best period is $>50$ days, we run a grid of Gaussian process fits selecting the fit that gives the minimum Akaike information criterion. We use the kernel in equation~\eqref{equation::kernel} with $N_\mathrm{O}=(1,2)$, $N_\mathrm{E}=(0,1)$, $\ln c_1=(-3,-10)$ (the damping of the oscillation) and initialized with the top three periods from the Fourier fit. When two periodic terms are used, we take the period associated with the higher amplitude kernel term as the primary period. If, however, the higher amplitude period is over $1000$ days, we use the lower-amplitude period as the primary period (provided it is under $1000$ days). In Appendix~\ref{appendix:period_comparison} we demonstrate the quality of the period recovery using this procedure on a set of literature sources with VIRAC2 data.

In Fig.~\ref{fig:example_light curves} we show four example light curves for Mira variables in our final sample. We see that the modelling is able to capture the primary periodic signal whilst also having the flexibility to model cycle-to-cycle variations either through a longer periodic component or through a stochastic random-walk component. Fig.~\ref{fig:example_light curves} also illustrates the range in number of $K_s$ epochs we have for each source. We show the distribution of number of $K_s$ epochs in our final Mira variable sample in Fig.~\ref{fig:nepochs}. The range in number of detections arises in part due to the VVV and VIRCAM observing strategy (some sources are in the overlaps of the VIRCAM pointings necessary to fill each tile) and in part due to varying observing quality leading to varying levels of blending and saturation and hence unreliable detections not included in our light curve processing.

\subsection{Mira variable selection}\label{sec::mira_cuts}

From our candidate list of long period variables, we remove all sources with a log-likelihood difference between the Gaussian process model and a constant model with an additional variance less than $10$\footnote{Using the Akaike or Bayesian information criteria instead results in minor differences in the final sample. Choosing a cut of $\Delta$AIC$<-10$ results in $1783$ stars in the final sample and $\Delta$BIC$<-10$ results in $1776$ stars, compared to our default log-likelihood cut resulting in $1784$ stars.}. This initial cut results in $23496$ candidates. We then primarily select Mira variables using the period--amplitude diagram as shown in the left panel of Fig.~\ref{fig:selection}. Here the amplitude, $\Delta K_s$, is the difference between the $95$th and $5$th percentile for the model computed over one period centred on and averaged over each light curve datum. Guided by the sample of \citetalias{Matsunaga2009}, we find that the Mira variables lie on a sequence that begins about $P\approx80$ day with $\Delta K_s\approx0.5$, runs horizontally to about $P\approx 300$ day before increasing in amplitude with increasing period up to about $\Delta K_s\approx3$ at $P\approx1000$ day. We see that the density of stars changes below $\Delta K_s\approx0.35$ as this is likely the Mira variable boundary and objects with $\Delta K_s\lesssim0.35$ are semi-regular variables. We adopt the broad selection box shown in Fig.~\ref{fig:selection} to match the selection of \citetalias{Matsunaga2009}. Within this selection box, we find $2876$ stars.

We also employ a number of cuts that remove a further $\sim40\percent$ of the sample. Contaminants include young stellar objects \citep[YSOs, see][]{Guo2022}, other fainter giant stars and blended photometry not properly handled in our calibration step. We have found that a small fraction of aliases that appear around $1$ year period also have significant parallaxes $\varpi$ measured in VIRAC2. We therefore remove anything with $|\varpi/\sigma_\varpi|>5$ ($186$ stars). Furthermore, we employ several cuts based on Wesenheit magnitude (as the Mira variables follow period--luminosity relations), removing stars with
$W_{K_s,H}=K_s-1.328(H-K_s)>5\times10^{-5}(P-300\,\mathrm{days})^2+8.7$ ($826$ stars) or $W_{K_s,J}=K_s-0.482(J-K_s)>2\times10^{-5}(P-300\,\mathrm{days})^2+8.7$ ($744$ stars) or $W_{K_s,[4.5]}=K_s-1.6(K_s-[4.5])>10.5-P/150\,\mathrm{days}$ ($475$ stars) where the extinction coefficients are from \cite{Fritz2011}. In Appendix~\ref{appendix:wesenheit_period_cuts} and Fig.~\ref{fig:wesenheit_period_cuts} we display the impact of these cuts. We use the mean from the light curve fits for $K_s$, the inverse-variance-weighted mean magnitudes for $J$ and $H$ and the $[4.5]$ GLIMPSE/GALCEN measurement (for GLIMPSE-II this is the average of two epochs separated by six months). These cuts remove potential YSO contaminants unless they are very nearby and bright. We are left with a sample of $1784$ stars as shown in the central panel of Fig.~\ref{fig:selection}. We assign the stars a running index and name them `VIRAC Mira \#'. We append to this list the $40$ \citetalias{Matsunaga2009} sources with periods that are in VIRAC2 but do not enter our final sample (for reasons discussed in Section~\ref{section::selection}).

\subsection{Multi-band light curves}
\begin{figure*}
    \centering
    \includegraphics[width=0.97\textwidth]{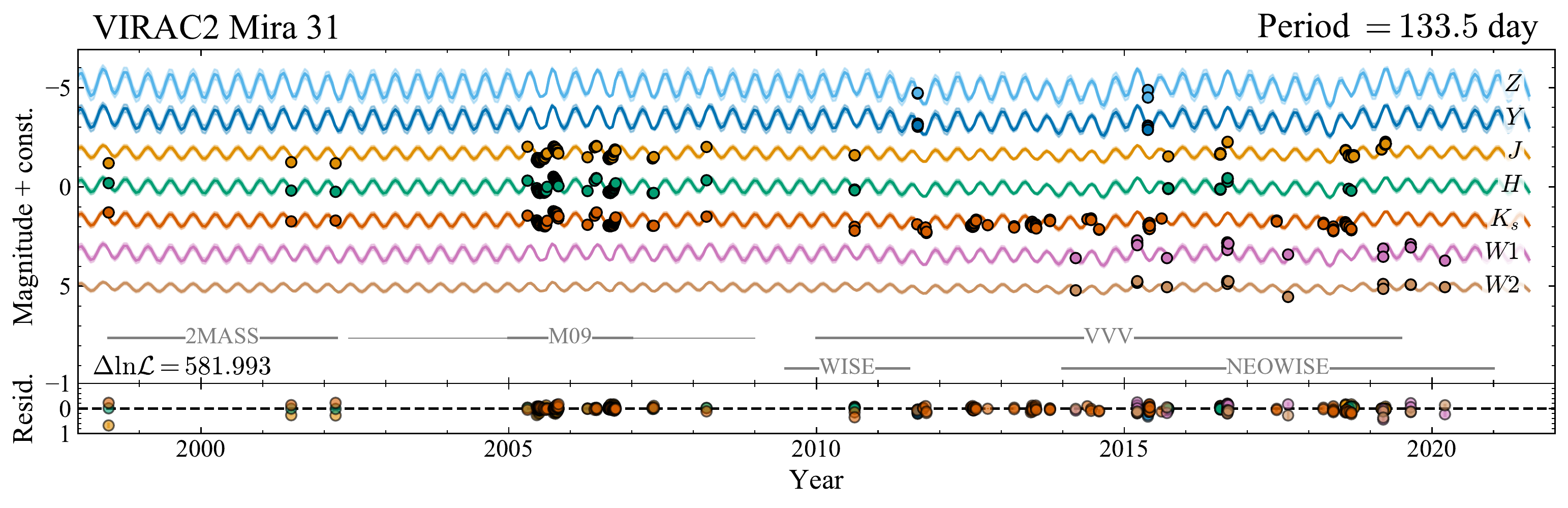}
    \includegraphics[width=0.97\textwidth]{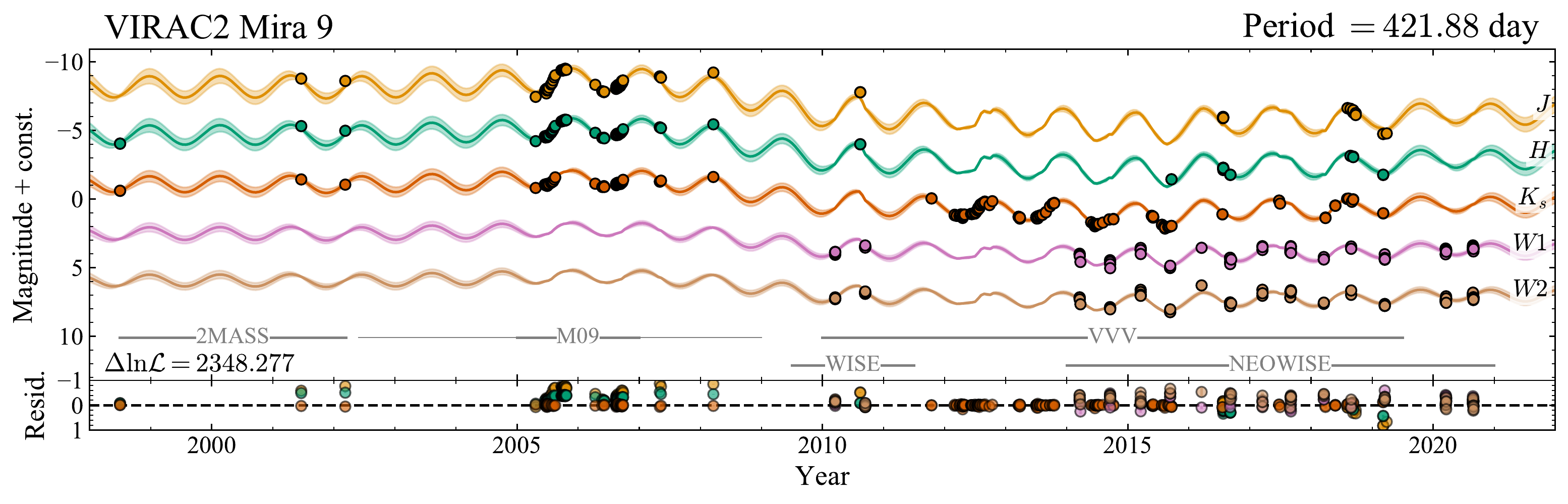}
    \includegraphics[width=0.97\textwidth]{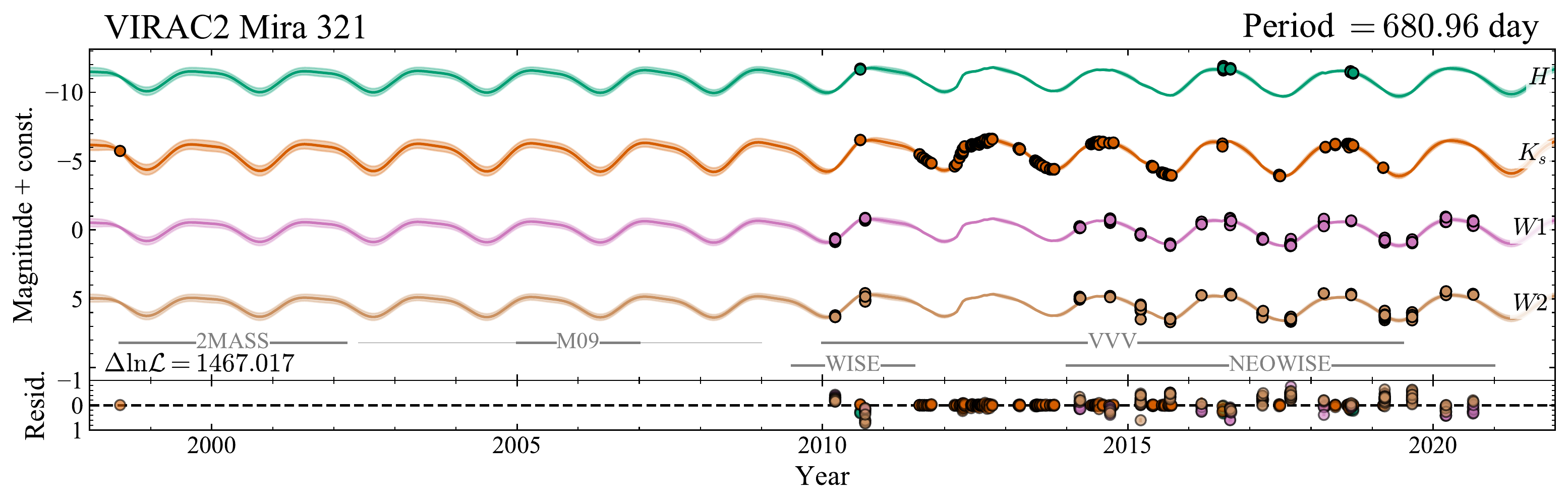}
    \includegraphics[width=0.97\textwidth]{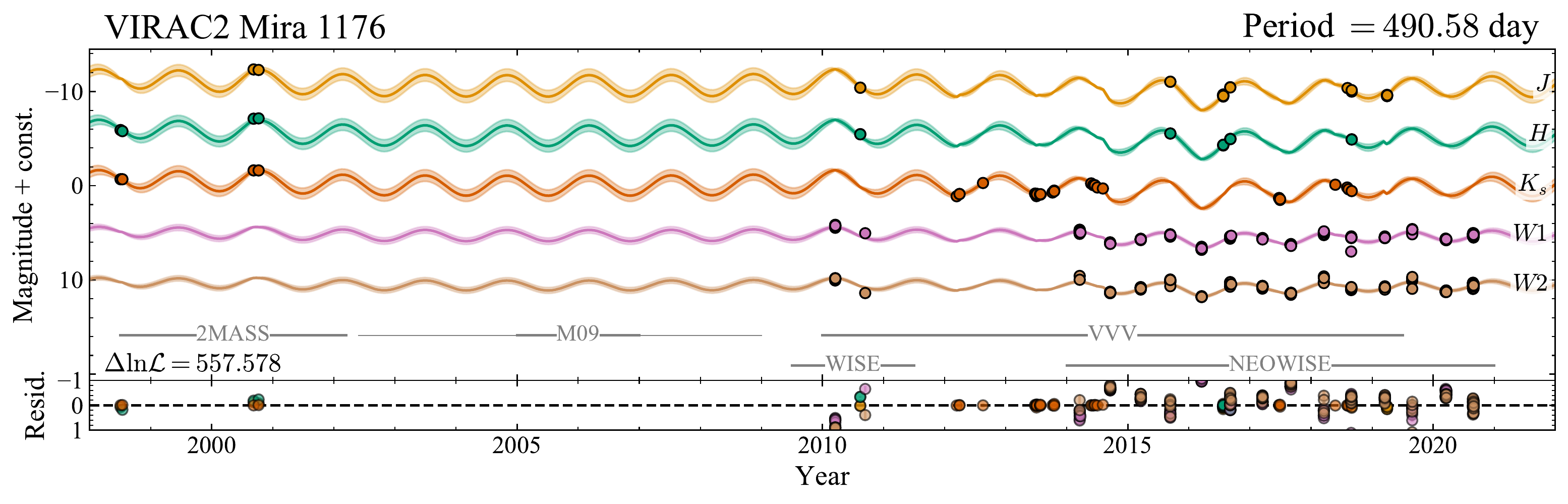}
    \caption{Example multi-band light curves along with a Gaussian process model. The top example has measurements in $7$ photometric bands. The second example has observations from \protect\cite{Matsunaga2009}. The third and fourth examples are the dense and sparse long-period examples in the second and fourth panels of Fig.~\ref{fig:example_light curves}.}
    \label{fig:multiband_examples}
\end{figure*}
Although the multi-epoch coverage from the VVV survey is primarily in the $K_s$ band, there are also multi-epoch observations available in $ZYJH$. Where available, we further complement the $JHK_s$ epochs with the data from \citetalias{Matsunaga2009} (including for those sources without measured periods in \citetalias{Matsunaga2009}) and the multi-epoch data from 2MASS (from the IRSA tables fp\_psc and pt\_src\_rej). The three photometric systems (VISTA, SIRIUS and 2MASS) are slightly different. However, this is only a minor concern as the light curve modelling method can accommodate small magnitude shifts. The WISE satellite initially surveyed every region of the sky over two epochs separated by approximately half a year before exhausting its coolant, and from 2013 was re-purposed for the NEOWISE survey which takes two groups of observations per year in $W1$ and $W2$. We take the observations from the IRSA allwise\_p3as\_mep and neowiser\_p1bs\_psd tables selecting high quality detections with \texttt{moon\_mask} $=0$, \texttt{saa\_sep} $\geq5$ and \texttt{qi\_fact} $\geq0.5$. The photometry is corrected as per footnote\footnoteref{footnote1}.

\begin{figure}
    \centering
    \includegraphics[width=\columnwidth]{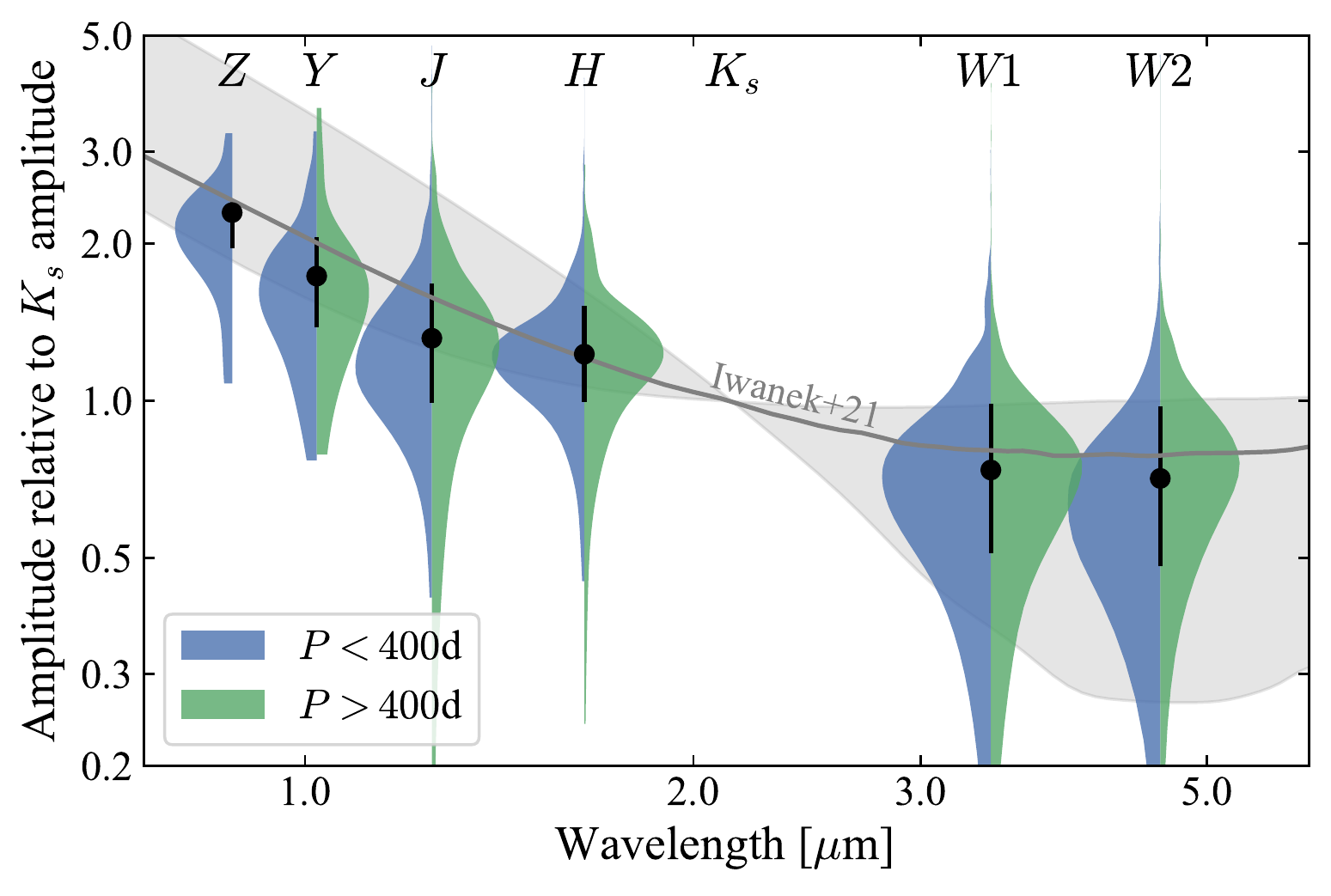}
    \caption{Measured light curve amplitude in the VVV $ZYJH$ bands and WISE $W1$ and $W2$ bands with respect to the $K_s$ band amplitude. The black errorbars show the medians with $\pm1\sigma$ estimated from the percentiles whilst the violins show the full distributions split by period (left blue $<400$ day, right green $>400$ day). The relation derived by \protect\cite{Iwanek2021} is shown in grey.}
    \label{fig:scaling}
\end{figure}

From VVV, 2MASS, the \citetalias{Matsunaga2009} dataset and WISE, we can construct up to $7$-band light curves for our sample which we fit using a 2d generalization of the \textsc{celerite} Gaussian process model \citep{ForemanMackey2017} as described in Appendix~\ref{appendix::2dLC}. This is particularly useful for measuring the mean magnitudes in each band (properly corrected for phase variation) as well as measuring the amplitudes in each band. We show some of the example multi-band light curves in Fig.~\ref{fig:multiband_examples}. We have chosen a case where all seven bands are observed, one where we have data from 2MASS, \citetalias{Matsunaga2009} and VVV, and then two examples from Fig.~\ref{fig:example_light curves}. It is somewhat evident from these examples that the amplitude of variability decreases with increasing wavelength. In Fig.~\ref{fig:scaling}, we show the distributions of the amplitudes in each photometric band relative to that in the $K_s$ band. Our sample follows the O-rich relation derived by \cite{Iwanek2021} which, it should be noted, was used as a weak prior on the amplitude ratios (see Appendix~\ref{appendix::2dLC}).

With the multi-band light curves in hand, we go through a final visual inspection stage of our sample. We assess three aspects: (i) whether the multiband light curve has perhaps produced spurious results (possibly due to contaminated WISE observations) in which case we resort to the results from the single $K_s$ light curve fit, (ii) whether the light curve fits have evidence of some periodicity but no clear Mira-like oscillations in which case we flag the star as `unreliable' and (iii) whether there is no clear evidence for periodicity in which case we remove the star from our sample. This visual inspection stage removes $42$ stars from our sample and flags $91$ as unreliable. $\sim75\percent$ of the unreliable stars have $K_s$ amplitudes below $0.4$ suggesting they are semi-regular variables \citepalias[e.g.][]{Matsunaga2009}. This procedure suggests the contamination in our full sample is between $5$ and $10\percent$.

Our final sample contains $1782$ Mira variable candidates of which $1691$ are deemed `reliable', $342$ have $K_s$ amplitudes less than $0.4$ so are potentially semi-regular variables and $272$ stars are in the sample of \citetalias{Matsunaga2009} ($209$ have periods reported by \citetalias{Matsunaga2009}). The catalogue is temporarily available at \url{https://www.homepages.ucl.ac.uk/~ucapjls/data/mira_vvv.fits}.

\section{Properties of our Mira variable sample}\label{section::properties}

\begin{figure*}
    \centering
    \includegraphics[width=\textwidth]{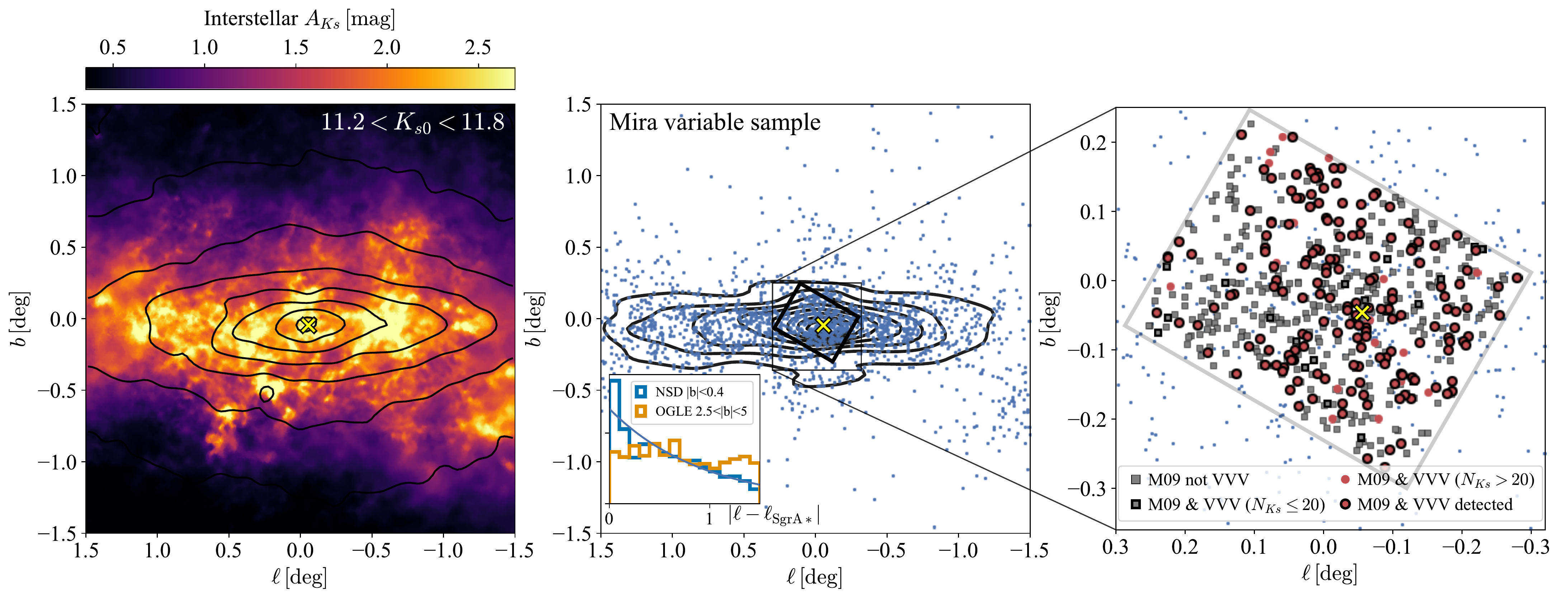}
    \caption{
    Nuclear stellar disc traced by Mira variables: the \textit{left panel} shows the $K_s$ interstellar extinction overlaid with the logarithmically-spaced density contours for stars with $11.2<K_{s0}<11.8\,\mathrm{mag}$ which exhibit a disc morphology. The \textit{central panel} shows the VVV Mira variable sample (including both reliable and unreliable identifications) with their corresponding density given by the contours. This sample also traces the NSD and approximately follows the distribution in the left panel but is subject to selection effects (see Section~\ref{section::selection}). The inset shows the distribution in absolute Galactic longitude (with respect to the location of Sgr A*) of our sample with amplitude $>0.4$ and $|b|<0.4\,\mathrm{deg}$ in blue and the OGLE Mira variables between $2.5\,\mathrm{deg}<|b|<5\,\mathrm{deg}$ from \protect\cite{Soszynski2013} in orange. The line shows an exponential fit with scalelength $130\,\mathrm{pc}$. In the \textit{right panel} we show a zoom-in of the region surveyed by \protect\citetalias{Matsunaga2009}. Grey squares show Mira variables with periods from \protect\citetalias{Matsunaga2009} not in VVV, grey squares outlined in black those with periods in \protect\citetalias{Matsunaga2009} and VVV but with $20$ or fewer epochs, red circles those with periods in \protect\citetalias{Matsunaga2009} and VVV with more than $20$ epochs but not detected picked up by our search, and finally red circles outlined in black are those in our final catalogue that are also in \protect\citetalias{Matsunaga2009} with periods.}
    \label{fig:sample}
\end{figure*}

With the Mira variable sample in hand, we now turn to inspecting some of its properties. 

\subsection{Are they nuclear stellar disc members?}

\begin{figure*}
    \centering
    \includegraphics[width=\textwidth]{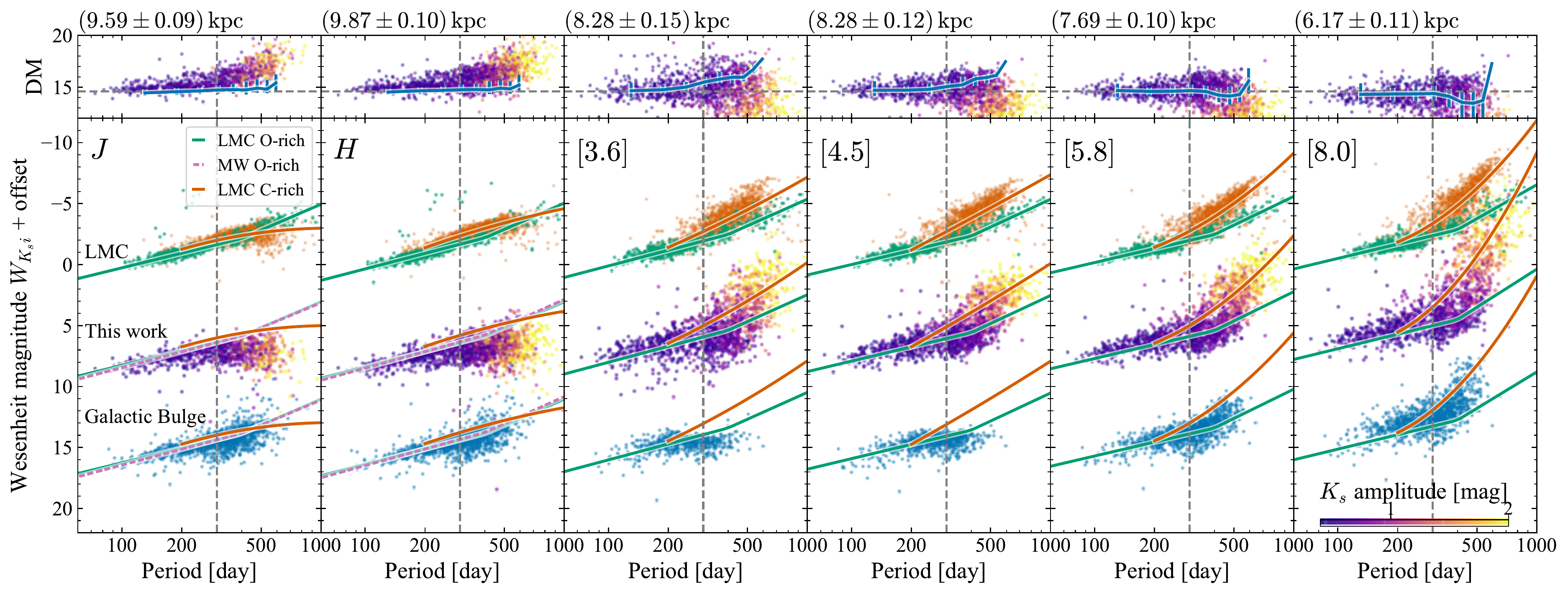}
    \caption{Period--luminosity relations for LMC Mira variables (green O-rich, orange C-rich), inner bulge Mira variables (blue) and our Mira variable sample (coloured by amplitude and limited to reliable sources with $\Delta K_s>0.4$). In all panels we show the Wesenheit magnitudes $W_{Ks,i}=K_s-A_{Ks}/(A_i-A_{Ks})(i-K_s)$ to account for extinction (the LMC row uses the \citealt{WangChen2019} coefficients whilst the other rows use the \citealt{Fritz2011} extinction law applied to the O-rich model spectra described in Section~\ref{section::photo}). The green and orange lines are the fits to the LMC O-rich and C-rich populations and the pink-dashed lines are fits to solar neighbourhood O-rich Mira variables (Sanders, in prep., not shown in LMC row as they overlap the LMC relation). The small top panels show the distance modulus of our sample computed using the O-rich period--luminosity relations (MW for $JHK_s$ and LMC otherwise, with the number above the plot reporting the median distance for stars with periods less than $300$ days) and the blue errorbars show similar for the inner bulge sample. 
    Note the truncation in the $[3.6]$ and $[4.5]$ distributions for the inner bulge sample is due to saturation.
    }
    \label{fig:wesenheit}
\end{figure*}

\begin{figure*}
    \centering
    \includegraphics[width=\textwidth]{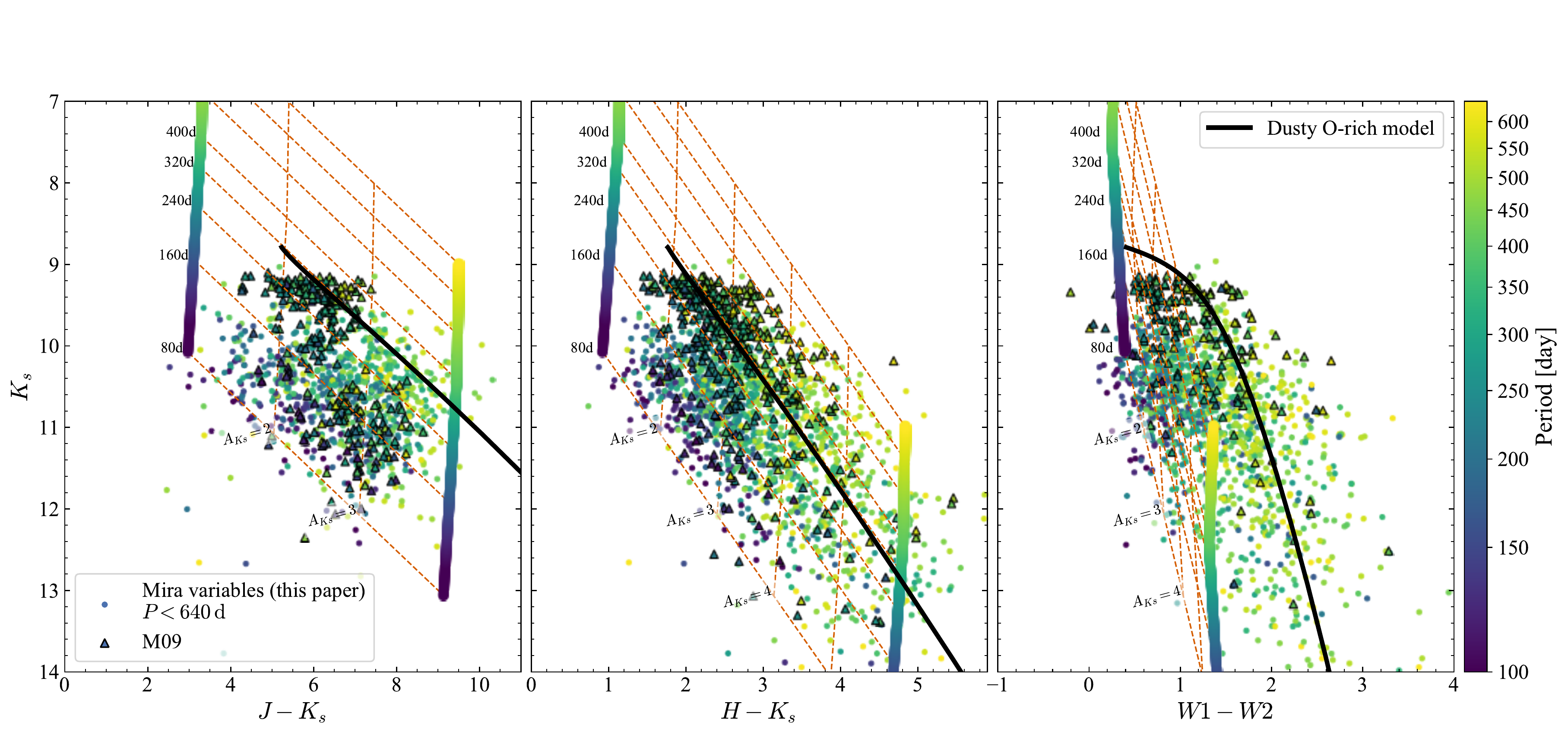}
    \caption{Colour--magnitude diagrams of the Mira variable sample presented in this work (dots, restricting to reliable sources with $\Delta K_s>0.4$ and periods less than $640$ day) with the sample from \protect\citetalias{Matsunaga2009} shown in black outlined triangles (we use the matched VIRAC2 photometry for this sample where the SIRIUS photometry is unavailable). Both samples are coloured by the period. The mesh shows lines of constant period (diagonal) assuming the O-rich LMC period--luminosity relations from Table~\ref{tab:plr} and a Galactic Centre distance, and lines of constant extinction (vertical) assuming the extinction law from \protect\cite{Fritz2011}. The black line shows a sequence of dusty O-rich models of increasing opacity.}
    \label{fig:colour_magnitude_diagram}
\end{figure*}

One crucial question regarding the presented sample is whether in fact the Mira variables are genuinely part of the NSD. 
As a comparison sample, we take all reliable VIRAC2 stars with unextincted $K_s$ (denoted $K_{s0}$) between $11.2$ and $11.9$ (encompassing the asymptotic giant branch bump at the Galactic Centre distance). We use the $E(H-[4.5])$ interstellar extinction maps from \cite{Sanders2022}. From Fig.~\ref{fig:sample}, it is clear that the on-sky distribution of the Mira variable sample is flattened as per the comparison sample. However, as we will discuss shortly, this is in large parts due to the larger in-plane extinction that makes Mira variables faint enough for reliable detection in VVV. We can see from Fig.~\ref{fig:sample} that very approximately the dust is mostly a function of Galactic latitude. In the small inset in the central panel of Fig.~\ref{fig:sample} we display the absolute Galactic longitude (with respect to the location of Sgr A*) of our sample with $|b|<0.4\,\mathrm{deg}$ and amplitude $>0.4$ along with the equivalent distribution of OGLE Mira variables with $2.5\,\mathrm{deg}<|b|<5\,\mathrm{deg}$ \citep{Soszynski2013}. We see that our sample is more centrally concentrated than the nearly flat OGLE `bar-bulge' distribution and has an exponential scalelength of $130\,\mathrm{pc}$ \citep[assuming the population is at $8.275\,\mathrm{kpc}$][]{GravityCollaboration2021}.
This is similar to the scalelength of $\sim90\,\mathrm{pc}$ found in the models of \cite{Sormani2022} although we have not attempted to separate out the bar-bulge contamination. In the models of \cite{Sormani2022}, the projected densities of the NSD and the background bar/bulge are equal along an elliptical contour intersecting $(|\ell|,|b|)\approx(1.5,0)\,\mathrm{deg}$ and $(|\ell|,|b|)\approx(0,0.4)\,\mathrm{deg}$. The NSD is then dominant for smaller $|\ell|$ and $|b|$ increasing to around $80\percent$ of the total projected density around the Galactic centre \citep[see table 2 of][]{Sormani2022}. This implies that on-sky location is only a weak indicator of NSD membership. 

\begin{table}
    \caption{Period--luminosity relations adopted for Fig.~\ref{fig:wesenheit} and for estimating the distance to our Mira variable sample. The period--luminosity relations are $M=a+b(\log_{10}P-2.3)$ for $\log_{10}P<2.6$ and $M=a+0.3b+c(\log_{10}P-2.6)$ for $\log_{10}P>2.6$ where $P$ is the period in days. $e$ gives the extinction coefficient $A_{K_s}/E(i-K_s)$ for the $i$th band used for computing the Wesenheit magnitudes.}
    \centering
    \begin{tabular}{lccccc}
        System & Band & $a$ & $b$ & $c$ & $e$\\
        \hline
        MW&$J$      &$-5.73$&$-3.45$&$-3.87$&$+0.483$\\
        MW&$H$      &$-6.46$&$-3.69$&$-4.75$&$+1.345$\\
        MW&$K_s$    &$-6.89$&$-4.04$&$-5.86$&-\\
        LMC&$J$     &$-5.93$&$-3.35$&$-6.70$&$+0.483$\\
        LMC&$H$     &$-6.67$&$-3.49$&$-6.81$&$+1.345$\\
        LMC&$K_s$   &$-7.01$&$-3.79$&$-6.93$&$-$\\
        LMC&$[3.6]$ &$-7.40$&$-4.05$&$-7.11$&$-2.180$\\
        LMC&$[4.5]$ &$-7.51$&$-3.85$&$-7.35$&$-1.660$\\
        LMC&$[5.8]$ &$-7.68$&$-3.80$&$-7.63$&$-1.522$\\
        LMC&$[8.0]$ &$-7.86$&$-3.87$&$-8.51$&$-1.900$\\
    \end{tabular}
    \label{tab:plr}
\end{table}

A further check of the Mira variable candidates' possible NSD membership is if their distance distribution coincides with the Galactic centre distance of $\sim8.275\,\mathrm{kpc}$ from \cite{GravityCollaboration2021}. Again the background bar/bulge density distribution is also expected to peak here but will have a higher line-of-sight dispersion. In Fig.~\ref{fig:wesenheit} we show the Wesenheit magnitudes ($W_{Ks,x}=K_s-A_{Ks}/(A_x-A_{Ks})(x-K_s)=x-A_x/(A_x-A_{Ks})(x-K_s)$ for band $x$ where $A_x$ is the extinction in this band) computed using the (mean) VVV and GLIMPSE bands and using extinction coefficients, $A_x/A_{Ks}$, found from the median of a grid of O-rich AGB spectra (described in Section~\ref{section::photo}) combined with the extinction curve from \cite{Fritz2011}. This accounts for the potential non-linearity of the extinction coefficients by using an estimate of the $K_s$ band extinction from 2d $E(H-[4.5])$ interstellar extinction maps \citep{Sanders2022}. Wesenheit magnitudes are useful reddening independent measures that can account for reddening from both interstellar and circumstellar dust. However, the choice of the coefficient is key and any misalignment between the interstellar and circumstellar extinction vectors will give rise some spread in the Wesenheit magnitudes \citep{Ita2011}. In Table~\ref{tab:plr} we report the extinction coefficients $A_{Ks}/(A_x-A_{Ks})=A_{Ks}/E(i-K_s)$ at the median $A_{Ks}$ averaged over the model grid. In Fig.~\ref{fig:wesenheit} we also compare with Mira variables in the inner Galactic bulge \citep[$|\ell|<2.5\,\mathrm{deg}$ and $|b|<2.5\,\mathrm{deg}$ from OGLE and Gaia,][]{Soszynski2013,Mowlavi2018} using the $G$-band amplitude cut from \citet[][see Appendix~\ref{appendix:period_comparison}]{Grady2019} and the LMC Mira variables from \cite{Soszynski2009} (again using the $G$-band amplitude cut).

Given the Wesenheit magnitude, we compute the distance using O-rich period--luminosity relations as given in Table~\ref{tab:plr} (Sanders, in prep.). For the Spitzer bands, we consider period--luminosity relations derived for the LMC whilst for the VVV bands we use relations derived for the solar neighbourhood \citep[transformed from the 2MASS system to the VVV system,][]{GonzalezFernandez2018} from a sample of Gaia DR2 O-rich Mira variables defined using the $G$-band amplitude cut \citep[][Sanders, in prep.]{Mowlavi2018,Watson2006}. The distance modulus is shown in the top row in Fig.~\ref{fig:wesenheit}. The short period Mira variables trace a tight period--luminosity relation that is in approximate agreement with the expectation that the Mira variables are situated at the expected Galactic Centre distance and have a similar period--luminosity relation to the solar-neighbourhood and LMC O-rich Mira variables. There is some discrepancy in the derived median distance (top panels of Fig.~\ref{fig:wesenheit}), particularly for $J$, $H$ and $[8.0]$ that could reflect shortcomings in the extinction correction (due to the high extinction for the NSD Mira variables, the Wesenheit magnitudes are quite sensitive to small differences in the coefficients $A_x/(A_x-A_{Ks})$), particularly as there is very little observed variation in the near-infrared period--luminosity relations for O-rich Mira variables \citep[][Sanders, in prep.]{Whitelock2008,Goldman2019} although \citetalias{Matsunaga2009} find quite different period--luminosity relations in the Spitzer bands than \cite{Ita2011} do for the LMC data. In particular, the $[8.0]$ extinction coefficient is quite unreliable as it is very sensitive to the source spectrum and the total extinction. Furthermore, the larger scatter about the period--luminosity relations could also reflect the increased presence of circumstellar dust for these (potentially significantly more metal-rich) stars.

For a different projection of the data, we display three colour--magnitude diagrams in Fig.~\ref{fig:colour_magnitude_diagram} compared to a grid of LMC O-rich period--luminosity relations (Table~\ref{tab:plr}, assuming the distance to the Galactic Centre and using the GLIMPSE bands as approximations of the WISE photometry) and lines of constant extinction \citep{Fritz2011}. No extinction correction has been applied to the data. We see again the sample is approximately consistent with being at the Galactic Centre distance behind approximately $A_{Ks}=2-4$ magnitudes of extinction (which could be a combination of both interstellar and circumstellar dust). The WISE colour--magnitude diagram does not trace the extinction vector due to the increased dustiness of the Mira variables at long periods i.e. the misalignment of the interstellar and circumstellar extinction vectors.

\begin{figure}
    \centering
    \includegraphics[width=\columnwidth]{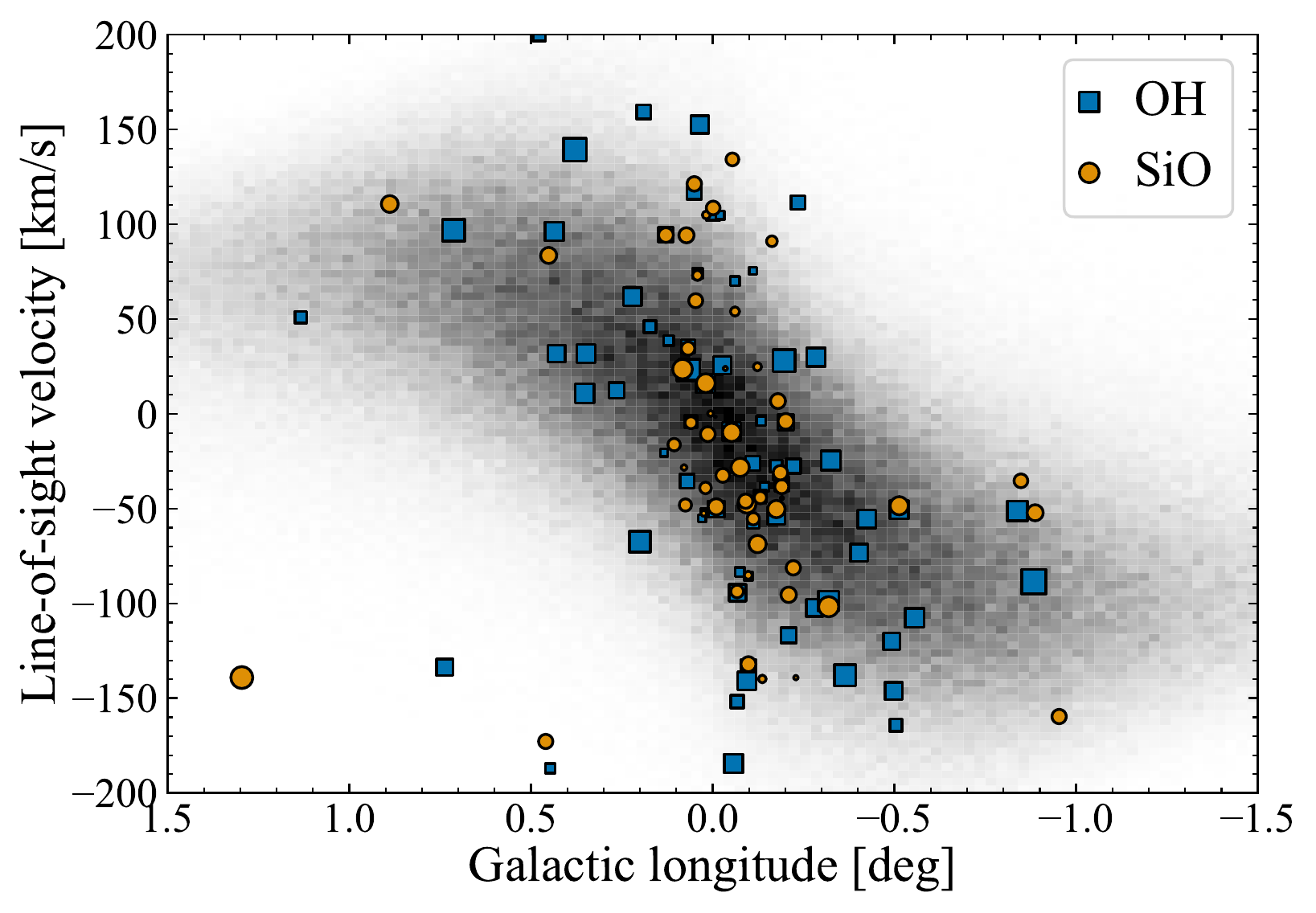}
    \caption{Line-of-sight velocities for our reliable Mira variable sample ($|b|<0.4\,\mathrm{deg}$, $\Delta K_s>0.4$) from maser observations. The background greyscale shows the density (linearly-scaled) from the \protect\cite{Sormani2022} dynamical model (only the NSD component). The blue points show the OH maser observations and orange the SiO. The point size scales linearly with the period of each object.}
    \label{fig:maser}
\end{figure}

Further evidence for the NSD membership of the presented sample can be found through their kinematics. We have proper motion data for all stars in the sample, which we will analyse in a follow-up to this paper. However, there is a subsample of our stars with maser observations (\citealt{Engels2015} for OH masers and \citealt{Messineo2002,Messineo2004}, \citealt{Deguchi2004} and \citealt{Fujii2006} for SiO masers). We show the $\ell$--$v_\mathrm{los}$ distribution of the stars with $\Delta K_s>0.4$ and $|b|<0.4\,\mathrm{deg}$ compared with the dynamical NSD model from \cite{Sormani2022}. In accord with the results of \cite{Habing1983} and \cite{Lindqvist1992}, there is clearly net rotation in this sample reaching the $\sim100\,\mathrm{km\,s}^{-1}$ level in line with the dynamical model giving confidence that a high fraction of our long-period sample are indeed NSD members. However, the Mira variables with maser observations are limited to periods $\gtrsim400$ days. Therefore, based on maser observations alone, we are unable to conclude anything on the NSD membership of the short-period Mira variables.

\subsection{The sample selection function}\label{section::selection}
\begin{figure*}
    \centering
    \includegraphics[width=0.9\textwidth]{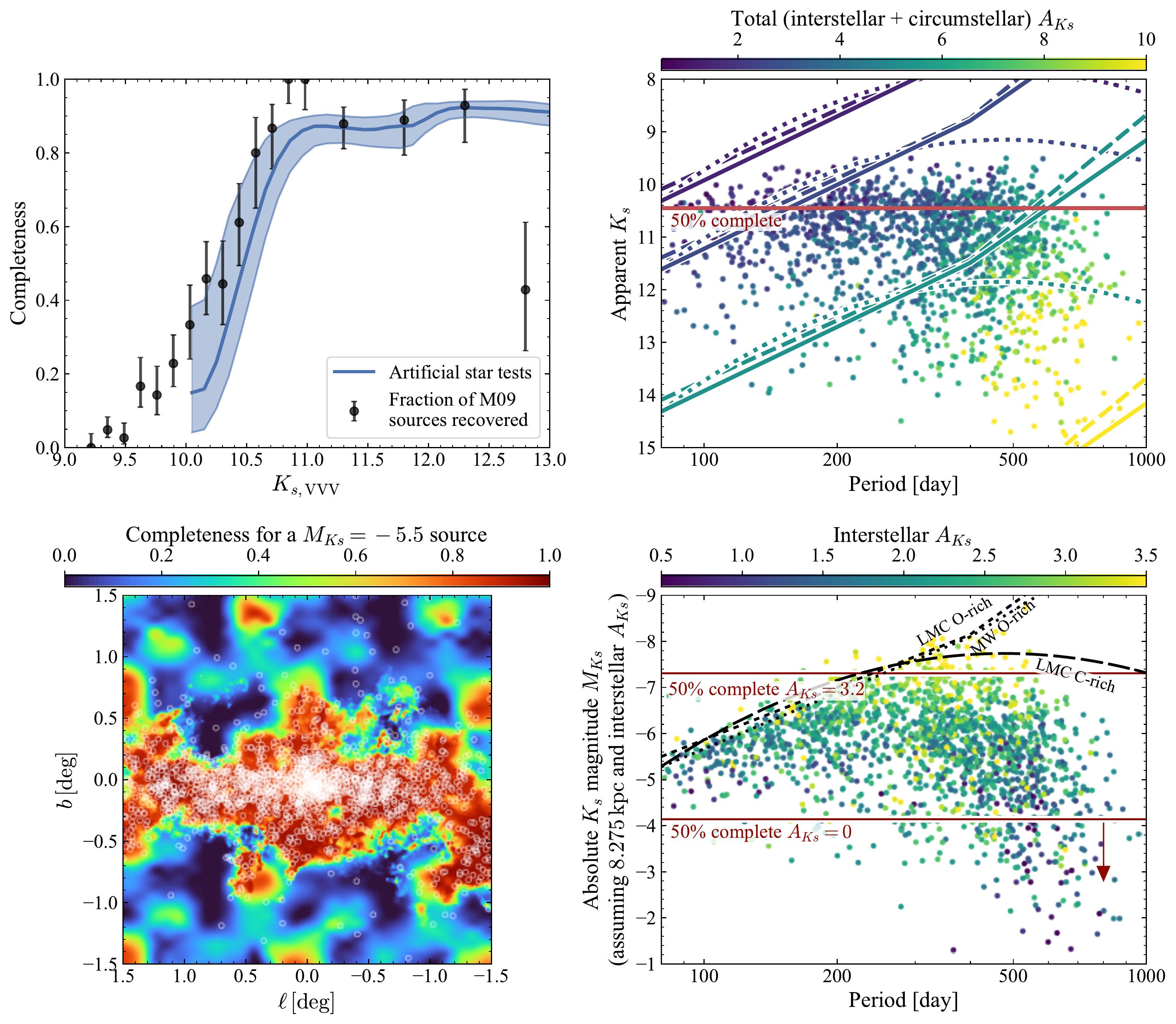}
    \caption{
    Selection effects for our sample: in the top left panel we show in blue a comparison of the completeness computed from artificial star tests (the line and shaded bracket give the median and $\pm1\sigma$ over the region surveyed by \protect\citetalias{Matsunaga2009}) and the fraction of sources with periods (predominantly Mira variables) in \protect\citetalias{Matsunaga2009} recovered by this work. The top right panel shows the apparent $K_s$ magnitude vs. period for our reliable sample coloured by their extinction estimated from the colour excess with respect to the $K_s-[4.5]$ period colour relation (a proxy for the interstellar \emph{plus} circumstellar extinction). The horizontal red line shows the $50\percent$ completeness from the top left panel. Period-luminosity relations (as labelled in the lower right panel) are shown for stars at the Galactic centre distance extincted by different amounts ($A_{Ks}=1,2.3,5,10$ according to their colours). The bottom right panel shows the same sample in absolute magnitude vs. period (assuming all sources are at the Galactic Centre distance -- more distant sources could also enter the selection). The $K_s$ magnitudes are dereddened using solely the interstellar extinction estimates. The horizontal lines show the $50\percent$ completeness limit assuming zero and $3.2$ magnitudes (the $90$th percentile of the sample) of extinction. The bottom left panel shows the completeness for sources with absolute magnitude of $-5.5$ with the sample overlaid in white. Note the correlation between the sample and the completeness map.}
    \label{fig:selection_function_plot}
\end{figure*}

As already highlighted, the presented Mira variable sample appears to be flattened on the sky and so traces the NSD. However, this effect is  driven in part by a combination of observational limitations (e.g. arising from the magnitude range of the VVV survey) and the distribution of interstellar dust towards the NSD. VVV begins to saturate around $K_s\approx11.5$. Consider an unreddened Mira variable with period $\sim100\,\mathrm{day}$ at the Galactic Centre. This is typically the lowest period reached by Mira variables and hence is the faintest source we might consider. This star will have $K_s\sim9$ and will hence be too bright for VVV. However, as the reddening towards the Galactic Centre is significant,  $A_{Ks}\approx2\,\mathrm{mag}$, typically this type of source will have $K_s\sim11$ and so be just measurable by VVV. In the top right panel of Fig.~\ref{fig:selection_function_plot} we display the distribution of our sample in apparent $K_s$ magnitude vs. period. Due to saturation effects, we have very few stars with $K_s\lesssim10$. Comparison with the period-luminosity relations from Table~\ref{tab:plr} for a star at the Galactic centre distance reddened by different amounts demonstrates clearly the necessity of some reddening to make the Mira sample faint enough for reliable VVV observations. As we will discuss further, this can be due to either interstellar \emph{or} circumstellar extinction.

We can assess the effect of incompleteness on our sample more quantitatively using the completeness calculations presented by \cite{Sanders2022}. These were based on artificial star tests requiring that the artificial stars are recovered in $20\percent$ of the observations. An example of a typical result from this calculation is shown in the top left panel of Fig.~\ref{fig:selection_function_plot}. Combining these calculations with the 2d $E(H-[4.5])$ interstellar extinction map presented in \cite{Sanders2022}, we find the on-sky completeness as shown in the lower left panel of Fig.~\ref{fig:selection_function_plot}. Clearly, we are only nearly complete in the high-density regions near the mid-plane. The completeness map lines up very well with the on-sky source density. 

In the lower right panel of Fig.~\ref{fig:selection_function_plot} we show the absolute $K_s$ magnitudes computed assuming a distance of $8.275\,\mathrm{kpc}$ \citep{GravityCollaboration2021} and using extinctions derived in the next subsection (\ref{section::spectral_fits}). These extinction estimates only account for the interstellar extinction and are biased by our prior estimates from the 2d $E(H-[4.5])$ interstellar extinction map. We note that in this projection a high fraction of our sample are too faint to be consistent with the LMC/solar-neighbourhood period--luminosity relations despite the results of Fig.~\ref{fig:wesenheit}. Using these extinction estimates then suggests that the objects are more distant than the NSD. However, another explanation is that the average extinction estimates we are using here are inappropriate for our sample which may be more dust-obscured than the average bulge giant star in each region of the sky. The extinction varies significantly over very small scales in this part of the sky and we are biased towards fainter objects. It could also be that in high extinction regions we are losing the bulge/NSD red giants in $(H-[4.5])$ as they are too faint and the resulting extinctions are biased towards the extinction of more foreground objects \citep[as discussed in][in the comparison of {$E(H-[4.5])$} from giant branch stars compared with $E(H-K_s)$ from red clump stars]{Sanders2022}. The faintness may also arise due to the stars having more circumstellar dust than expected. This may be the case at the long-period end ($\gtrsim400\,\mathrm{day}$) but such dust appears comparatively rare in short-period objects \citep[][and subsection~\ref{section::spectral_fits}]{Ita2011}. Other effects, such as metallicity variation, may give rise to intrinsic differences in Mira variables independent of the dust properties although there is little evidence for this being such a significant effect in the near-infrared \citep{Whitelock2008}. Finally, as already noted, we are working at the bright limits of VVV and saturation could cause stars to be fainter than they are.

To test this discrepancy further, we follow \cite{Matsunaga2009} and estimate the total colour excess with respect to a period-colour relation (from Table~\ref{tab:plr}). This estimate includes contributions from both interstellar and circumstellar extinction (or more precisely any additional circumstellar extinction relative to the reference population for the period-colour calibration). The points in the top right panel of Fig.~\ref{fig:selection_function_plot} are coloured by the $A_{Ks}$ implied by the $(K_s-[4.5])$ colour excess. For the bluer objects this colour excess estimate agrees with the 2d $E(H-[4.5])$ extinction map but for most sources it is higher (as seen approximately by comparing the colourbars of the upper and lower right panels of Fig.~\ref{fig:selection_function_plot}), even for short-period objects. Therefore, there is evidence that the 2d $E(H-[4.5])$ interstellar extinction map produces underestimates to these objects. For longer-period objects it is more difficult to say what is happening as it requires disentangling the interstellar and circumstellar extinction. These stars have evidence of significant circumstellar extinction but they are also younger objects, probably embedded in more dusty environments. We note that this mismatch between the Mira variable colour excesses and the interstellar extinction maps was also found by \cite{Nikzat2022} who attributed it to the Mira variables being at larger distances (possibly in the background disc) and so behind more dust than the bulge stars. This may well be the case for their higher latitude sample, but Fig.~\ref{fig:wesenheit} demonstrates that the distances of our stars are consistent with NSD/bulge membership. 

Returning to what the extinction means for the completeness of our sample, we have marked with horizontal lines the values at which samples with different amounts of interstellar reddening would be $50\percent$ complete i.e. we would typically see all stars fainter than these limits. We see that our sample truncates at around the limit for the $90$th percentile of extinction $A_{Ks}=3.2$ for our sample. We have no way to probe sources intrinsically brighter than this. Typically these would be the longer period sources, but we notice that many of the long-period sources actually fall significantly under the expected period--luminosity relations (either as a result of circumstellar dust or underestimated interstellar extinction) as already highlighted above. This is fortuitous for our purposes. 

Finally, as a validation of the selection effects affecting the catalogue, the left panel of Fig.~\ref{fig:selection_function_plot} compares the fraction of \citetalias{Matsunaga2009} sources with reported periods we recover together with the expectation from the artificial star completeness tests. We have approximated $K_{s,\mathrm{VVV}}$ as the mean SIRIUS $K_s$ magnitudes from \citetalias{Matsunaga2009} which are $\sim0.05$ fainter than the $K_s$ flux-means reported by \citetalias{Matsunaga2009}. The artificial star tests are for constant sources and no adjustment has been made to consider how an artificial variable source might be recovered. However, the high degree of correspondence suggests our catalogue is as complete as can be expected given the quality of the photometry, and our algorithms for selecting and processing variable stars have not artificially removed any genuine Mira variables. Of the $549$ \citetalias{Matsunaga2009} long period variables with periods we detect $169$ ($\sim30\percent$). $212$ are in VIRAC2 but only $192$ have more than the $20$ epochs required to enter our high-quality catalogue (the missing $20$ all satisfy the Wesenheit period cuts so enter the combined catalogue). Of the remaining $23$ with a sufficient number of epochs but that we failed to detect, $3$ fail our Wesenheit-period cuts (hence why we end up with an additional $40$ sources in the combined catalogue), $2$ stars are missed by the initial candidate selection (visible in Fig.~\ref{fig:mira_selection}), and then $7$ fail the period cuts to be further processed using the Gaussian process model. $6$ of the remaining $11$ stars fall outside the period--amplitude selection (as shown in Fig.~\ref{fig:selection}) and the other $5$ fail the parallax and Wesenheit-period cuts.

Although the completeness is well understood, the contamination of the catalogue is much more difficult to assess. Through visual inspection we assessed the contamination at the $5$ to $10\percent$ level before removal of potential contaminants. One potential contaminant is young stellar objects. Our cuts on Wesenheit magnitude have sought to eliminate these intrinsically dimmer contaminants. We have checked whether any of our objects are in the young stellar object catalogues of \cite{Guo2021} and \cite{Guo2022}. One source in our catalogue (VIRAC Mira 68) has been identified by \citet[][VVV\_PB\_122]{Guo2022} as a potential periodic outbursting young stellar object. This star is also in the \citetalias{Matsunaga2009} catalogue with an associated period and $K_s$ amplitude of $0.55$. It is one of the brightest objects in the \cite{Guo2022} catalogue so we therefore think it likely that it is a long-period variable but it is unclear.

With these considerations, there is significant scope for improving the completeness of the Mira variable NSD sample. It may be possible to utilise $(Z,Y,J,H,W1,W2)$ photometry without the reliance on $K_s$ -- unfortunately in the VVV reduction we are using a reliable $K_s$ detection is a necessary requirement although this isn't a fundamental limitation. Furthermore, the presence of dark lanes in infrared images suggests we may be missing some Mira variables embedded in or shrouded by such thick dust that our searches need to go fainter to probe the full Mira variable population. In future, the PRime-focus Infrared Microlensing Experiment (PRIME)\footnote{\url{http://www-ir.ess.sci.osaka-u.ac.jp/prime/index.html}} and the JASMINE satellite \citep{Gouda2020}\footnote{\url{http://jasmine.nao.ac.jp/index-en.html}} are expected to provide better coverage of this area particularly at the brighter magnitude end.

\subsection{Photometric characterisation of our sample}\label{section::photo}
\begin{figure}
    \centering
    \includegraphics[width=\columnwidth]{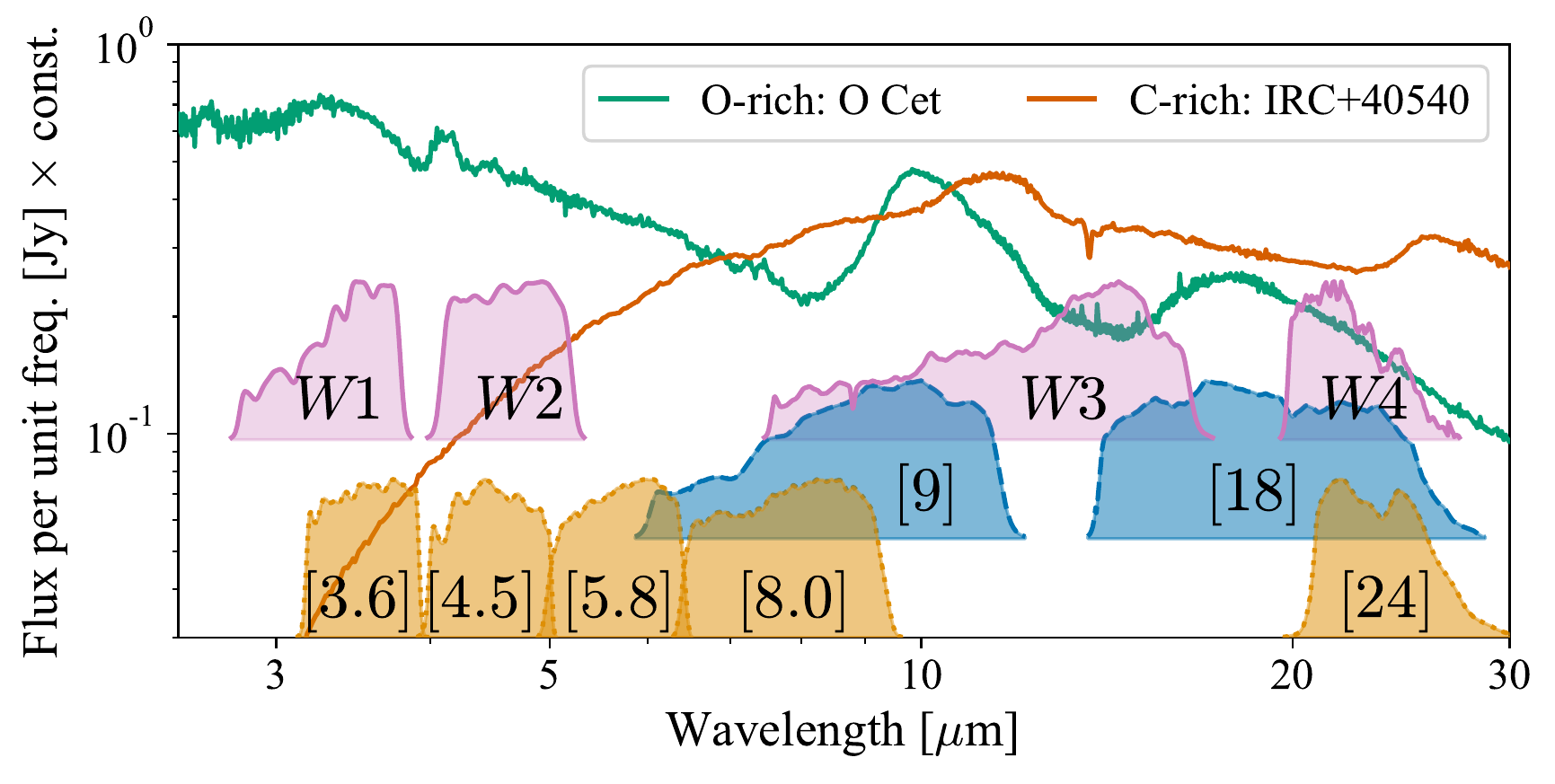}
    \caption{Mid-infrared spectra of typical O-rich (green) and C-rich (orange) Mira variables (o Cet -- Mira itself, and IRC+40540) from \protect\cite{Sloan2003}. Also displayed are the relative spectral response functions for WISE (pink), AKARI (blue) and Spitzer IRAC/MIPS (orange). Note the strong silicate features in the O-rich spectrum at $10$ and $18\,\mu\mathrm{m}$ and the SiC feature at $11\,\mu\mathrm{m}$ in the C-rich spectrum.}
    \label{fig::spectra}
\end{figure}

\begin{figure*}
    \centering
    \includegraphics[width=\textwidth]{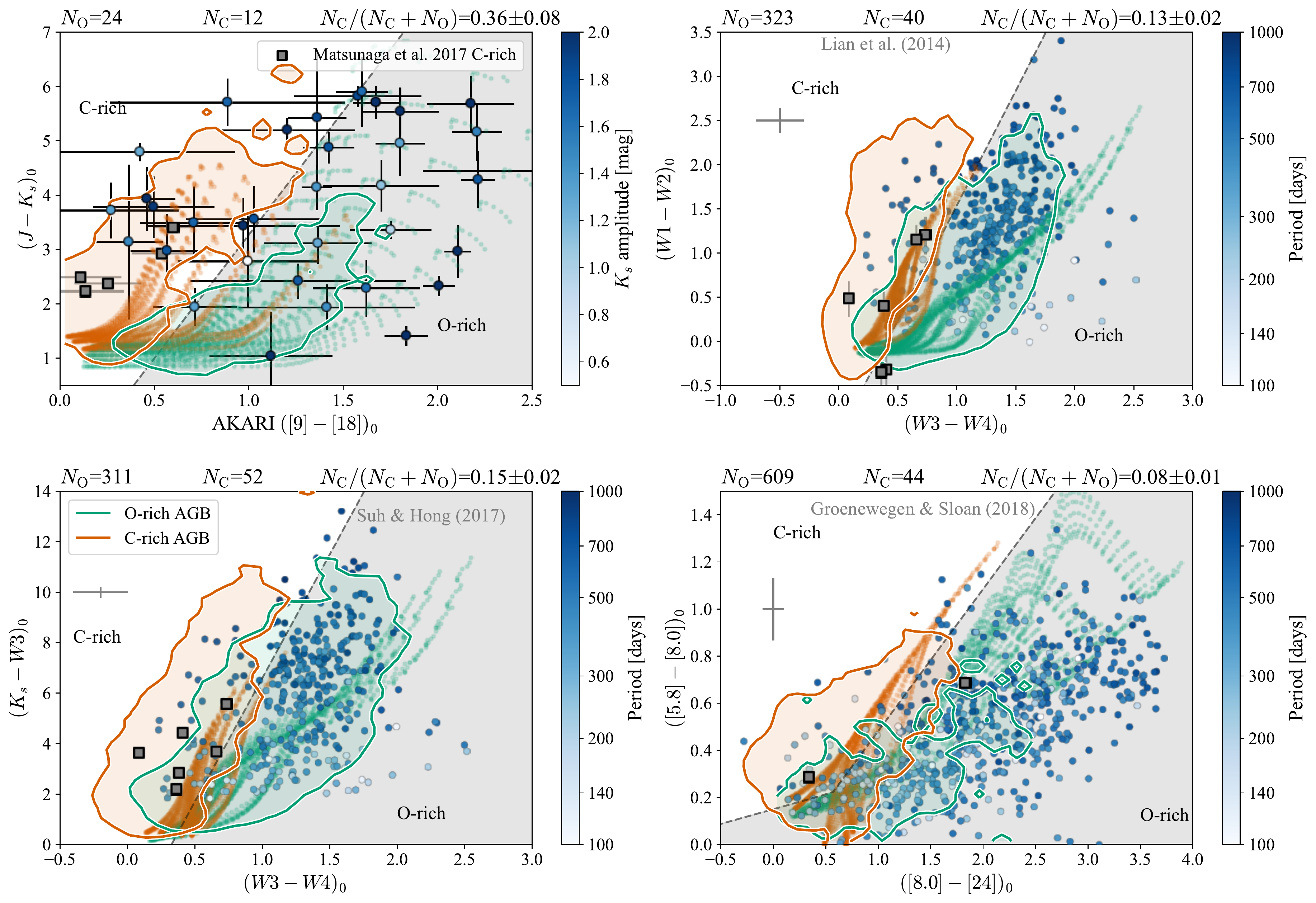}
    \caption{Compositions of the reliable Mira variables: four colour-colour diagrams in which O-rich and C-rich AGB stars separate. The O-rich silicate feature at $10\,\mu\mathrm{m}$ is covered by AKARI $[9]$, WISE $W3$ and GLIMPSE $[8.0]$ allowing for separation of O- and C-rich. Blue circular points show the Mira variable sample presented here ($90\percent$ of the sample have uncertainties smaller than the displayed grey errorbars), grey squares are the OGLE C-rich Mira variables from \protect\cite{Matsunaga2017}, the green and orange shaded regions show the extent of the O-rich and C-rich AGB samples from \protect\citet[][except for the lower right panel where Mira variables in the LMC are shown]{SuhHong2017} and the green and orange points are theoretical models of dusty O-rich and C-rich AGB stars. 
    The \emph{top left} panel shows the distribution of the sample in dereddened $(J-K_s)$ against dereddened $([9]-[18])$ from AKARI coloured by amplitude. We show the line defined by \protect\cite{Ishihara2011} to separate C-rich and O-rich Mira and display the six OGLE Mira classified as C-rich by \protect\cite{Matsunaga2017}.
    The \emph{top right} panel shows the distribution in dereddened WISE colours, as suggested by \protect\cite{Lian2014}, now coloured by period.
    The \emph{bottom left} panel shows similar but for a combination of $K_s$ and WISE magnitudes similar to that suggested by \protect\cite{SuhHong2017}. 
    The \emph{bottom right} panel shows a combination of Spitzer IRAC/MIPS colours used by \protect\cite{Groenewegen2018}.
    Above each panel we display the number of Mira classified as C- and O-rich. From this, we estimate that $\lesssim10\percent$ of our sample are C-rich Mira.
    }
    \label{fig:crichorich}
\end{figure*}
AGB stars are typically classified by their C and O compositions, with C-rich C stars having $[\mathrm{C/O}]>1$, O-rich M stars having $[\mathrm{C/O}]<1$ and the S stars having $[\mathrm{C/O}]\approx1$ \citep{Hofner2018}. The $[\mathrm{C/O}]$ ratio governs the dominant molecular species in the atmospheres and dusty circumstellar envelopes of AGB stars (e.g. silicates for O-rich and carbonaceous species, e.g. C$_2$ or C, for C-rich), and, as such, governs the observed colours, and the mass loss rate of AGB stars. The observed $[\mathrm{C/O}]$ ratio is determined by the third dredge-up, which mixes more central C-rich material into the outer envelope, and is a function of both mass and metallicity of the star. Typically C-rich AGB stars are associated with younger and/or more metal-poor systems. For this reason, we typically find the bulge of the Milky Way occupied by O-rich Mira variables, whilst the outer disc has a greater number of C-rich stars \citep{Blanco1984}. Until the recent discovery of $5$ C-rich Mira variables in the bulge by \cite{Matsunaga2017}, it was believed the bulge consisted almost solely of O-rich Mira variables. However, it is not clear whether these stars are associated with younger or more metal-poor populations, or are the result of binary evolution. It is important to classify the types of Mira variables we are dealing with, not just because it gives a reflection of the age and metallicity of population they belong to, but also because O-rich and C-rich Mira variables follow quite different period--luminosity relations. 

For comparison to our Mira variable sample, we have run two grids of dusty AGB models, one O-rich and one C-rich. We utilize the \textsc{DUSTY} code \citep{IvezicElitzur1997} using the analytic radiatively-driven wind solution. The source spectra are taken from the synthetic libraries of \cite{Aringer2016} and \cite{Aringer2019} where we use the solar mass models with $[\mathrm{C}/\mathrm{O}]=0.55$, solar metallicity $Z=Z_\odot$ and $\log g=0$ for the O-rich sources and the solar mass models with $[\mathrm{C}/\mathrm{O}]=1.1$, $Z=Z_\odot$ and $\log g=-0.4$ for C-rich. These choices are similar to those of \cite{Lian2014} but as discussed by \cite{Aringer2009} and \cite{Kucinskas2005} the variation of the broadband infrared colours with mass, surface gravity and $[\mathrm{C}/\mathrm{O}]$ is weak. For O-rich circumstellar dust we take the O-rich interstellar warm silicate optical constants from \cite{Suh1999} and for C-rich we take the amorphous carbon optical constants from \cite{Suh2000}. We run grids of models parametrized by the opacity at $10\,\mu\mathrm{m}$, $\tau_{10}$, (up to a maximum of $\tau_{10}=10$ for the O-rich models and $\tau_{10}=0.5$ for the C-rich), inner dust temperature $T_\mathrm{in}$ (between $600\,\mathrm{K}$ and $1400\,\mathrm{K}$ for O-rich and $600\,\mathrm{K}$ and $1800\,\mathrm{K}$ for C-rich) and the effective temperature $T_\mathrm{eff}$ of the source spectrum \citep[see][for other model grids]{Goldman2020}. The radiatively-driven wind solution is self-similar \citep{ElitzurIvezic2001} allowing for simple rescaling for different central luminosities and gas-to-dust ratios. We have displayed the sequence of models with $T_\mathrm{eff}=2600\,\mathrm{K}$ and $T_\mathrm{in}=600\,\mathrm{K}$ in Fig.~\ref{fig:colour_magnitude_diagram}. We see that the circumstellar dust acts like the interstellar extinction vector in the $(J-K_s)$ and $(H-K_s)$ vs. $K_s$ colour--magnitude diagrams. However, in $(W1-W2)$ the models are aligned quite differently to the interstellar extinction vector and indeed it appears that the dusty models are necessary to match the data distribution.

As a further comparison to our sample, a large catalogue of O- and C-rich AGB stars was presented by \cite{SuhHong2017} who compiled classifications from the literature which were primarily based upon maser observations (e.g. OH masers are associated with an O-rich AGB star) or low resolution spectroscopy \citep[see][for a significantly expanded, more recent catalogue]{Suh2021}. As discussed in \cite{Suh2021}, the two types can also be more approximately separated with photometry. Colour selections are most effective in the near- and mid-infrared where there are a number of molecular features. In Fig.~\ref{fig::spectra} we show the mid-infrared spectra of typical O-rich and C-rich Mira variables from the catalogue of \cite{Sloan2003}. Several authors have suggested colour combinations in which the C-rich and O-rich Mira variables separate, which mostly rely on the strong silicate feature at $10\,\mu\mathrm{m}$ which is covered by the AKARI $[9]$ band and WISE W3. \cite{Lebzelter2018} demonstrated that a combination of Gaia and 2MASS colours could be employed to effectively separate Mira variables in the LMC. Unfortunately, for our purposes, Gaia photometry is unavailable for all but a handful of our Mira variables due to high extinction. Here we explore the following four near- and mid-infrared colour-colour selections as shown in Fig.~\ref{fig:crichorich}:
\begin{enumerate}
    \item \cite{Ishihara2011} showed how AKARI photometric bands $[9]$ and $[18]$ covered the O-rich silicate features at $10$ and $18\,\mu\mathrm{m}$. We employ the selection in $(J-K_s)$ against $([9]-[18])$ where C-rich stars have redder $([9]-[18])$ at fixed $(J-K_s)$. This is colour-colour combination in which \cite{Matsunaga2017} searched for C-rich Mira variables in the bulge. The line separating O- and C-rich is 
    \begin{equation}
        (J-K_s)=4.55([9]-[18])-1.27.
    \end{equation}
    \item Both \cite{Lian2014} and \cite{Nikutta2014} demonstrated 
    that O-rich have redder $(W3-W4)$ at fixed $(W1-W2)$ than C-rich. The line separating O- and C-rich is \begin{equation}
    (W1-W2)=2.35(W3-W4)-0.84.
    \end{equation}
    \item In a very similar vein, \cite{SuhHong2017} show using IRAS photometry the populations separate in $(K_s-[12])$ against $([12]-[25])$, again with O-rich redder in $([12]-[25])$ than C-rich. $[12]$ and $[25]$ are similar to $W3$ and $W4$ respectively. The line separating O- and C-rich is 
    \begin{equation}
        (W3-W4)=0.102(K_s-W3)+0.427.
    \end{equation}
    \item \cite{Groenewegen2018} demonstrated the clear separation of O-rich and C-rich AGB stars in the $([5.8]-[8.0])$ against $([8.0]-[24])$ plane where $[24]$ is from the MIPS instrument on Spitzer \citep[see also][]{Kastner2008}. The main separating line is
    \begin{equation}
        ([5.8]-[8.0])=0.603([8.0]-[24])-0.121,
    \end{equation}
    although there is a slight break at the blue end.
\end{enumerate}
All of the colour-colour selections rely on dereddened photometry. We use extinction coefficients computed as the median over the O-rich model grid at each star's total interstellar extinction using the extinction law from \cite{Fritz2011} and the total extinction is derived from the $(H-[4.5])$ excess of all giant stars \citep{Majewski2011,Sanders2022}. Using the (interstellar \emph{plus} circumstellar) extinction estimated from the period-colour relations \citep{Matsunaga2009} does not significantly alter the results.
\begin{figure}
    \centering
    \includegraphics[width=\columnwidth]{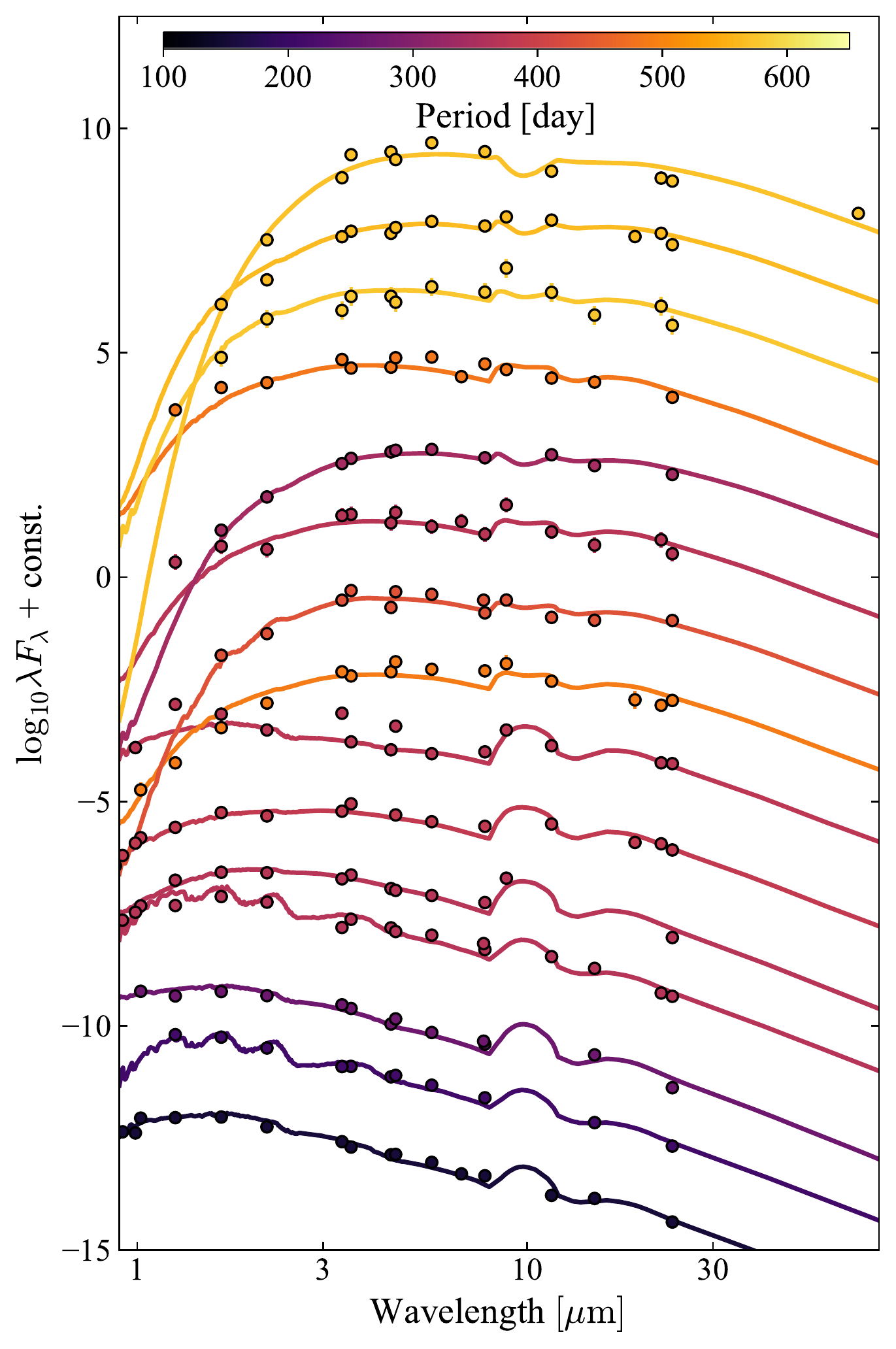}
    \caption{Multiband photometry fit with O-rich dusty AGB models. The stars are ordered such that their derived mass loss rates (and hence approximately the optical depths) increase upwards and are coloured by their period. The points are photometric measurements corrected by the best-fit interstellar extinction.}
    \label{fig:spectral_stack}
\end{figure}
There are two caveats when using WISE photometry: (i) AGB stars are bright in WISE bands so likely to be saturated. However, reliable magnitudes are still extracted for sources fainter than $(W1,W2,W3,W4)=(2,1.5,-3,-4)$ provided we correct the photometry as per footnote\footnoteref{footnote1}; (ii) the $W3$ and $W4$ angular resolutions are $6.3$ and $12.0\,\mathrm{arcsec}$ respectively. In the highly-crowded Galactic centre regions it is likely $W4$ measurements are contaminated. We have found that sources with $W4$ reduced chi-squared $>5$ are likely blended as they lie at significantly higher magnitudes than expected from comparison with both models and other data samples. We therefore adopt this as a quality cut.

In Fig.~\ref{fig:crichorich} we show our Mira variable sample in the four highlighted colour-colour spaces. We also overplot the \cite{SuhHong2017} AGB stars (those with WISE measurements greater than the previously quoted bright limits) and the two families of dusty AGB models. We observe that the majority of the sample are consistent with O-rich chemistry. Simply counting the number of stars in each region of the plots, we find $(36\pm8)\percent$, $(13\pm2)\percent$, $(15\pm2)\percent$ and $(8\pm1)\percent$ C-rich stars using the cuts (i), (ii), (iii) and (iv). There are very few stars with AKARI photometry and the uncertainties in the photometry are large, so we are more inclined to follow the GLIMPSE- and WISE-based cuts which suggest $\sim10\percent$ C-rich. Note though that the uncertainties in the photometry are significant (mostly coming from extinction uncertainties) and our simple cuts are not perfect. Indeed, we see from the AGB and LMC samples that there is contamination from O-rich within the defined C-rich region. Additionally, contamination in the longer wavelength bands, which might be expected for these crowded fields and large point-spread functions, could artificially move objects to redder colours and may be the cause of the disagreement between the dusty AGB models and the data (although this may also reflect shortcomings of the modelled dust composition). Finally, there is some sensitivity to the choice of extinction coefficients, particularly for $[8.0]$ which is quite sensitive to source spectrum and total extinction, and so possibly explains the small disagreement between the data and the model spectra. In conclusion, we find that the fraction of C-rich Mira variables in our sample is $\lesssim10\percent$ although it appears there are a few good candidates for genuine C-rich bulge stars as found by \cite{Matsunaga2017}.

\subsection{Spectral fits}\label{section::spectral_fits}
To further characterise the properties of the Mira variable sample, we fit the dusty AGB models described in the previous section to the broadband photometry. Based on the considerations above, we restrict to only considering the O-rich models. For each model we have computed the model flux $\tilde f=\lambda F_\lambda$ using the filters provided by the SVO filter service \citep{svo1,svo2}. For each star, we then minimise
\begin{equation}
    \sum_{i\in\mathrm{bands}}\frac{(f_i-\tilde f_i(T_\mathrm{eff},T_\mathrm{in},\tau_{10},A_{0},\mathcal{N})^2)}{\sigma_{fi}^2}+\frac{(A_{0}-0.7E(H-[4.5])^2}{(0.7\sigma_{E(H-[4.5]})^2},
\end{equation}
with respect to the model parameters $(T_\mathrm{eff},T_\mathrm{in},\tau_{10},A_{0},\mathcal{N})$.
The model grid is parametrized by $(T_\mathrm{eff},T_\mathrm{in},\tau_{10})$: the effective temperature of the star, the temperature of the dust at the inner edge and the optical depth at $10\,\mu\mathrm{m}$. $\mathcal{N}$ is a free normalization that can either be interpreted as distance or luminosity information.
We extinct the model spectra using the extinction law from \citet{Fritz2011} normalized by the reddening at $\lambda=2.149\,\mu\mathrm{m}$, $A_0$. This procedure accounts for possible non-linearities in the extinction coefficients. A prior is placed on $A_0$ based on the 2d $E(H-[4.5])$ interstellar extinction maps from \cite{Sanders2022}. However, as $A_0$ is free, it can in theory include contributions from both interstellar \emph{and} circumstellar extinction, although we expect in large part the circumstellar extinction is handled by the dust intrinsic to the models ($\tau_{10}$). Previously we have suggested the interstellar extinction towards these sources is underestimated. Relaxing the prior on $A_0$ produces degeneracies with $\tau_{10}$ but has little impact on the derived mass loss rates. The data fluxes, $f_i$, are from VVV ($Z,Y,J,H,K_s$), DECAPS ($z,y$), GLIMPSE ($[3.6],[4.5],[5.8],[8.0]$), WISE ($W1,W2,W3,W4$), ISO ($[7],[15]$), AKARI ($[9],[15]$) and MIPS ($[24],[70]$), and are found using the zeropoints provided by the SVO filter service. The uncertainties $\sigma_{fi}$ are a quadrature sum of the photometric uncertainties, the amplitude given the measured $K_s$ amplitude and the \cite{Iwanek2021} O-rich amplitude relation (many of the measurements are based on multi-epoch observations so the scatter is expected to be smaller than this but this will only marginally affect the results) and an error floor term, $\sigma_0$, to capture any additional variation. We interpolate our model grid fluxes for each choice of $(T_\mathrm{eff},T_\mathrm{in},\tau_{10},A_{0},\mathcal{N})$. After an initial fit, we remove any datum that differs by more than $5\sigma$ from the best-fit and repeat the fit. In Fig.~\ref{fig:spectral_stack} we show some example fits. The most pronounced spectral feature is the $10\,\mu\mathrm{m}$ silicate feature that transitions from in emission at low opacity to in absorption at high opacity \citep{Suh2021}.

\begin{figure*}
    \centering
    \includegraphics[width=\linewidth]{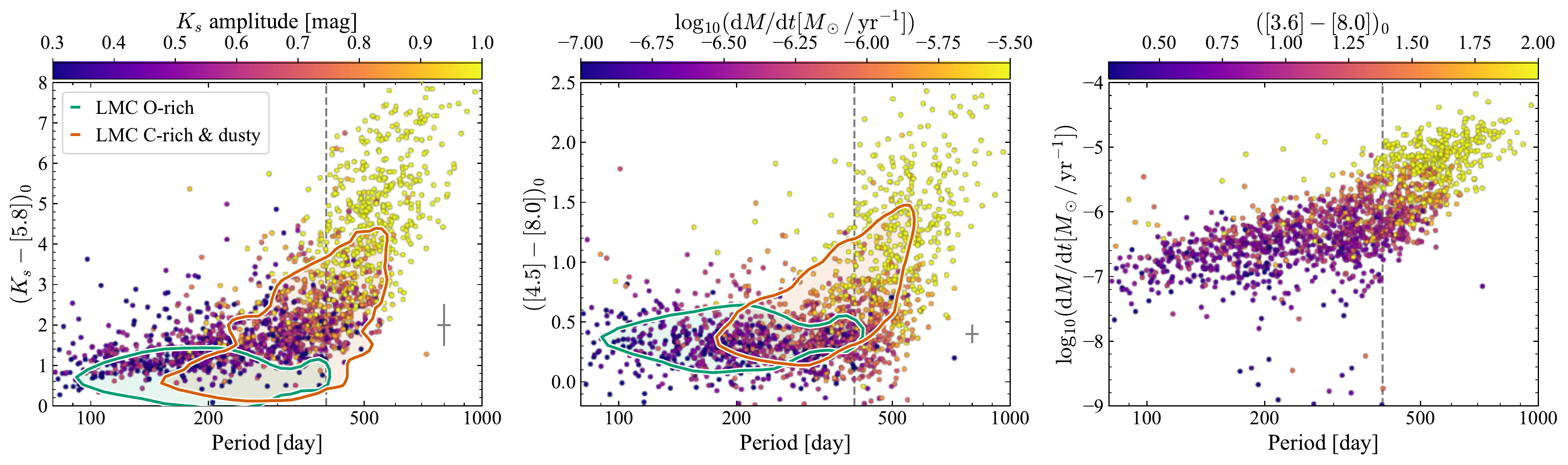}
    \caption{Period--colour diagrams for our reliable Mira variable sample and derived mass loss rates: the leftmost panel shows the distribution in period against $(K_s-[5.8])$ colour (dereddened using a 2D interstellar extinction map derived from the $H-[4.5]$ colour excess of giant stars) coloured by amplitude. $90\percent$ of the sample have errors smaller than the displayed grey errorbar. We overlay the distribution of O-rich (green) and C-rich or dusty (red) LMC Mira variables using the classification from \protect\cite{Lebzelter2018}. The second panel shows similar but for $([4.5]-[8.0])$ colour and coloured by the modelled mass loss rate assuming O-rich composition. There is a clear transition around 400 days (marked by the vertical line) where the Mira variables become increasingly dust-dominated. The third panel shows the modelled mass loss rate against period coloured by dereddened $([3.6]-[8.0])$.}
    \label{fig:period_colourr}
\end{figure*}

In Fig.~\ref{fig:period_colourr} we show the distribution of our sample in colour vs. period space and period vs. mass loss rate. The expansion velocity and mass loss rate can be computed from the radiatively-driven wind model through scaling relations \citep{ElitzurIvezic2001} assuming a luminosity and gas-to-dust ratio. Here we assume all stars are located at $8.275\,\mathrm{kpc}$ \citep{GravityCollaboration2021} to convert the normalization $\mathcal{N}$ into a luminosity and we set the gas-to-dust ratio at $r_\mathrm{gd}=45$ \citep{Goldman2017} such that the expansion velocities approximately match the expansion velocities measured for the OH maser sources ($v_\mathrm{exp}\propto r_\mathrm{gd}^{-1/2}$ and $\dot M\propto r_\mathrm{gd}^{1/2}$ if other gas-to-dust ratios are required). Note that this procedure produces large expansion velocities at short period so it is likely in reality $r_\mathrm{gd}$ increases for shorter period objects. This introduces a factor of a few uncertainty in the mass loss rates. We see that at short period ($<400$ day) the dereddened colours are flat with period, and the sources get significantly redder beyond $400$ day period due to circumstellar dust \citep{Whitelock2008}. There is a small offset in the $(K_s-[5.8])_0$ colours of our low-period ($\lesssim400\,\mathrm{day}$) Mira variables with respect to the O-rich LMC Mira variables possibly related to metallicity effects or underestimates of the interstellar extinction. The derived expansion velocities and mass loss rates have a similar structure to the period--colour diagrams where the distributions are relatively flat for $<400\,\mathrm{day}$ (at around the $3\times10^{-7}M_\odot\,\mathrm{year}^{-1}$ level for the mass loss) before evolving rapidly for larger periods. The mass loss rate for the longest period Mira variables reaches $\sim2\times10^{-5}M_\odot\,\mathrm{year}^{-1}$.

\subsection{Period and age distribution}
\begin{figure}
    \centering
    \includegraphics[width=\columnwidth]{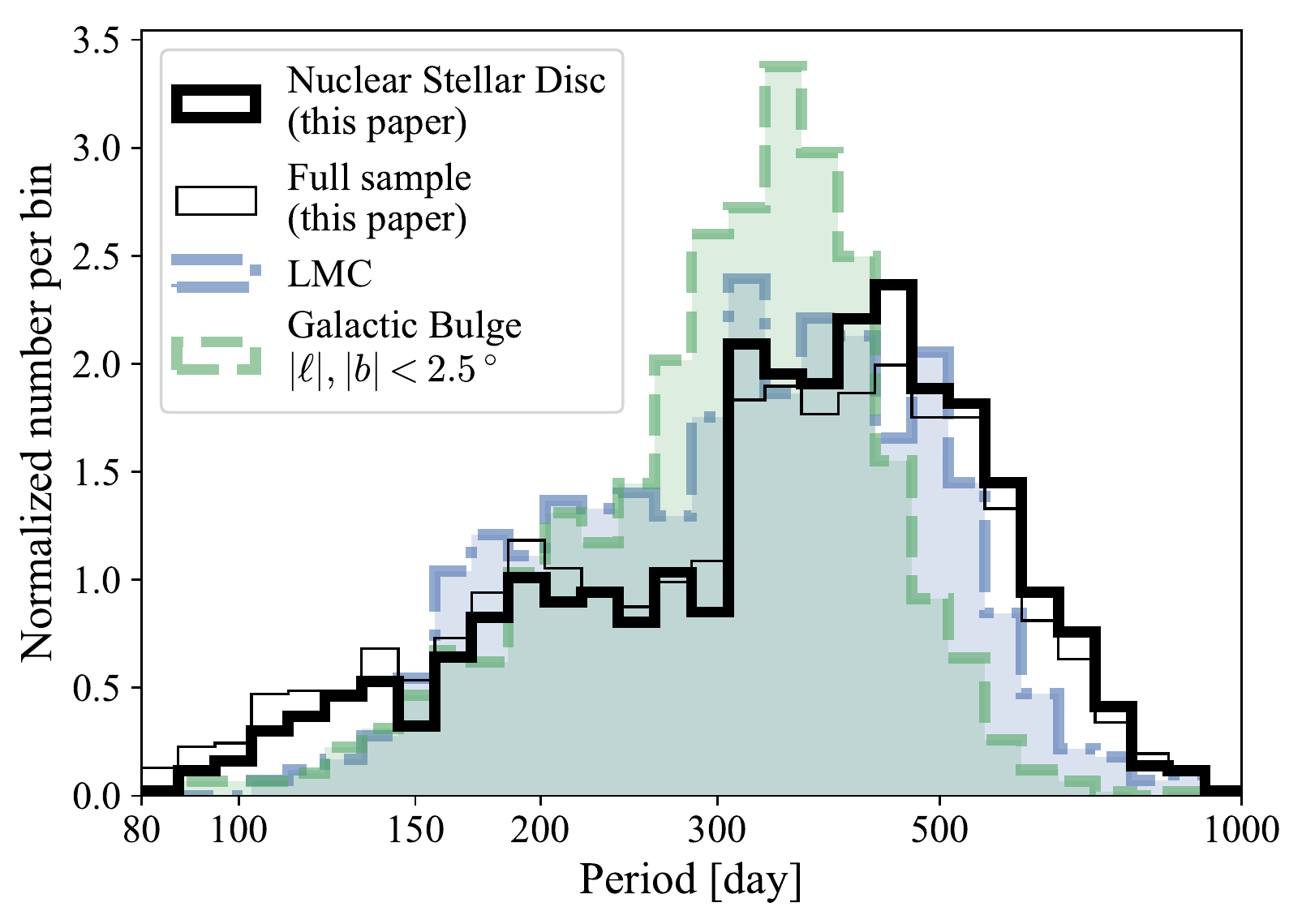}
    \caption{The distribution of periods of our Mira variables (the thick black line shows reliable sources with $|b|<0.4\,\mathrm{deg}$ and $\Delta K_s>0.4$ whilst the thin black line shows the full reliable sample). We also display the period distributions of the LMC Mira variables (blue dot-dash) and those in the central Galactic bulge from Gaia and OGLE ($|\ell|<2.5\,\mathrm{deg}$, $|b|<2.5\,\mathrm{deg}$, green dash).}
    \label{fig:period_distribution}
\end{figure}

We close our investigation of the sample by returning to the main purpose of investigating Mira variables in the NSD -- what is their age distribution and what does this tell us about the formation epoch of the Galactic bar?
In Fig.~\ref{fig:period_distribution} we show the period distribution of our sample alongside that of the LMC Mira variables and a reference Galactic bulge sample formed from all Gaia and OGLE Mira variables within a $5\,\mathrm{deg}\times5\,\mathrm{deg}$ box centred on the Galactic centre. We observe that the overall shapes of the distributions are quite similar with an increase in Mira variables around $300$ days. However, the Mira sample presented here has a broader long and short period tail suggesting more very young and very old stars than the LMC and the inner bulge. The LMC has a slightly broader long period tail than the inner bulge consistent with more recent star formation. 

\begin{figure*}
    \centering
    \includegraphics[width=\textwidth]{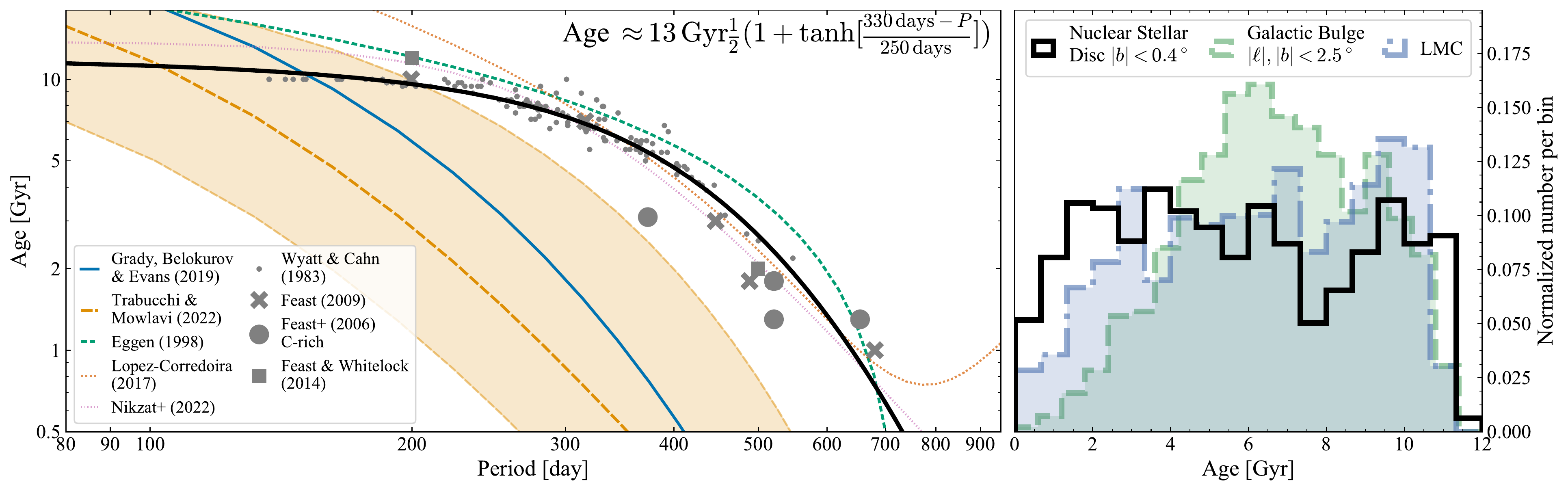}
    \caption{The age distribution of the nuclear stellar disc: the left panel shows the theoretical period--age relations whilst the right panel shows the derived age distributions assuming an `average' of the theoretical results of age $\approx 13\,\mathrm{Gyr}\tfrac{1}{2}(1+\tanh((330-P/\mathrm{day})/250)$. The black distribution corresponds to those reliable Mira variables presented here in the midplane $|b|<0.4^\circ$ with $\Delta K_s>0.4$, the blue distribution corresponds to Mira variables in the LMC and the green distribution corresponds to Mira variables in the inner Galactic bulge taken from OGLE and Gaia with $|\ell|<2.5^\circ$ and $|b|<2.5^\circ$. The long orange-dashed line is from table B1 of \protect\cite{Trabucchi2022} and the corresponding orange shaded area gives the region over which their model density is $>25\percent$ of the maximum.}
    \label{fig:age_dist}
\end{figure*}

As noted in the introduction, there are relatively few studies of the period--age relation for Mira variables. Most studies are based on empirical period-kinematic calibrations from \cite{FeastWhitelock1987}, \cite{FeastWhitelock2000}, \citet[][based on C-rich stars]{Feast2006}, \cite{Feast2009} and \cite{FeastWhitelock2014} that are approximately calibrated against the age--kinematic relations observed in the solar neighbourhood. \cite{WyattCahn1983}, \cite{FeastWhitelock1987} and \cite{Eggen1998} provide more theoretical investigations into the period--age relation, and produce relations that agree well with the kinematic calibrations (see Fig.~\ref{fig:age_dist}). Both \cite{LopezCorredoira2017} and \cite{Nikzat2022} have fitted the collection of these calibrations with simple analytic forms (also shown in Fig.~\ref{fig:age_dist}). There is a more recent theoretical investigation from \cite{Trabucchi2022} using the non-linear pulsation computations from \cite{Trabucchi2019}. Their calibration predicts significantly shorter periods at fixed age than the kinematically-calibrated relations (see Fig.~\ref{fig:age_dist}). Interestingly, this agrees well with the calibration from \cite{Grady2019} that was based upon Mira variables in LMC and Milky Way clusters. The study of \cite{Trabucchi2022} also highlights that there is a significant spread in age at each period as the Mira variables undergo thermal pulsations. Utilising this latter relation for stars in the Galactic bulge \citep{Catchpole2016} would mean the Galactic bulge contains significant populations of stars with ages $\lesssim1\,\mathrm{Gyr}$. The bulge is typically considered to be composed of old stars \citep{Zoccali2003,Bovy2019,Hasselquist2020} although there is evidence for a small number of younger ($\sim5\,\mathrm{Gyr}$) stars \citep{Bensby2013,Bernard2018}. Possibly there are complications related to the period--metallicity relation \citep{FeastWhitelock2000P}. However, some old globular clusters e.g. NGC 5927 \citep[$\gtrsim10\,\mathrm{Gyr}$,][]{Dotter2010,VandenBerg2013} have Mira variables with $\sim300\,\mathrm{day}$ periods \citep{Feast2002} suggesting $<300\,\mathrm{day}$ Mira variables are associated with $\gtrsim8\,\mathrm{Gyr}$ old populations. Based on these considerations we use the kinematically-calibrated results. In Fig.~\ref{fig:age_dist} we display a simple analytic fit to the period--age ($P$--$\tau$) results of
\begin{equation}
    \tau = 13\,\mathrm{Gyr}\frac{1}{2}\Big(1+\tanh\Big[\frac{330\,\mathrm{day}-P}{250\,\mathrm{day}}\Big]\Big).
\end{equation}
Using this relation, we display the age distribution of our Mira variable sample (restricting to those reliable stars with $|b|<0.4$ and amplitudes $\Delta K_s>0.4$), compared to the inner Galactic bulge sample and the LMC O-rich sample. We see that the NSD sample contains significant numbers of long-period stars indicative of recent star formation \citep{Morris1996}. There are also significantly older stars than in the Galactic bulge sample although this may reflect incompleteness in the Gaia/OGLE samples at the short period end due to extinction. An interesting feature of the age distribution is the lack of stars around $8\,\mathrm{Gyr}$. Visually this looks similar to the models of \citet[][their figure 5]{Baba2019} where an old nuclear bulge is included giving rise to an old peak before a younger peak due to the NSD formation. Another interpretation is that the NSD formation is contributing to the older peak and perhaps the younger peak is a secondary burst due to heightened gas accretion, or perhaps due to bar destruction and reformation at an early epoch. Either way, the distribution suggests an old ($\gtrsim8\,\mathrm{Gyr}$) bar formation. Our sample is naturally contaminated with foreground Galactic bulge stars \citep[e.g. the models of][suggest the relative on-sky density of bulge and NSD stars is only $\gtrsim1$ for $|b|<0.4\,\mathrm{deg}$]{Sormani2022} so it is difficult to conclude which stars are genuine NSD members and we reserve their full investigation to a follow-up publication.

\section{Conclusions}\label{section::conclusions}
We have described a methodology for discovering Mira variable stars from the VVV multi-epoch infrared data and presented a sample of $1782$ Mira variable candidates across the NSD of the Milky Way. Our study was motivated by Mira variables being bright intermediate-age indicators that have the potential to characterise the NSD's star formation history and in turn the epoch of bar formation in the Milky Way. 

We have demonstrated that although our sample spatially traces the NSD, it is subject to significant selection effects. In the absence of extinction, Mira variables at the Galactic Centre distance are too bright for VVV so we are only able to study Mira variables in highly-extincted regions. However, the completeness of the sample is in excellent agreement with results from artificial star tests so the impact of the selection effects can be modelled. Furthermore, we have demonstrated that:
\begin{enumerate}
    \item the sample is dominated by stars with oxygen-rich chemistry ($\gtrsim90\percent$),
    \item the short-period ($\lesssim400$ day) sample follows period--luminosity relations that locate the stars approximately at the distance of the Galactic Centre with period--Wesenheit relations based on $K_s$, $[3.6]$ and $[4.5]$ appearing to produce the most reliable distance measurements,
    \item longer period variables have significant circumstellar dust producing increasingly red colours and high mass loss rates (up to $\sim2.5\times10^{-5}M_\odot\,\mathrm{yr}^{-1}$) and fall well under the O-rich solar neighbourhood and O-rich LMC $K_s$ period--luminosity relations suggesting large quantities of circumstellar dust,
    \item and finally the age distribution shows two peaks at $10\,\mathrm{Gyr}$ and $\lesssim8\,\mathrm{Gyr}$ either of which we could tentatively associate with NSD formation although the contamination by bulge stars is expected to be significant.
\end{enumerate}

We have only briefly touched upon what is possible using this catalogue. The VIRAC2 reduction of the VVV photometry also provides proper motion measurements for all stars. Although, as we have evidenced, the selection effects of our catalogue are well understood, modelling of kinematic data is often significantly simpler than modelling of spatial data as proper motion observations of bright stars are typically not subject to significant selection effects. This makes identifying NSD membership more straightforward. This will be the subject of the second paper on this sample. We also envisage the catalogue being useful for broader searches and studies of Mira variable stars. Indeed, an earlier version of the catalogue presented here has already been used as part of a long-period variable training set by \cite{Molnar2022}. 

\section*{Acknowledgements}
We thank the referee for their insightful comments that helped improve the presentation of the results.
JLS acknowledges the support of the Royal Society (URF\textbackslash R1\textbackslash191555) and the Leverhulme and Newton Trusts. 
D.M. gratefully acknowledges support by the ANID BASAL projects ACE210002 and FB210003, and Fondecyt Project No. 1220724. We thank Fatemeh Nikzat for their useful comments.
This paper made used of the Whole Sky Database (wsdb) created by Sergey Koposov and maintained at the Institute of Astronomy, Cambridge by Sergey Koposov, Vasily Belokurov and Wyn Evans with financial support from the Science \& Technology Facilities Council (STFC) and the European Research Council (ERC). This paper made use of
\textsc{numpy} \citep{numpy},
\textsc{scipy} \citep{scipy}, 
\textsc{matplotlib} \citep{matplotlib}, 
\textsc{seaborn} \citep{seaborn} and
\textsc{astropy} \citep{astropy:2013,astropy:2018}. We acknowledge the use of Jake VanderPlas' NFFT package (\href{https://github.com/jakevdp/nfft}{https://github.com/jakevdp/nfft}). This work has made use of data from the European Space Agency (ESA) mission
{\it Gaia} (\url{https://www.cosmos.esa.int/gaia}), processed by the {\it Gaia}
Data Processing and Analysis Consortium (DPAC,
\url{https://www.cosmos.esa.int/web/gaia/dpac/consortium}). Funding for the DPAC
has been provided by national institutions, in particular the institutions
participating in the {\it Gaia} Multilateral Agreement. Based on data products from observations made with ESO Telescopes at the La Silla or Paranal Observatories under ESO programme ID 179.B-2002. 
This research has made use of the International Variable Star Index (VSX) database, operated at AAVSO, Cambridge, Massachusetts, USA. This research is based on observations with AKARI, a JAXA project with the participation of ESA. This research has made use of the SVO Filter Profile Service (\url{http://svo2.cab.inta-csic.es/theory/fps/}) supported from the Spanish MINECO through grant AYA2017-84089.

\section*{Data Availability}
The resulting catalogue of Mira variables will be made available via Vizier (temporary link available at \url{https://www.homepages.ucl.ac.uk/~ucapjls/data/mira_vvv.fits}).


\bibliographystyle{mnras}
\bibliography{bibliography} 



\appendix

\section{Time series modelling}\label{appendix::period}
To extract a sample of Mira variables from the VVV VIRAC2 data, we use both standard periodogram methods as well as more flexible Gaussian process methods. In this appendix, we describe in detail our implementations of these methods. We describe our methods in terms of a general light curve of $N$ magnitudes $\bs{y}$ with covariance matrix $\Sigma$ (where we assume uncorrelated errors so $\Sigma=\bs{\sigma}^2\mathsf{I}$) at times $\bs{t}=(t_1,t_2,\cdots,t_N)$.

\subsection{Periodogram methods}
The standard method for characterising the periodicity of a variable source is using the Lomb-Scargle method \citep{Lomb1976,Scargle1982}, which has been generalised to account for a floating mean \citep{Zechmeister2009}. As discussed by \cite{Vanderplas2018}, this method is equivalent to fitting the first three terms of a Fourier series $K_s(t)=
a_0+a_1\sin(\omega t)+b_1\cos(\omega t)$ and so performs well for near-sinusoidal light curves. For non-sinusoidal, but periodic, light curves with long-term trends (as observed in some Mira variables), a general Fourier series plus polynomial of the form 
\begin{equation}
    K_{s}(t)
    =a_0+\sum_{k=1}^{N_\mathrm{F}}\Big[ a_k\sin(k\omega t)+b_k\cos(k\omega t)\Big]+\sum_{k=1}^{N_\mathrm{P}}c_k(t-t_0)^k,
\label{eqn::fourier}
\end{equation}
can be fitted to find $a_k, b_k$ and $c_k$ using linear least-squares \citep{Palmer2009}. Here $t_0$ is an arbitrary zeropoint not fitted for. For computational speed, non-equispaced Fast Fourier transforms (NFFT) are used to evaluate the resulting sums over trigonometric functions \citep[times polynomial terms,][]{PressRybicki,NFFT}. These operate by `extirpolating' the non-equispaced data onto a regular grid using sets of basis functions e.g. Lagrange polynomials \citep{PressRybicki} or a set of localised Gaussians where sparsity can be enforced as a further approximation \citep{NFFT}. We use Jake VanderPlas' implementation of the algorithms presented in \cite{NFFT}\footnote{\url{https://github.com/jakevdp/nfft}}. To avoid overfitting higher order Fourier terms, regularization is introduced by including a term $\bs{\theta}^T\bs{\Lambda}\bs{\theta}$ in $\chi^2$ where $\bs{\theta}$ is the parameter vector, $\bs{\theta}^\mathrm{T}=(a_0,a_1,b_1,a_2,b_2,\cdots)$
\citep{Vanderplas2015}. We choose $\bs{\Lambda}=\lambda\mathsf{Tr}\Sigma^{-1}\bs{k}^2=\lambda\mathsf{Tr}\Sigma^{-1}(0,1,1,4,4,\cdots)$, which minimises the curvature of the Fourier fits. As a default, we set $\lambda=0.01$. No regularization is employed for the polynomial coefficients $c_k$.

When finding the best-fitting period, we use a regular grid of test frequencies from the \textsc{astropy} \texttt{autofrequency} routine which implements the guidelines in \citet[][\citealt{astropy:2013,astropy:2018}]{Vanderplas2018}. The best-fit period is that with the lowest $\chi^2$ and the uncertainty is estimated by the local curvature of $\chi^2$. The covariance of the Fourier coefficients are found by summing the covariance at the best period with the covariance from linear propagation of the period error (calculated by finite-differencing the Fourier coefficients on the period grid).

\subsection{Gaussian process models}
For quasi-periodic light curves such as those of Mira variables, we require a more flexible fitting procedure that can account for stochastic variations. Several authors \citep{He2016,Zinn2017,Yuan2018} have demonstrated the power of Gaussian process models for Mira variable light curves. A Gaussian process is a prior on the space of functions $y(\bs{x})$, where a finite set of $N$ random variables $\{y_i\}$ with uncertainties $\{\sigma_i\}$ indexed by $D$-dimensional points $\{\bs{x}_i\}$ is distributed as a multidimensional Gaussian distribution \citep{RasmussenWilliams2006}. The Gaussian process is characterised by a kernel $\mathsf{K}_{ij}\equiv k(\bs{x}_i,\bs{x}_j)$, which describes the degree of correlation between points. Here we will use stationary kernels i.e. those that depend only on the Euclidean distance $\tau_{ij}=|\bs{x}_i-\bs{x}_j|$. Kernel parameters are optimally set by maximising the likelihood of $\{y_i\}$ given $\{\bs{x}_i\}$.

Despite their simplicity, Gaussian processes are ill-suited for large datasets due to the requirement of performing the $\mathcal{O}(N^3)$ matrix inversion. However, for one-dimensional data \cite{Kelly2014}, \cite{Ambikasaran2015} and \cite{ForemanMackey2017} have demonstrated that for kernels composed of sums of stationary exponential kernels, the kernel matrix $\mathsf{K}$ is semi-separable and the matrix inversion can be evaluated with an $\mathcal{O}(N)$ algorithm. The algorithm is implemented in the \textsc{celerite} python package \citep{ForemanMackey2017}. In the \textsc{celerite} scheme, the kernel must be positive-definite and of the form
\begin{equation}
    k(\tau) = \sum_{j=1}^J \frac{1}{2}\left[
    (a_j + i\,b_j)\,e^{-(c_j+i\,d_j)\,\tau} +
    (a_j - i\,b_j)\,e^{-(c_j-i\,d_j)\,\tau}
\right],
\end{equation}
where $a_j$, $b_j$, $c_j$ and $d_j$ are real numbers and $\tau$ the separation between points. Positive definiteness is ensured by requiring $|b_j d_j| < a_j c_j$ for all $j$. \cite{ForemanMackey2017} show how a number of popular kernels can be approximated with this kernel. In particular, both damped random walk (or the Ornstein-Uhlenbeck kernel) $k_\mathrm{DRW}(\tau)$ ($b_j=d_j=0$) and (quasi-)periodic kernels are permitted. For periodic kernels, \cite{ForemanMackey2017} suggest working with the damped harmonic oscillator kernel $k_\mathrm{SHO}(\tau)$ described by the power spectral density
\begin{equation}
    S(\omega) = \sqrt{\frac{2}{\pi}}\frac{S_0\omega_0^4}{(\omega^2-\omega_0^2)^2+\omega_0^2\omega^2/Q^2},
\end{equation}
where $S_0$ is related to the amplitude of oscillation, $\omega_0$ the frequency and $Q$ is the quality factor describing the damping. These parameters are related to $a_j$, $b_j$, $c_j$ and $d_j$.

We have found it useful to work with a kernel of the form
\begin{equation}
    k(\tau) = \sum_i^{N_\mathrm{O}}k_{\mathrm{SHO},i}(\tau)k_{\mathrm{DRW},i}(\tau)+\sum_j^{N_\mathrm{E}}k_{\mathrm{DRW},j}(\tau).
\label{equation::kernel}
\end{equation}
The first term gives a sum of damped harmonic oscillators -- we choose to multiply each SHO term by a DRW and initialize $Q$ as a large value. Although the DRW multiplying the SHO duplicates the parameters in the SHO we have found it helps with convergence of the models. The second term is a pure random walk.
We additionally include a white noise term in the kernel so $\mathsf{K}\leftarrow \mathsf{K}+\sigma_w^2\mathsf{I}$.
Such a model is appropriate for Mira variables as it encompasses (i) multiple periodic signals, (ii) slow changes in the period and (iii) stochastic changes. For small datasets, \cite{He2016} opt for a hybrid Fourier and Gaussian process method where the residuals with respect to a Fourier series fit are described by a Gaussian process with a DRW kernel. We find this method gives similar results to only using a Gaussian process with a SHO plus DRW kernel, but is more sensitive to the initial choice of period and does not allow the small changes in the period (without employing a more complex kernel).

\subsection{Multi-band light curves}\label{appendix::2dLC}
For the modelling of multi-band light curve data, the Gaussian process models can be generalized to include effective wavelength as a second dimension by using a 2D kernel \citep[e.g.][for supernova light curve modelling]{Fakhouri2015}. This method produces correlations in the phase across different photometric bands. However, we then lose the ability to use the rapid algorithms such as \textsc{celerite}, applicable for 1D. For higher dimensionality data, \cite{george} have described how covariance matrices can be inverted approximately by hierarchical factorisation in $\mathcal{O}(N\log^2N)$ time.
However, 
the overheads become costly for periodic kernels. A cheaper method is to combine all the individual photometric bands into a single light curve using a band-dependent scale and shift. The combined light curve can then be considered as a 1D problem and modelled with a \textsc{celerite} kernel \citep{celerite2d}. The disadvantage of this procedure is we are unable to introduce a smooth correlation scale in the wavelength dimension to perhaps reflect the fact that variability in close bands are related. 

We utilise the same kernel as given in equation~\eqref{equation::kernel} and maximise the likelihood to find the kernel parameters additionally fitting for the white noise $\bs{\sigma_w}$, the mean $\bs{\mu}$ and scaling $\bs{\alpha}$ (relative to the $K_s$ band) in each photometric band. Furthermore, we place priors on the scaling amplitudes $\bs{\alpha}$ based on the O-rich amplitude ratios derived from synthetic spectral energy distributions from fits to multi-band light curves by \citet[][reported relative to $I$ band amplitude]{Iwanek2021}. We use their table 1 to find the mean $\mathrm{amp}(x)/\mathrm{amp}(K_s)$ and variance $\sigma^2_{\mathrm{amp}(x)/\mathrm{amp}(K_s)} = (\mathrm{amp}(I)/\mathrm{amp}(K_s))^2\Big(\sigma^2_{\mathrm{amp}(x)/\mathrm{amp}(I)}+(\mathrm{amp}(x)/\mathrm{amp}(K_s))^2\sigma^2_{\mathrm{amp}(K_s)/\mathrm{amp}(I)}\Big)$. Due to the uncertainty in  $\mathrm{amp}(K_s)/\mathrm{amp}(I)$ from \cite{Iwanek2021} this is quite a generous prior.

\section{Blended photometry}\label{appendix::blended}
\begin{figure}
    \centering
    \includegraphics[width=\columnwidth]{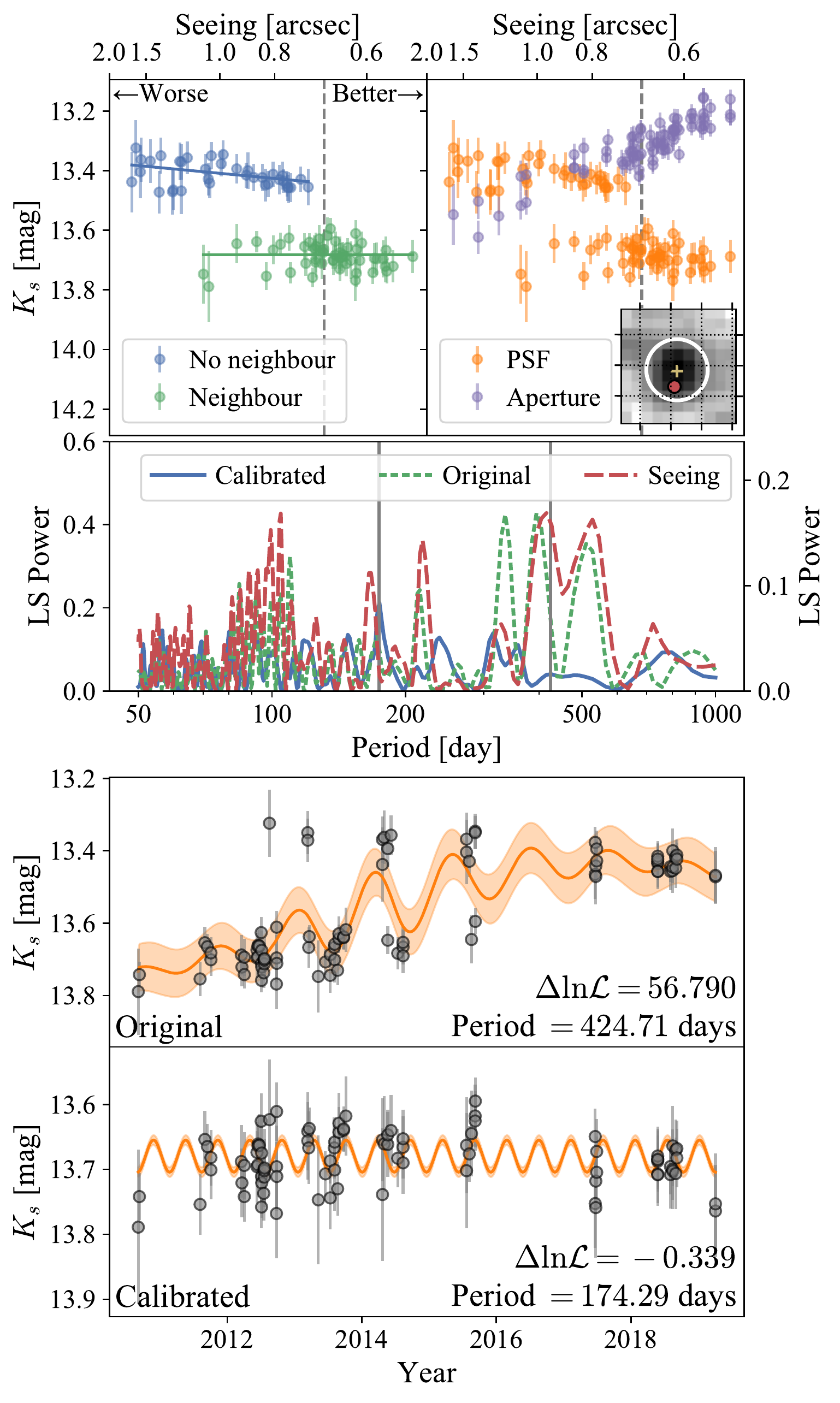}
    \caption{Example of a light curve for a blended source showing spurious variability and periodicity. The \emph{top two panels} show the variation of magnitude with (inverse) seeing: in the \emph{left panel} we colour the points by whether a neighbour (green) or no neighbour (blue) is detected within $1\,\mathrm{arcsec}$ at the same epoch (see small deep-stack image inset of source [yellow cross] and neighbour [red circle] with a white $1\,\mathrm{arcsec}$ circle) and display the mean trend models adopted; the \emph{right panel} shows the variation of PSF (orange and corresponding to all points in the left panel) and aperture (purple) photometry with seeing. The \emph{second panel} shows periodograms of the calibrated light curve (blue solid), original light curve (green short-dashed) and seeing (red long-dashed) (note the seeing periodogram corresponds to the right axis). The \emph{bottom two panels} show the light curves with Gaussian process models where the two periods are marked on the periodogram panel.}
    \label{fig:blending}
\end{figure}

In our early Mira variable candidate lists, we found significant numbers of variable sources clumped in period and with on-sky distributions tracing the VIRCAM footprint. One of these sources is shown in Fig.~\ref{fig:blending}, which illustrates the following discussion and which we will discuss in full below. The most notable of these spurious sources had periods of $100$ or $\sim215$ days (and its multiple at $\sim430$ days and were concentrated in the region $0<\ell<1.5\,\mathrm{deg}$ and $-1.5\,\mathrm{deg}<b<0$. The likely explanation for a clumping in period is due to aliases arising from the cadence of the observations. However, the periodogram of an example light curve with the magnitudes replaced by noise \citep{Vanderplas2018} did not show peaks at the observed periods. Inspecting some example light curves of these spurious sources revealed that many of these sources had almost `bimodal' light curves, where the magnitude was one of two values. This bimodality corresponded to two distinct sequences in magnitude against seeing (and in turn, magnitude against \textsc{DoPhot} $\chi^2$). For low seeing observations, the magnitudes were consistently fainter. These spurious sources regularly had other VIRAC detections within $1\,\mathrm{arcsec}$, and furthermore, the two distinct magnitude sequences corresponded to the epochs when the neighbour was detected or not. For low seeing, the blended source is resolved and a fainter magnitude assigned, whilst higher seeing results in all the light being attributed to a single source. Curiously, periodicity in the seeing, possibly due to seasonal variations, is then transferred into periodicity of the source. This is confirmed by taking a periodogram of the seeing which has exactly the observed peaks of $100$, $\sim215$ and $\sim430$ days.

The described issues with blended sources are a limitation of the employed photometric reduction procedure. \cite{dophot} demonstrate that both \textsc{DoPhot} and \textsc{DaoPhot} \citep{daophot} exhibit the same systematics with blending, so the problem is likely to occur if we employed a different photometric algorithm. However, a cleaner route around the problem is to use a list of sources from the best seeing epochs to perform forced photometry on the poorer seeing epochs. Furthermore, the source list can be supplemented with the Gaia, DECAPS or Pan-STARRS catalogues. However, here we will adopt a simpler procedure by a posteriori correcting any trends in magnitude with seeing when the source is suspected to be contaminated by blending. It should be said that the problem was more significant in early versions of the VIRAC2 catalogues that did not include information on whether detections were astrometric outliers. With this information the problem has been somewhat alleviated.

For each candidate variable source, we take the raw light curve and remove those observations with $\chi^2>10$ for $K_s<13.2$ (the basic quality cut described in Section~\ref{Section::Data}). We then check if there are any Gaia, DECAPS or reliable VIRAC sources within $1\,\mathrm{arcsec}$ of the source (for Gaia and DECAPS we require two sources within $1\,\mathrm{arcsec}$ -- the source itself and potential neighbours). Here, reliable means non-duplicate, detected at least $10$ times and detected in at least $20\percent$ of the covering observations as explained in Section~\ref{sec:primary_lcdata}. We then separate the light curve observations into those without neighbours detected at the same epoch, and those with different combinations of neighbours detected (e.g. if two neighbours, there are four groups: 1. no neighbours, 2. neighbour 1, 3. neighbour 2 and 4. neighbours 1 and 2). For each group of observations, we fit a simple Gaussian process model to $1/$seeing against magnitude using an exponential squared kernel and a white noise term \citep[using the basic solver in \textsc{george},][]{george}. The observations are then shifted by the fitted model trend and the overall magnitude is selected as the model value at the $10$th percentile in seeing for the group with the most neighbours detected (provided there are more than four observations in this group, in which case we use the group with the second most neighbours detected etc.). We select the best seeing observations with most neighbours for the overall shift as this is anticipated to produce the most accurate photometry for a source. However, this is a likely source of systematic biases in the magnitudes.

In Fig.~\ref{fig:blending} we show an example blended source in the VIRAC2 catalogue (at (RA,Dec)=$(267.64689,-28.36941)\,\mathrm{deg}$). This source has median magnitude $K_s=13.5$ and a single neighbour $0.52\,\mathrm{arcsec}$ away with $K_s=14.2$. The observations where the neighbour is detected separate clearly in magnitude and seeing from those where no neighbour is detected. In the worse seeing observations, the neighbour is not detected and the magnitude is measured as $\sim13.4\,\mathrm{mag}$, whilst the better seeing observations allow resolution of the two sources and a magnitude measurement of $\sim13.7\,\mathrm{mag}$. The two clumps separate in seeing around the `Nyquist limit' ($2\times$ pixel size) of $0.678\,\mathrm{arcsec}$ for VIRCAM. Interestingly, the aperture photometry shows the opposite trend in photometry with seeing. For poor seeing, more light falls outside the aperture and for good seeing, less light. In both cases, this is not corrected for correctly. The original light curve (third panel) shows the bimodal distribution of magnitudes, which are seemingly well fitted by a periodic signal with $\sim430$ days. This period corresponds to the highest peak in the Lomb Scargle periodogram. The errors introduced by the seeing produce a clear periodicity. This can be seen plainly by the periodogram of seeing which traces the light curve periodogram. Correcting the light curve by the procedure described above removes the $430$ day peak and reduces the amplitude of variability. The model now fits a period of $174$ days which may still be a systematic error (the peak corresponds to the slight wing in the seeing periodogram) but this periodicity is not deemed significant.

\section{Comparison with known Mira periods}\label{appendix:period_comparison}

\begin{figure*}
    \centering
    \includegraphics[height=.29\textwidth]{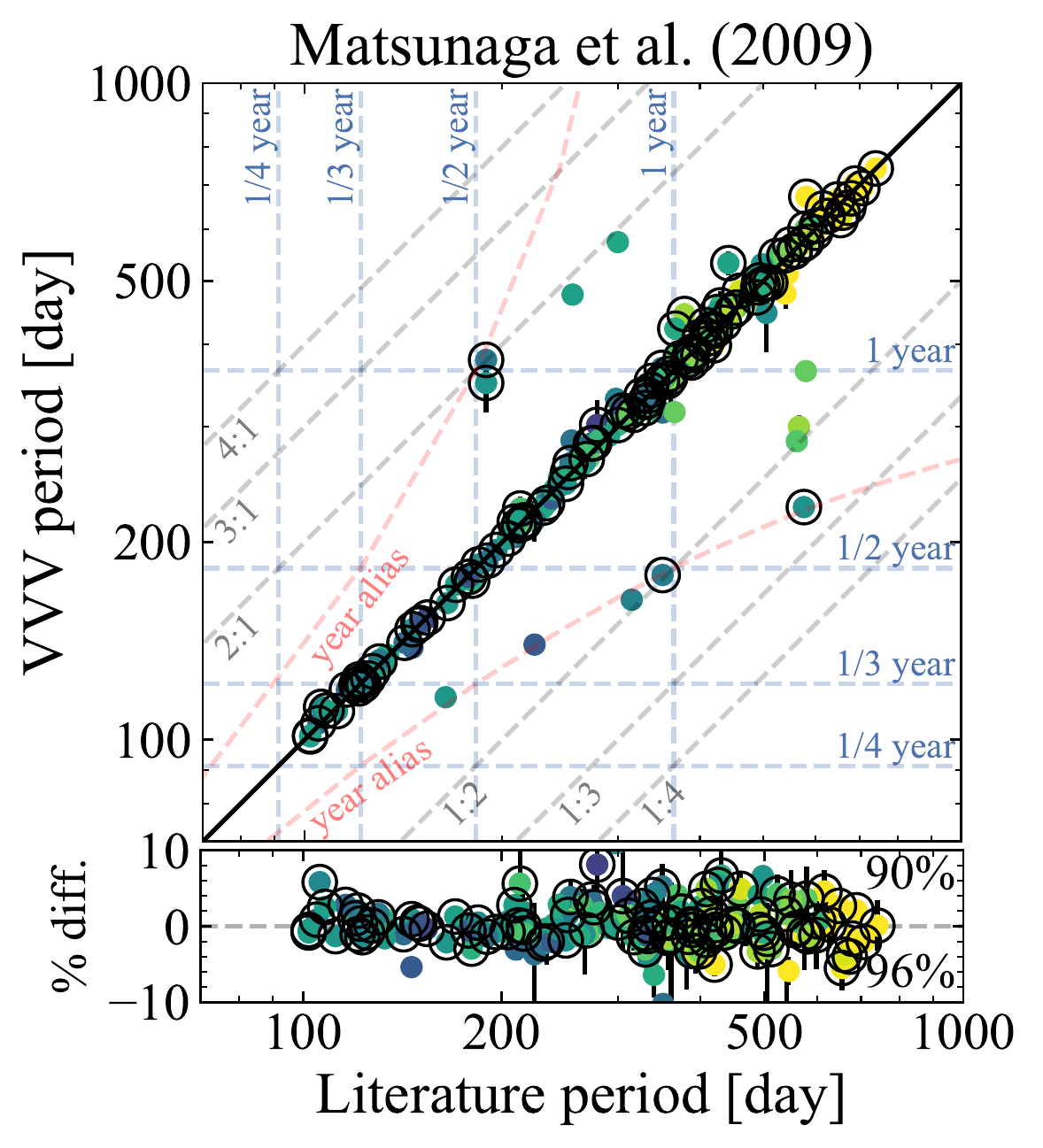}
    \includegraphics[height=.29\textwidth]{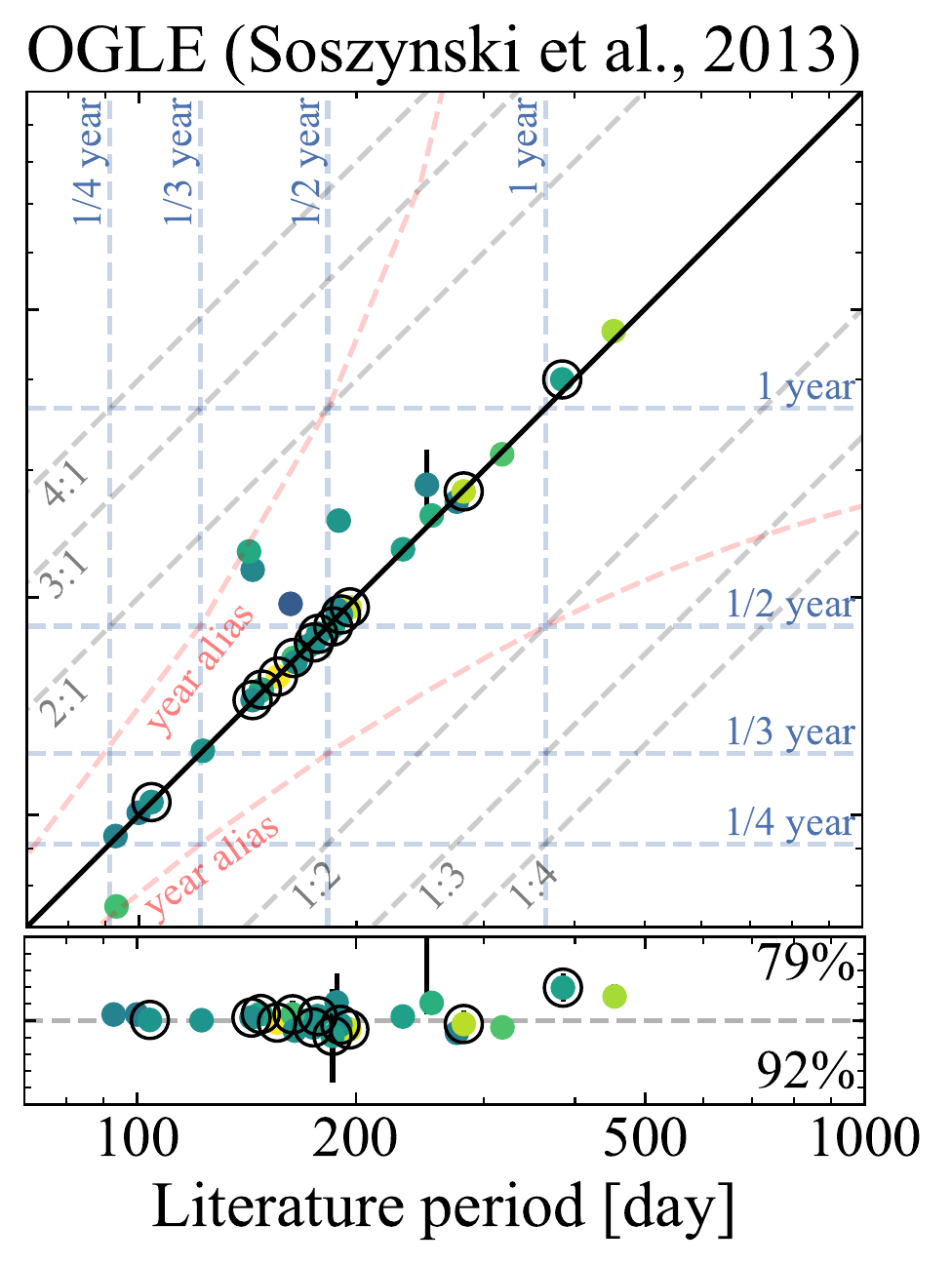}
    \includegraphics[height=.29\textwidth]{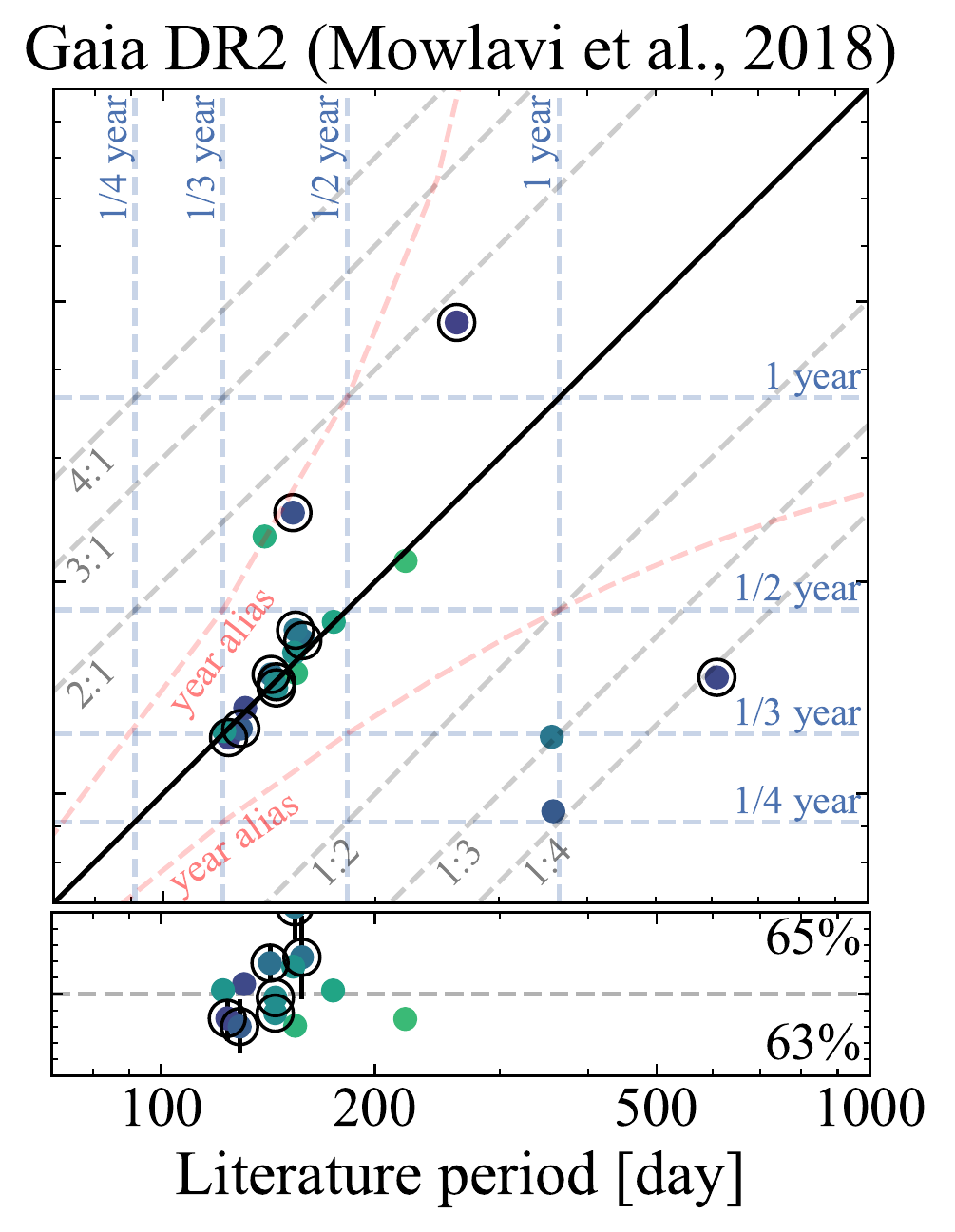}
    \includegraphics[height=.29\textwidth]{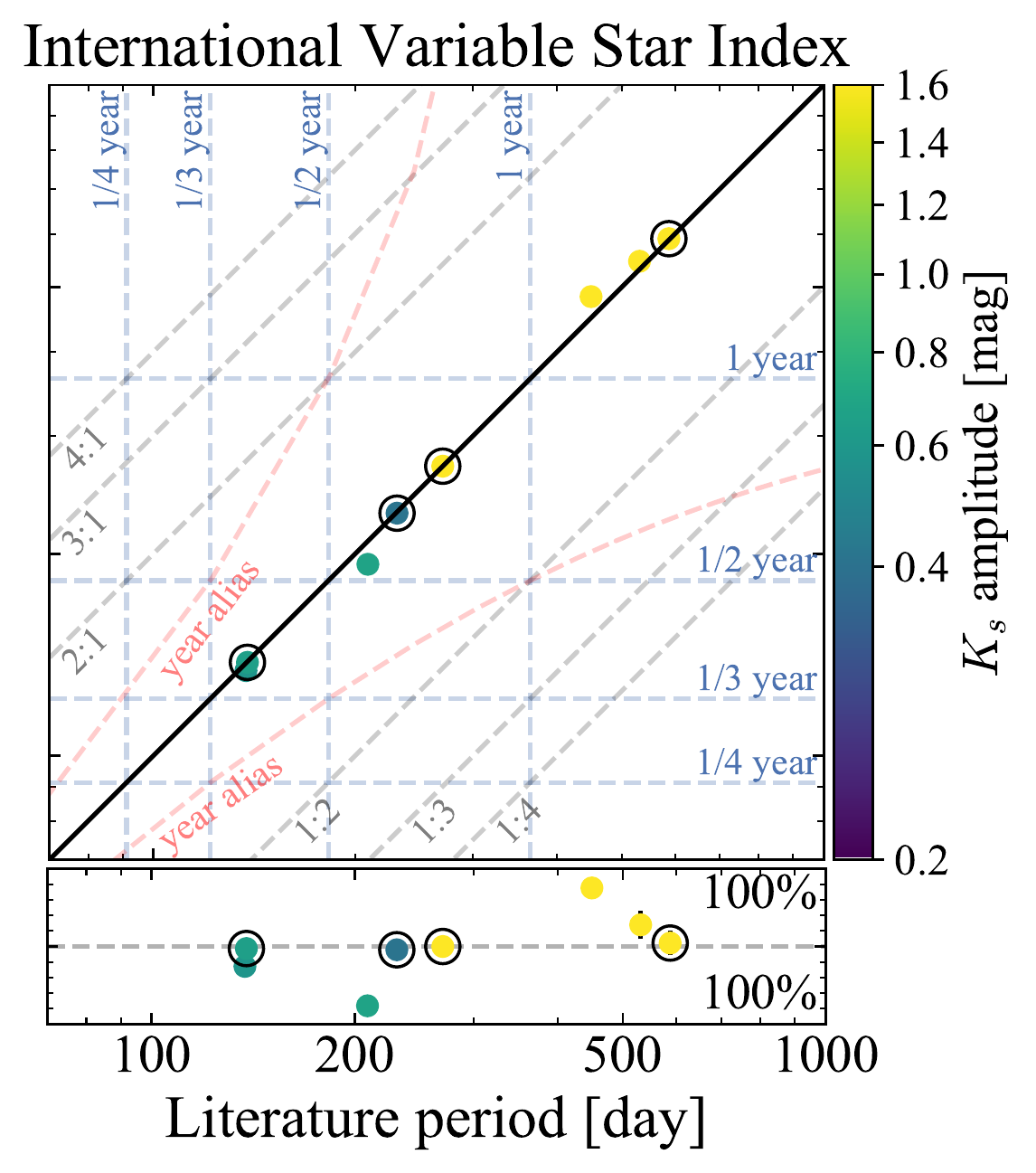}
    \caption{Period recovery using $K_s$-band VVV data for known Mira variables observed by VVV. We show all stars with periods provided by \protect\cite{Matsunaga2009}, OGLE \protect\citep{Soszynski2013}, Gaia \protect\citep{Mowlavi2018} and the International Variable Star Index (VSX), and with more than $20$ valid $K_s$ epochs and a $K_s$ amplitude of $>0.2$ in VVV (circled points have more than $100$ valid epochs). The points are coloured by $K_s$ amplitude. The black solid line is $1:1$ whilst dashed grey lines are $4:1$, $3:1$, $2:1$, $1:2$, $1:3$ and $1:4$. The red dashed lines are $1/(1/\mathrm{Period}\pm1/365\,\mathrm{days})$ -- likely aliases given the VVV observing strategy. Horizontal and vertical blue dashed lines show fractions of a year. The upper (lower) numbers in the lower panels give the percentages of sources with $>20$ ($>100$) $K_s$ epochs for which our period agrees within $25\percent$ with the literature period.}
    \label{fig:period_recovery}
\end{figure*}

We validate our procedure for measuring the periods of Mira variables using those previously discovered and studied. Unfortunately, the VVV overlap with known Mira variables is relatively small. Mira variables are intrinsically very red $(G-K_s)\sim 6$ and those in the bulge can be heavily extincted. This means the Mira variables are typically at the bright end of infrared surveys and the faint end of optical surveys. Non-linearity in the VVV photometry begins brighter than $K_s\sim11.5\,\mathrm{mag}$ and the photometry completely saturates for $\sim9\,\mathrm{mag}$ meaning the bright well-studied Mira variables in low extinction regions are too bright for VVV. Similarly, those Mira variables that are well covered by VVV are typically too faint for the optical surveys. However, a sufficient number overlap for testing purposes, particularly as we are fortunate to have the infrared Mira variable catalogue of \citetalias{Matsunaga2009}. For cross-matching to other catalogues, we employ a cross-match radius of $0.4\,\mathrm{arcsec}$. For OGLE, Gaia and VSX, we also cross-match to 2MASS with a $1\,\mathrm{arcsec}$ radius and ignore the source if 2MASS $K_s<9$ (as it is likely saturated in VVV) or VVV and 2MASS $K_s$ differ by more than $2$ magnitudes. In these cases any VVV cross-match is likely erroneous.

\subsection{\protect\cite{Matsunaga2009} Mira variable catalogue}
\citetalias{Matsunaga2009} used the SIRIUS near-IR camera on the Infrared Survey Facility (IRSF) 1.4 m telescope to image in $JHK_s$ a $20\times30\,\mathrm{arcmin}^2$ area around the Galactic centre. The authors first select sources with $3$ times more variable photometry than the median expectation as a function of magnitude, and then measure the periods by fitting sinusoids in a least squares sense (equation~\eqref{eqn::fourier} with $N_\mathrm{F}=1$) to identify $1364$ long period variable candidates, of which $549$ were assigned periods. The majority of the sources with periods are likely Mira variables, although as acknowledged by \citetalias{Matsunaga2009} the exact definition of a Mira variable from infrared photometry is awkward. For certainty in their analysis, these authors consider stars non-Mira variables if the amplitude in any of $J$, $H$ or $K_s$ is less than $0.4$ although this removes only $\sim10\percent$ of the stars. The \citetalias{Matsunaga2009} sample is most similar to the VVV data so ideal for checking the VVV results, except that (i) the \citetalias{Matsunaga2009} photometry is non-linear brighter than $K_s=9\,\mathrm{mag}$ so many brighter Mira variables are better studied than in VVV and (ii) variability and periodicity is characterised in $J$, $H$ and $K_s$ (although only all three for $\sim20\percent$ of the data) whilst in VVV we must essentially rely on $K_s$ alone (there is more limited coverage from $J$ and $H$ photometry). Of the $549$ long period variables with periods in \citetalias{Matsunaga2009}, we successfully cross-match $212$ in VIRAC2, of which $192$ have more than $20$ valid $K_s$ epochs in their cleaned light curve.

\subsection{OGLE long period variables in the bulge}
The Optical Gravitational Lensing Experiment (OGLE) is a long term variability study of the Magellanic Clouds and the Galactic bulge and disc from mainly $I$ (and some $V$) photometry using the 1.3m Warsaw telescope at the Las Campanas Observatory. The OGLE-II and OGLE-III surveys took place between 1997 and 2009 and produced catalogues of long period variables in both the LMC and the Galactic bulge \citep{Soszynski2009,Soszynski2013}. Periods are measured in the range $5-2000$ days using a $N_\mathrm{F}=3$-term Fourier series least-squares fit which is subtracted and iteratively repeated to measure up to five periods (of which three are reported). From the Galactic bulge catalogue of \cite{Soszynski2013}, $6528$ Mira variables were discovered, the majority of which lie within the VVV footprint. Although the $I$ band is near optimal for surveying unextincted Mira variables, within the Galactic bulge its use becomes limited in high extinction regions ($|b|\lesssim2\,\mathrm{deg}$). However, the dense time-sampling of the OGLE survey makes the period determination highly reliable. Of the $6528$ Mira variables, we successfully cross-match $43$ in VVV, of which $39$ have more than $20$ valid $K_s$ epochs.

\subsection{Gaia DR2 long period variables}

A catalogue of long period variables (LPVs) with periods was produced as part of the Gaia DR2 data release \citep{Mowlavi2018}. Based on an average of $26$ observations per star over $668$ days, LPV candidates were identified using a random-forest classifier trained on photometric attributes, of which red colours and long term variability are the most useful \citep{Holl2018,Rimoldini2019}. A further cut to retain LPVs with $G$-band amplitudes ($5-95\%$) greater than $0.2\,\mathrm{mag}$ was applied, producing a list of $\sim150,000$ candidates. The generalized Lomb-Scargle method was employed to search for periods in the range $10$ to $1000$ days and only those $\sim90,000$ LPVs with periods $>60$ days have published periods. Comparison with all-sky ASAS-SN observations and OGLE observations in the LMC and bulge demonstrated the good recovery of the Gaia periods for overlapping sources although due to the scanning strategy periods around $\sim190$ days and below $120$ days are susceptible to aliasing.

We limit the Gaia DR2 LPV catalogue to likely Mira variables using the cut on $G$-band amplitude from \cite{Grady2019}: $\mathrm{Amp}(G)=\log_{10}(\sqrt{N_\mathrm{obs}}\sigma_{\bar{I}_G}/\bar{I}_G)>-0.55$ (where $\bar{I}_G$ is the mean $G$ flux and $\sigma_{\bar{I}_G}$ its error). \cite{Mowlavi2018} also provide frequency uncertainties $\Delta f$ which we transform to period uncertainties under a linear approximation: $\Delta T = \Delta f/f^2$. Cross-matching to VVV, we find $39$ matches of which $35$ have more than $20$ valid $K_s$ epochs.

\subsection{VSX}
The AAVSO International Variable Star Index \citep{Watson2006} is a compilation of variables initially constructed from the General Catalogue of Variable Stars \citep{GCVS}. It includes all variables from the ASAS-SN Variable Stars Database \citep{Shappee2014,Jayasinghe2018,Jayasinghe2019a,Jayasinghe2019b} We downloaded all variables labelled as type `M' or `M:' (`M:' are uncertain Mira classifications) on 5th August 2020, removed objects from \citetalias{Matsunaga2009}, OGLE-III and Gaia which we have already considered (the full Gaia LPV catalogue was not included in the VSX catalogue version we used), and cross-match to VIRAC2. We successfully cross-match $11$ Mira variables with periods to sources in VVV, of which $10$ have more than $20$ valid $K_s$ epochs.

\subsection{Mira variable period recovery}
In Fig.~\ref{fig:period_recovery} we show the periods from the four literature sources plotted against the periods measured from the VVV data. We employ our Gaussian process method using a range of different initial period guesses and kernels (as described previously). We only show results for light curves with more than $20$ valid $K_s$ epochs and with a $K_s$ amplitude greater than $0.25\,\mathrm{mag}$ (corresponding to our lower selection limit in Fig.~\ref{fig:selection}). We have found that light curves outside these cuts have quite poor period recovery and this motivates our choice of epoch number cut for the full catalogue. From Fig.~\ref{fig:period_recovery}, we find a high fraction of measurements lie on the $1:1$ line, particularly those circled sources with $>100$ valid $K_s$ epochs. Likely alias frequencies for VVV light curves are $n\,\mathrm{year}^{-1}$ ($n=1,2,3,\cdots$, due to the observing strategy), and linear combinations of the true frequency and $1\,\mathrm{year}^{-1}$ i.e. the measured period is $1/(1/P\pm1\,\mathrm{year})$ for true period $P$. Also, we might expect for near sinusoidal Mira light curves, the measured period may be mis-measured as a harmonic of the true period. We see there is a tendency for the VVV period to lie along the alias lines. However, in general the period recovery is very good. We deem periods to agree if they are within $25\percent$ of each other (as this encompasses all sources on the one-to-one in the \citetalias{Matsunaga2009} comparison). We find upwards of $\sim90\percent$ agree with the literature samples for those sources with $>100$ $K_s$ epochs and this decreases to only $\gtrsim80\percent$ for those with $>20$ $K_s$ epochs. For the Gaia DR2 sample the agreement is less good ($\sim65\percent$). Restricting to only those in the bulge region ups this to $\sim77\percent$ possibly. We also see two sources have year periods from Gaia. This may be genuine and then missed by VVV or they could be aliases in Gaia. Only $14$ of the cross-matched DR2 LPVs are in the new Gaia DR3 LPV catalogue from \cite{Lebzelter2022} of which we find $75\percent$ matching periods or $80\percent$ restricting to those with $>100$ $K_s$ epochs. This suggests some of the matched Gaia DR2 LPVs are spurious or possibly contaminating YSOs in the VVV disc region \citep{Mowlavi2018}.

\section{Wesenheit magnitude--period cuts}\label{appendix:wesenheit_period_cuts}
\begin{figure*}
    \centering
    \includegraphics[width=\textwidth]{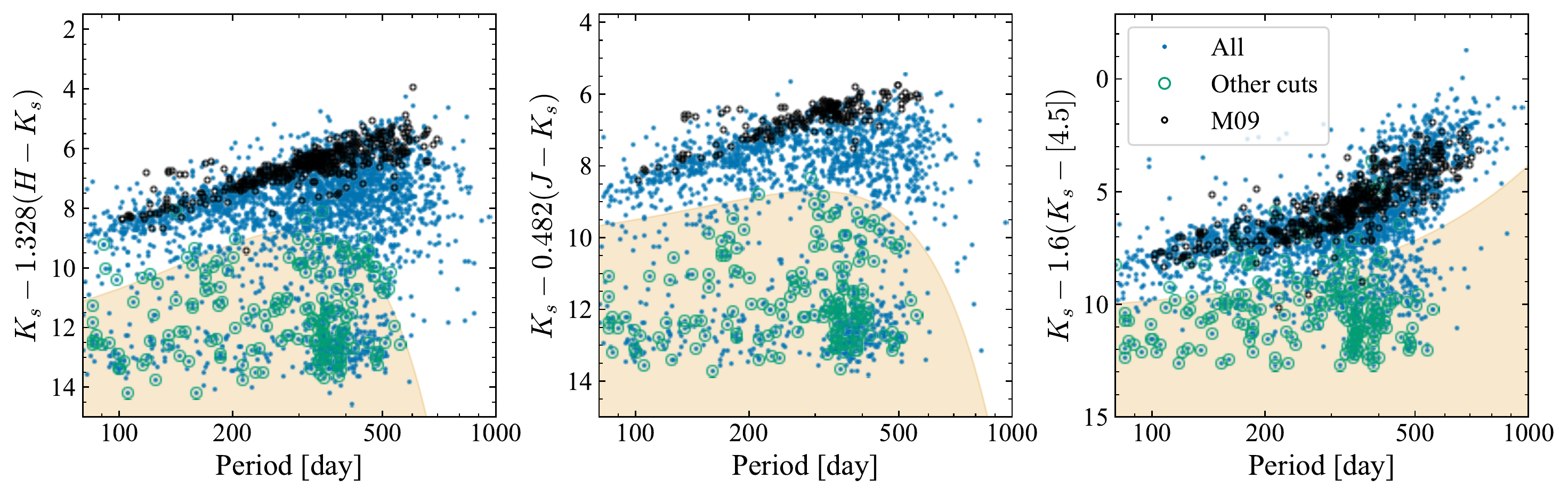}
    \caption{Illustrations of the cuts performed to isolate the likely Mira variable stars. Each panel shows a different Wesenheit magnitude--period distribution. The blue points are the full set of Mira variable candidates without any Wesenheit magnitude--period cuts applied (they still satisfy the period--amplitude, parallax and delta log-likelihood cuts described in Section~\ref{sec::mira_cuts}). The orange shaded regions show the selection of sources we exclude. In each panel, the green circles show the sources that satisfy the other two cuts. The black points are the data from \protect\cite{Matsunaga2009}.}
    \label{fig:wesenheit_period_cuts}
\end{figure*}
In Section~\ref{sec::light_curve_modelling} we employ a series of cuts for isolating reliable Mira variables in our sample. The primary cuts are in delta log-likelihood with respect to a non-periodic model, period--amplitude space (as shown in Fig.~\ref{fig:selection}), parallax signal-to-noise and Wesenheit magnitude--period space (as Mira variables are known to follow period--luminosity relations). In Fig.~\ref{fig:wesenheit_period_cuts} we display the Wesenheit magnitude--period cuts described in Section~\ref{sec::mira_cuts}. There is a population of contaminants around $K_s-1.328(H-K_s)\approx13$ that is particularly prominent around $1\,\mathrm{year}$ periods. These sources appear to fall significantly under the expected Wesenheit magnitude against period relation (as traced approximately by the \citetalias{Matsunaga2009} sources) in all projections.


\bsp	
\label{lastpage}
\end{document}